\documentclass[12pt]{article}
\pdfoutput=1

\usepackage[utf8]{inputenc}

\usepackage{draft}
\usepackage{putex}
\usepackage{xcolor}
\usepackage[all]{xy}

\colorlet{Generic}{black!5}
\definecolor{Displacement1}{RGB}{155, 93, 229}
\definecolor{BrokenRot1}{RGB}{241, 91, 181}
\definecolor{BrokenCurrent1}{RGB}{0, 187, 249}
\definecolor{BrokenSUSY1}{RGB}{254, 228, 64}
\colorlet{BrokenCurrent}{BrokenCurrent1!30!white}
\colorlet{Displacement}{Displacement1!30!white}
\colorlet{BrokenSUSY}{BrokenSUSY1!30!white}
\colorlet{BrokenRot}{BrokenRot1!30!white}
\definecolor{Reject}{RGB}{230, 57, 70}
\definecolor{Accept}{RGB}{29, 53, 87}

\usepackage{amssymb,amsmath,amsthm}

\newtheorem{theorem}{Theorem}

\usepackage{stackrel}

\usepackage[all]{xy}
\usepackage{graphicx}
\usepackage{caption}
\usepackage{amsmath,bm}
\usepackage{array}
\usepackage{subcaption}
\usepackage{epstopdf}
\usepackage{enumerate}
\usepackage{cite}
\usepackage{tensor}
\usepackage{slashed}
\usepackage[utf8]{inputenc}
\usepackage{rotating}
\usepackage{bigfoot}
\usepackage[
colorlinks=true,
linkcolor=blue,
urlcolor=blue,
filecolor=black,
citecolor=red,
]{hyperref}

\usepackage{wasysym}
\usepackage{tikz-cd}

\def\XXint#1#2#3{{\setbox0=\hbox{$#1{#2#3}{\int}$ }
		\vcenter{\hbox{$#2#3$ }}\kern-.6\wd0}}

\numberwithin{equation}{section}

\def\<{\langle}
\def\>{\rangle}
\def\pa{\partial}

\def\ep{\epsilon}

\allowdisplaybreaks[1]

\newcommand{\leftrarrows}{\mathrel{\raise.75ex\hbox{\oalign{%
				$\scriptstyle\leftarrow$\cr
				\vrule width0pt height.5ex$\hfil\scriptstyle\relbar$\cr}}}}
\newcommand{\lrightarrows}{\mathrel{\raise.75ex\hbox{\oalign{%
				$\scriptstyle\relbar$\hfil\cr
				$\scriptstyle\vrule width0pt height.5ex\smash\rightarrow$\cr}}}}
\newcommand{\Rrelbar}{\mathrel{\raise.75ex\hbox{\oalign{%
				$\scriptstyle\relbar$\cr
				\vrule width0pt height.5ex$\scriptstyle\relbar$}}}}

\makeatletter
\def\leftrightarrowsfill@{\arrowfill@\leftrarrows\Rrelbar\lrightarrows}
\newcommand{\xleftrightarrows}[2][]{\ext@arrow 3399\leftrightarrowsfill@{#1}{#2}}
\makeatother

\newcommand{\so}{{\mathfrak{so}}}
\newcommand{\su}{{\mathfrak{su}}}
\newcommand{\psu}{{\mathfrak{psu}}}
\newcommand{\usp}{{\mathfrak{usp}}}
\newcommand{\msp}{{\mathfrak{sp}}}
\newcommand{\msl}{{\mathfrak{sl}}}
\newcommand{\osp}{{\mathfrak{osp}}}
\newcommand{\uu}{{\mathfrak{u}}}

\newcommand{\acceptCirc}{\begin{tikzpicture}[baseline=-\the\dimexpr\fontdimen22\textfont2\relax]
    \node [draw,circle,thick,color=Accept,fill=blue!5,minimum width=0.25 cm] at (0,0){};\end{tikzpicture}}
\newcommand{\sampleCirc}{\begin{tikzpicture}[baseline=-\the\dimexpr\fontdimen22\textfont2\relax]
    \node [draw,circle,thick,color=black,fill=black!5,minimum width=0.25 cm] at (0,0){};\end{tikzpicture}}
\newcommand{\rejectCirc}{\begin{tikzpicture}[baseline=-\the\dimexpr\fontdimen22\textfont2\relax]
    \node [draw,circle,thick,color=Reject,fill=red!5,minimum width=0.25 cm] at (0,0){};\end{tikzpicture}}
\newcommand{\acceptCross}{\begin{tikzpicture}[cross/.style={path picture={ 
  \draw[Reject]
(path picture bounding box.south east) -- (path picture bounding box.north west) (path picture bounding box.south west) -- (path picture bounding box.north east);
}},
baseline=-\the\dimexpr\fontdimen22\textfont2\relax]
      \node [draw,cross,circle,thick,color=white,minimum width=0.25 cm] at (0,0){}; 
\end{tikzpicture}}
\newcommand{\rejectCross}{\begin{tikzpicture}[cross/.style={path picture={ 
  \draw[Reject]
(path picture bounding box.south east) -- (path picture bounding box.north west) (path picture bounding box.south west) -- (path picture bounding box.north east);
}},
baseline=-\the\dimexpr\fontdimen22\textfont2\relax]
      \node [draw,cross,circle,thick,color=white,minimum width=0.25 cm] at (0,0){}; 
\end{tikzpicture}}
\newcommand{\sampleCross}{\begin{tikzpicture}[cross/.style={path picture={ 
  \draw[black]
(path picture bounding box.south east) -- (path picture bounding box.north west) (path picture bounding box.south west) -- (path picture bounding box.north east);
}},
baseline=-\the\dimexpr\fontdimen22\textfont2\relax]
      \node [draw,cross,circle,thick,color=white,minimum width=0.25 cm] at (0,0){}; 
\end{tikzpicture}}
\newcommand{\acceptBoth}{\begin{tikzpicture}[cross/.style={path picture={ 
  \draw[Reject]
(path picture bounding box.south east) -- (path picture bounding box.north west) (path picture bounding box.south west) -- (path picture bounding box.north east);
}},
baseline=-\the\dimexpr\fontdimen22\textfont2\relax]
      \node [draw,cross,circle,thick,color=Accept!60,fill=blue!5,minimum width=0.25 cm] at (0,0){}; 
\end{tikzpicture}}
\newcommand{\rejectBoth}{\begin{tikzpicture}[cross/.style={path picture={ 
  \draw[purple!60]
(path picture bounding box.south east) -- (path picture bounding box.north west) (path picture bounding box.south west) -- (path picture bounding box.north east);
}},
baseline=-\the\dimexpr\fontdimen22\textfont2\relax]
      \node [draw,cross,circle,thick,color=red!60,fill=red!5,minimum width=0.25 cm] at (0,0){}; 
\end{tikzpicture}}
\newcommand{\sampleBoth}{\begin{tikzpicture}[cross/.style={path picture={ 
  \draw[black]
(path picture bounding box.south east) -- (path picture bounding box.north west) (path picture bounding box.south west) -- (path picture bounding box.north east);
}},
baseline=-\the\dimexpr\fontdimen22\textfont2\relax]
      \node [draw,cross,circle,thick,color=black,fill=black!5,minimum width=0.25 cm] at (0,0){}; 
\end{tikzpicture}}

\usetikzlibrary{arrows,shapes,positioning,decorations.markings,calc}
\tikzstyle directed=[postaction={decorate,decoration={markings,
    mark=at position .55 with {\arrow[arrowstyle]{stealth}}}}]
\usetikzlibrary{calc}
\tikzset{stretch/.initial=1}
\usepackage{tkz-euclide}
\usetikzlibrary{shadings}
\newcommand\drawloop[4][]%
   {\draw[shorten <=0pt, shorten >=0pt,#1]
      ($(#2)!\pgfkeysvalueof{/tikz/stretch}!(#2.#3)$)
      let \p1=($(#2.center)!\pgfkeysvalueof{/tikz/stretch}!(#2.north)-(#2)$),
          \n1= {veclen(\x1,\y1)*sin(0.5*(#4-#3))/sin(0.5*(180-#4+#3))}
      in arc [start angle={#3-90}, end angle={#4+90}, radius=\n1]%
   }
\usetikzlibrary{calc}
\tikzset{stretch/.initial=1}
\newcommand{\AxisRotator}[1][rotate=0]{%
    \tikz [x=0.25cm,y=0.60cm,line width=.2ex,-stealth,#1] \draw (0,0) arc (-150:150:1 and 1);%
}

\begin{document}
	
	\preprint{}

 	\institution{HU}{Jefferson Physical Laboratory, Harvard University,
		Cambridge, MA 02138, USA}
	\institution{CMSA}{Center of Mathematical Sciences and Applications, Harvard University, Cambridge, MA 02138, USA}

	\title{
		Classifying Superconformal Defects \quad\quad\quad\quad\quad in Diverse Dimensions
		Part I:  ~~~~~~~~~~~~
		Superconformal Lines
	}

	\authors{Nathan B. Agmon\worksat{\HU}, Yifan Wang\worksat{\HU,\CMSA}}

	\abstract{
	We initiate the classification of unitary superconformal defects in unitary superconformal field theories (SCFT) of diverse spacetime dimensions $3\leq d \leq 6$. Our method explores general constraints from the defect superconformal symmetry, unitarity, as well as consistency conditions of local bulk-defect couplings. Such features are common to all superconformal defects regardless of any Lagrangian description. In particular, modified Ward identities of conserved currents in the presence of the defect induce a distinguished set of conformal primary operators on the defect worldvolume, which includes the universal displacement operator associated with broken translations transverse to the defect. Consistency with the preserved superconformal symmetry and unitarity requires that such operators arrange into unitarity multiplets of the defect superconformal algebra, which in turn leads to nontrivial constraints on what kinds of defects are admissible in a given SCFT. We carry out the analysis explicitly for one-dimensional defects, namely superconformal lines, and leave the study of higher dimensional defects to forthcoming work. Along the way, we determine the superconformal algebras relevant for candidate lines and classify their unitary representations. For the allowed lines, we further investigate supersymmetric deformations induced by local defect operators found in the multiplet analysis. In SCFTs of $d>3$, we find that superconformal lines preserving transverse rotations  (or  sufficient supersymmetry) admit no relevant or marginal deformations. On the other hand, lines in 3d SCFTs have a much richer structure, permitting marginal and sometimes even relevant deformations. Interestingly, certain lines, such as the half-BPS line in 3d $\cN=8$ SCFTs and general lines that break continuous flavor symmetries, are \textit{required} to admit marginal deformations. We also comment on the implications of our results for one-form symmetries in SCFTs.
		}
	\date{}

	\maketitle
	
	\tableofcontents

	\pagebreak

\section{Introduction and Summary}

\subsection{Conformal Field Theory and Defects}

A central problem in theoretical physics is to delineate the space of consistent quantum field theories (QFT).
This is a difficult task since QFTs are generically strongly-interacting and conventional Lagrangian descriptions are often either nonexistent or not useful. To this end, conformal field theory (CFT) provides a powerful framework to classify and characterize universality classes of QFTs. CFTs govern the critical behavior of quantum systems near the fixed points of renormalization group (RG) flow, which often enjoy enhanced conformal symmetry. Many physical questions about QFTs, including their classification, can be addressed by identifying the CFTs that they are connected to by RG flows, and furthermore by understanding the associated observables and deformations thereof. One major advantage of this approach is that it completely bypasses the need for a perturbative Lagrangian description of the RG flow and its associated fixed points, owing to a nonperturbative definition of CFTs. 
The fundamental observables in any CFT are the correlation functions of local operators. They admit an operator-product-expansion (OPE) and are required to obey certain consistency conditions such as conformal symmetry, unitarity, and crossing invariance, which along with the spectrum of local operators define the CFT axiomatically. The general \textit{conformal bootstrap program} explores this framework to rule on the consistency of abstract CFTs beyond kinematic constraints, and in particular seeks to provide bounds on and determine fixed point data such as critical exponents and OPE coefficients. 
There has been lots of exciting progress along this line in recent years (see \cite{Poland:2018epd} and references therein).

There is yet a richer layer of critical phenomena in QFTs that arises from the universal behavior of  quantum systems in the presence of defects which break a part of the spacetime symmetry. This arises naturally in the condensed matter setup, where the defects come from quantum impurities in the system.\footnote{See \cite{1993cond.mat.11054A,Saleur:1998hq} for reviews on relation between quantum impurities and defect CFTs.} One well-known example is the Kondo model \cite{hewson_1993,10.1143/PTP.32.37}, which describes a single magnetic impurity surrounded by an electron gas in a non-magnetic metal (restricted to the s-wave); correspondingly, the critical phase is described by a conformal line defect in the two-dimensional $SU(2)_k$ Wess-Zumino-Witten CFT \cite{Affleck:1990zd,Affleck:1990by,Affleck:1995ge,Bachas:2004sy}. These kinds of defect critical phenomena also play an important role in string theory, where D-branes are described as boundary 
defects (Cardy states) of the worldsheet CFT \cite{Cardy:1989ir,Gaberdiel:2002my,Cardy:2004hm,Recknagel:2013uja}. The D-branes evolve nontrivially under worldsheet RG flows, induced by both boundary and bulk relevant perturbations, and are crucial in understanding both open string and closed string tachyon condensation \cite{POLCHINSKI1989123, COOPER1991132, Schmidhuber:1994bv,Sen:1998sm, Sen:1999mg,Harvey:2000na, Adams:2001sv,Harvey:2001wm,Sen:2002nu,Freedman:2005wx}. More generally, defects in QFT can be defined either as insertions of exponentiated integrals of defect densities $\cL$, which are composed of the \textit{bulk} elementary fields $\phi$, over a worldvolume $\Sigma$,
\ie
\cD = e^{i \int_\Sigma \cL(\phi)}\,,
\label{orderdef}
\fe
or as boundary conditions along $\Sigma$ in the path integral,
\ie
\la \cD \dots \ra = \int \left. D\phi\right|_{\phi(\Sigma)=\phi_0}( e^{i S(\phi)} \dots) \,.
\label{disorderdef}
\fe
The former is often referred to as the order-type defect, a canonical example of which is the Wilson line (loop) \cite{Wilson:1974sk}. The latter is known as the disorder-type defect, which includes the codimension-2 twist (monodromy) defect \cite{Dixon:1986qv,Gukov:2006jk}   and the 't Hooft line (loop) \cite{tHooft:1977nqb}. These two types of defects can be superimposed to define more general defects, such as Wilson-'t Hooft lines \cite{Kapustin:2005py}. Defects can also be modified by coupling them to degrees of freedom localized on their worldvolume $\Sigma$.\footnote{We do not consider \textit{trivial} defects that involve stacking a lower-dimensional CFT on a spacetime submanifold without direct coupling to the bulk CFT.}  Note that it is not uncommon for a single conformal defect to have multiple (UV) descriptions, such as order and disorder defects at the same time.\footnote{For an example of such IR dualities for defects in CFTs, we refer the reader to the discussion about surface operators in 4d $\cN=2$ SCFTs in \cite{Frenkel:2015rda}.}

The inclusion of defects ushers in a new classification program of \textit{extended} QFTs, and their corresponding universality classes by \textit{extended} CFTs (also known as defect CFTs or DCFTs for short). In fact, defects are essential for a complete description of QFTs in general. In particular, both the generators and charged objects for higher-form global symmetries are generally realized by extended defect operators, which play an important role in describing the phase diagram of the theory \cite{Gaiotto:2014kfa}. A familiar example  is given by  Wilson-'t Hooft line operators \cite{Kapustin:2005py} in four-dimensional gauge theories, which are charged under one-form symmetries and act as order parameters for confinement/de-confinement phase transitions. Furthermore, QFTs that are identical at the level of local observables can harbor significantly different defect operators depending on global structures and topological interactions of the theory (e.g. discrete $\theta$-angles) \cite{Aharony:2013hda}.

For simplicity, we focus on defects that are planar and extend  along a subspace $\Sigma$ in the bulk spacetime as $\Sigma=\mR^{1,p-1}\subset \mR^{1,d-1}$ with $1\leq p\leq d-1$.\footnote{The spherical conformal defects with $\Sigma=S^p\subset \mR^d$ preserve an isomorphic conformal subalgebra (after Wick rotation) and can be obtained from the planar defects by inversion.} In the critical phase, the defect preserves a subalgebra of the full $\mf{so}(2,d)$ conformal symmetry associated to the $d$-dimensional bulk fixed point. The preserved defect symmetry includes the conformal subalgebra $\mf{so}(2,p)$ of $\Sigma$, and hence defines a $p$-dimensional conformal defect $\cD$. Note that the commutant of $\mf{so}(2,p) \subset \mf{so}(2,d)$ is the rotation symmetry $\mf{so}(d-p)$ transverse to $\Sigma$, which may or may not be preserved by the defect $\cD$ itself due to either specific bulk-defect couplings or local couplings on the defect.

The presence of conformal defects greatly enlarges the set of observables in CFT, including critical exponents and OPE coefficients intrinsic to the defects. The axiomatic definition of the bulk CFT also has a natural extension that incorporates the defect observables, which leads to bootstrap conditions that tie together both defect and bulk CFT data \cite{Liendo:2012hy,Billo:2016cpy,Gadde:2016fbj,Lauria:2018klo}. By exploring the bootstrap constraints, one can hope to carve out the space of extended CFTs decorated by conformal defects, as well as solve for the defect observables in specific CFTs.\footnote{The $\mZ_2$ twist line defect is the 3d Ising model has been studied from the bootstrap approach in \cite{Billo:2013jda,Gaiotto:2013nva}.}

\subsection{Superconformal Defects and Defect Multiplets}

A large class of nontrivial conformal defects are known in superconformal field theories (SCFT) that preserve a fraction of the bulk supersymmetry, namely the \textit{superconformal defects}.\footnote{For a recent review on (super)conformal defects, we refer the readers to \cite{Andrei:2018die}.} In fact for $d>4$, the only known unitary interacting CFTs  are superconformal.\footnote{In \cite{BenettiGenolini:2019zth}  some evidence was presented for a potential non-supersymmetric 5d CFT obtained from deforming a well-known 5d SCFT of \cite{Seiberg:1996bd}.}
Supersymmetry provides us with extra tools to analyze defects and attack their classification problem. 
In one approach, we can analyze superconformal defects that admit explicit descriptions, either in terms of supersymmetric RG flows connected to a Lagrangian description (such as for Wilson loops in gauge theories), or in terms of supersymmetric defect branes in dual string/M-theory constructions. For these theories, a precision study of critical defect observables is possible in large part due to non-renormalization theorems. An indirect but more general approach is to follow the implications of superconformal symmetry (combining supersymmetry and conformal symmetry), which leads to strong constraints on what kinds of defects can exist in unitary SCFTs. These restrictions often turn out to be quite stringent even at the kinematical level of the putative superconformal algebras and their representation theory. One of the main goals of this paper (and subsequent work) is to systematically explore such constraints, which paves the way for a more complete classification through the dynamical information of the defect SCFT.

The constraining power of superconformal symmetry and its associated unitary representations is well-proven in the absence of defects. Indeed, there is an explicit classification of the superconformal symmetries $\mf{G}_s$ of (local) interacting SCFTs, which exist only between $d=2$ and $d=6$ and with at most 32 fermionic generators (supercharges and superconformal charges) \cite{Nahm:1977tg,Minwalla:1997ka,Cordova:2016emh}. A crucial part of the argument in \cite{Cordova:2016emh} is the synergy between the unitary representations of superconformal algebras and the existence of a local, conserved, traceless stress-tensor $T_{\m\n}$. For instance, requiring that the stress-tensor fits consistently in a unitary representation of $\mf{G}_s$ rules out otherwise admissible superconformal algebras (e.g. $\mf{osp}(8^*|\cN)$ for $\cN>2$ in $d=6$ and $\mf{su}(2,2|\cN)$ for $\cN>4$ in $d=4$).

Our strategy for classifying superconformal defects is similar in spirit to that of \cite{Cordova:2016emh}. The key ingredient giving rise to the nontrivial constraints used in \cite{Cordova:2016xhm,Buican:2016hpb,Cordova:2016emh} is unitarity. Here we rely on a natural extension of the usual notion of unitarity to the defect operator spectrum and the defect OPE. In other words, a unitary conformal defect in a unitary CFT is defined by a collection of bulk and defect operators whose quantum numbers obey unitarity bounds arising from the (super)conformal algebra, and whose correlation functions satisfy reflection-positivity (after Wick rotation to Euclidean signature) and crossing symmetry. In practice, this requires the order-type defect \eqref{orderdef} to come with a real defect density $\cL(\phi$) and the disorder-type defect \eqref{disorderdef} to be specified by boundary conditions in agreement with the reality conditions imposed on the fields.   

While we mostly focus on superconformal line defects in the main text, we present our strategy for the classification of general  superconformal defects. 
We first enumerate superconformal subalgebras $\mf{g}_s$ of the bulk superconformal algebra $\mf{G}_s$ that describe the superconformal symmetry of a putative $p$-dimensional defect $\cD$ in the $d$-dimensional SCFT. As we shall see, the list is rather short due to the rigid nature of superconformal algebras. For example, the fact that $F(4;2)$ is the only 5d superconformal algebra, and cannot be a subalgebra of any other superconformal algebra except itself \cite{DHoker:2008wvd}, implies that there cannot exist any superconformal defects of dimension $p=5$ (i.e. interface or boundary) in 6d SCFTs.

We then proceed to classify unitary representations of $\mf{g}_s$.\footnote{To be more precise, we are interested in unitary \textit{lowest weight} (positive-energy) representations whose scaling dimensions are bounded from below. This follows from the usual condition that the Hamiltonian is bounded from below in a physical QFT as applied to the radially-quantized CFT \cite{Mack:1975je}. In this paper, all unitarity representations are of the lowest weight (positive-energy) type unless  explicitly stated otherwise.} We note that some of these representation theory results have previously been  developed for different purposes and can be found scattered through the literature \cite{Gunaydin:1987hb,Gunaydin:1988kz,Gunaydin:1990ag, Liendo:2016ymz,Bianchi:2017ozk,Bianchi:2018scb,Bianchi:2018zpb,Liendo:2018ukf,Barns-Graham:2018xdd,Dorey:2018klg,Bianchi:2019umv,Gimenez-Grau:2019hez,Dorey:2019kaf,Lee:2019uen,Bianchi:2020hsz}. In this paper, we carry out the analysis systematically and explicitly following the Racah-Speiser (RS) algorithm for each instance of 1d superconformal algebras relevant for line defects.
A crucial difference between our analysis and that of \cite{Cordova:2016emh} is that defects generally do not support local conserved stress-tensors on their worldvolume $\Sigma$. Instead, the role of the bulk stress-tensor multiplet of $\mf{G}_s$ in \cite{Cordova:2016emh} is substituted by the displacement multiplet of $\mf{g}_s$ on the defect $\cD$. 

A general conformal defect $\cD$ breaks translation symmetry orthogonal to the defect worldvolume $\Sigma$, which leads to a modification of the Ward identity of the bulk stress-tensor by operator-valued contact terms on $\Sigma$ \cite{Billo:2016cpy},\footnote{Here we split the flat space coordinates as $x^\m=(x^a,x^i)$ with $a=0,\dots,p-1$ for longitudinal directions to the defect $\cD$ and $i=p,\dots,d$ for transverse directions. We also denote collective $x^a$ by $x_\parallel$ and $x^i$ by $x_\perp$.  }
\ie
&\pa_\m T^{\m i} (x) = -\D_\Sigma (x_\perp) {\mathsf D}^i (x_\parallel) \,.
\label{introward}
\fe
Here ${\mathsf D}^i$ is the displacement operator, named as such because its insertion on the defect $\cD$ effects the infinitesimal change of  $\cD$ under a transverse translation by $P_i$, as can be seen by integrating the above modified Ward identity. The displacement operator ${\mathsf D}^i$ is a universal defect conformal primary whose normalization and transformation properties under other symmetries of the bulk CFT are completely fixed by \eqref{introward}. Similar to how correlation functions of the stress-tensor encode universal quantities such as conformal anomalies (for $d$ even) and more generally central charges (e.g. the stress-tensor two-point function coefficient $C_T$) of the bulk CFT, the correlation functions of  ${\mathsf D}^i$ are related to new conformal anomalies localized on the defect worldvolume $\Sigma$  \cite{Graham:1999pm,Henningson:1999xi,Gustavsson:2003hn,Gustavsson:2004gj,Nozaki:2012qd,Jensen:2015swa,Herzog:2017xha,Herzog:2017kkj,Herzog:2018lqz,Mezei:2018url,Drukker:2020dcz,Drukker:2020swu,Herzog:2020wlo}. More broadly, the Ward identities for other conserved currents in the bulk can also be modified in the presence of the defect and give rise to conformal primaries on its worldvolume. Altogether, the bulk continuous symmetries broken by the defect $\cD$ lead to a distinguished set of defect conformal primaries which will be central to our analysis.

To set the notation straight, we denote the global symmetry algebra of the bulk SCFT by $\mf{G}_f$ so that the full set of conserved currents in the SCFT is characterized by the algebra 
\ie
\mf{G}= \mf{G}_s\oplus \mf{G}_f\,.
\label{bulksym}
\fe
The superconformal defect $\cD$ preserves a subalgebra $\mf{g}\subset \mf{G}$,
\ie
\mf{g}=\mf{g}_s\oplus \mf{g}_b\oplus \mf{g}_f
\label{defsym}
\fe
which contains the superconformal subalgebra $\mf{g}_s\subset \mf{G}_s$ on the defect worldvolume $\Sigma$, a bosonic symmetry algebra $\mf{g}_{b}\subset \mf{G}_s$ generated by (combinations of) R-symmetry and transverse rotations that preserve the defect $\cD$ and commute with $\mf{g}_s$,  and the residual flavor symmetry  $\mf{g}_f \subset\mf{G}_f$  preserved by the defect. Note that from the defect perspective, $\mf{g}_b$ and $\mf{g}_f$ are both thought of as (global) flavor symmetries. 

For a superconformal defect $\cD$, the defect conformal primaries are organized into multiplets of the defect superconformal algebra $\mf{g}_s$ and are related by acting with the supercharges $\cQ$ preserved by $\cD$. They also carry additional quantum numbers from the bosonic symmetries $\mf{g}_b\oplus \mf{g}_f$ that commute with $\mf{g}_s$. Among all the conformal primaries in a unitary superconformal multiplet, there is a \textit{unique}  bottom component of lowest scaling dimension (which transforms in a particular representation of the weight-zero subalgebra of $\mf{g}_s$), namely the superconformal primary. At the other end of the multiplet are top components whose $\cQ$ descendants are conformal descendants. We emphasize that a superconformal multiplet can have multiple top components \cite{Cordova:2016emh}. Some are \textit{manifest} top components in the sense that their $\cQ$-descendants must also be conformal descendants based purely on the allowed quantum numbers of conformal primaries.\footnote{Note that the conformal primary of the highest scaling dimension is a manifest top component. For a long multiplet (with no null states), this is the unique top component \cite{Cordova:2016emh}. However for short multiplets the converse is not true. A short multiplet may contain manifest top components that are not of the highest scaling dimension, named sporadic top components. A short multiplet may also contain accidental top components.} Much subtler are \textit{accidental} top components, which can appear in short multiplets that contain null states.   In such cases, the short multiplet contains conformal primaries (accidental top components) whose $\cQ$ descendants could describe conformal primaries in principle, as their existence is consistent at the level of quantum numbers, but are in fact conformal descendants.\footnote{A familiar example is the conserved $\mf{so}(6)_R$ singlet $U(1)$ current in the 3d $\cN=6$ stress-tensor multiplet \cite{Bashkirov:2011fr,Cordova:2016emh}.} This analysis hinges in particular on an explicit construction of the states in the representation.

By analyzing the modified Ward identities for bulk symmetries broken by the defect (e.g. \eqref{introward}), we can deduce simple structural theorems for the induced operators on the defect and how they sit inside superconformal multiplets. For example, the displacement operator has to be a top component in a distinguished multiplet, namely the displacement multiplet (see Section~\ref{sec:struc}).   

Since the quantum numbers of the displacement operators are completely fixed, we can verify whether such displacement multiplets are admissible among the unitary representations of the putative superconformal symmetry $\mf{g}_s$. This places strong constraints on whether $\mf{g}_s$ can be realized by a superconformal defect. Similarly, by analyzing possible realizations of other defect primaries associated with broken bulk symmetries in superconformal multiplets of $\mf{g}_s$, one can deduce constraints on what kinds of bulk symmetries can be broken in a way consistent with the preserved defect superconformal symmetry $\mf{g}_s$.

\subsection{Summary of Main Results}

Here we give a short summary of the main results in the paper\footnote{Superconformal lines in 2d SCFTs are distinguished by their enhanced super-Virasoro symmetry. As such, we will defer their analysis to future work.}
\begin{itemize}
	\item \textit{Enumeration of 1d superconformal subalgebras for line defects}:~
	\newline
	In Section~\ref{sec:subalg}, we list 1d superconformal subalgebras of the known superconformal algebras in $d=3,4,5,6$. These represent the superconformal symmetries of candidate superconformal line defects in the corresponding SCFT. Along the way, we correct some incorrect statements in the literature.
	\item \textit{Unitary representations for 1d superconformal subalgebras}:~
	\newline
	In Section~\ref{sec:ureps}, we present an explicit classification of the unitary representations of the 1d superconformal algebras, following the RS algorithm. This includes the unitarity bounds, constituent conformal primaries, and null states for short multiplets. We also discuss recombination rules for long multiplets at threshold (i.e. at the unitarity bound).
	\item \textit{Structural theorems for the displacement multiplet  and general broken current multiplets:}~
	\newline
	In Section~\ref{sec:brokencurrents}, we analyze the modified Ward identities of bulk conserved currents in the presence of a superconformal defect, and the corresponding defect conformal primaries induced by the broken bulk symmetries. In Section~\ref{sec:struc}, we deduce structural theorems for defect superconformal multiplets that host these distinguished defect operators, which include the displacement multiplet and more general  broken current multiplets. 
	\item \textit{Classification of superconformal line defects}:~
	\newline
	In Section~\ref{sec:classify}, we systematically classify potential line defects $\cD$ preserving the superconformal subalgebras $\mf{g}_s$ (and any flavor symmetries) found in Section~\ref{sec:subalg}. We do so by identifying the conformal substructure of their unitarity representations, as listed in Section~\ref{sec:ureps}, with the distinguished defect operators required by the structural theorems of Section~\ref{sec:struc}. We rule out potential line defects whenever the two sides are incompatible (see Section~\ref{sec:forbiddenLines} for details). In this way, we exhaust the kinematic constraints from superconformal symmetry and its unitary representations on superconformal line defects.
	
	\item \textit{Deformations of superconformal line defects}:~
	\newline
	An interesting byproduct of our representation theory analysis concerns possible supersymmetric deformations of line defects. These are associated to top components $\cO$ of defect superconformal multiplets integrated along the defect worldvolume $\Sigma$. Since the $\cQ$ descendants of $\cO$ are all defect conformal descendants (namely total derivatives), the deformation preserves all supercharges $\cQ \in \mf{g}_s$. 
	As we explain in Section~\ref{sec:defor}, the only instances of superconformal defects that admit relevant or marginal deformations preserving the supercharges arise when the defect superconformal symmetry is given by $\mf{osp}(1|2)$, $\mf{su}(1,1|1)$, $\mf{psu}(1,1|2)$ (centrally extended) and $\mf{su}(1,1|\cN)$ with $\cN=3,4$.  Relevant supersymmetric deformations are only possible in the first two cases, while marginal supersymmetric deformations are possible for all of them. In other words, the superconformal line defects outside this list are rigid and represent stable fixed points of supersymmetric defect RG flows. Furthermore, some defects are forced to have marginal couplings, which includes the half-BPS line defect in 3d $\cN=8$ SCFTs and more generally line defects that break (any of) the bulk flavor symmetry.

\end{itemize}
We discuss further applications of our classification results on superconformal lines and the unitary representations of their symmetries in Section~\ref{sec:discuss}.

\section{Superconformal Algebras and Defect Subalgebras}
\label{sec:subalg}

In this section, we analyze the 1d superconformal subalgebras of known superconformal algebras in $d=3,4,5,6$ dimensions. We begin with a brief review of super Lie theory and its consequences for SCFTs. 

\subsubsection*{Lie Superalgebras}

Recall that a (complex) Lie superalgebra $\mf{G}$ is a $\bZ_2$-graded associative algebra over $\bC$ equipped with a super-bracket that satisfies a generalized form of the Jacobi identity.\footnote{For a complete review of Lie superalgebras, see \cite{Kac:1977aa, Frappat:1996pb}. A particularly illustrative discussion of superconformal algebras and their associated real forms can be found in \cite{DHoker:2008wvd}.} It admits a unique decomposition $\mf{G} = \mf{G}_{\bar 0} \oplus \mf{G}_{\bar 1}$ that consists of a $\bZ_2$-even (bosonic) subspace $\mf{G}_{\bar 0}$, called the \textit{maximal bosonic subalgebra}, and a $\bZ_2$-odd (fermionic) subspace $\mf{G}_{\bar 1}$ that transforms in a (possibly reducible) representation under $\mf{G}_{\bar 0}$. The superalgebras most relevant to the study of SCFT are the (classical) \textit{basic} Lie superalgebras\footnote{Our discussion omits the strange series $P(n)$ and $Q(n)$, which do not play a role in the study of SCFTs and their defects.} 
\ie{}\label{eq:complexSuperAlg}
\osp(n|2m)\,, \quad \msl(m|n) \:\: (m \neq n), \quad \mf{psl}(m|m), \quad F(4), \quad G(3), \quad D(2,1;\lambda) \,.
\fe

For the purpose of constructing positive-energy unitary representations, it is necessary to impose various reality conditions on the bosonic and fermionic generators (e.g. by specifying their behavior under conjugation). On a more abstract level, we seek the real forms of the classical Lie superalgebras in \eqref{eq:complexSuperAlg}. Given a Lie superalgebra $\mf{G}$ over $\bC$ and a semi-linear (anti-linear) involution $\phi: \mf{G} \to \mf{G}$ that preserves the $\bZ_2$-grading as well the super-bracket, one can directly define a real superalgebra $\mf{G}^\phi \equiv \mf{G} + \phi(\mf{G})$. Intuitively, the map $\phi$ can be viewed as a sort of complex conjugation operation on complex superalgebras, under which $\mf{G}^\phi$ is manifestly invariant. It follows that all real Lie superalgebras can be obtained from this construction, including the special case where the real Lie superalgebra can simply be regarded as a complex superalgebra over $\bR$. Note that specifying the real form of the bosonic subalgebra $\mf{G}_{\bar{0}}$ uniquely determines the real form of $\mf{G}$. We list the relevant real forms of \eqref{eq:complexSuperAlg} in Table~\ref{table:realSCA}.
\begin{table}[htbp]
\small
\renewcommand{\arraystretch}{1.5}
\begin{center}
\begin{tabular}{|c|c|c|}
\hline
\textbf{Superalgebra} & \textbf{Bosonic Subalgebra} & \textbf{Fermionic Generators} \\  \hline\hline
$\osp(m|2n;\bR)$ & $\so(m) \oplus \msp(2n,\bR)$ & $\bf (m,2n)$  \\ \hline
$\osp(2m^*|2n)$  & $\so^*(2m) \oplus \usp(2n)$ & $\bf (2m,2n)$ \\ \hline
$\su(p,q|m)$  & $\su(p,q) \oplus \su(m) \oplus \uu(1)$ & $\bf (p+q,\overline{m}) \oplus (\overline{p+q},m)$ \\ \hline
$\psu(p,q|m)$, $p+q=m$  & $\su(p,q) \oplus \su(m)$ & $\bf (p+q,\overline{m}) \oplus (\overline{p+q},m)$ \\ \hline
$F(4;p)$, $p=0,1$ & $\so(p,7-p) \oplus \msl(2,\bR)$ & $ \bf (8_s,2)$\\ 
\qquad \quad\:\:$p=2,3$ & $\so(p,7-p) \oplus \su(2)$ & $ \bf (8_s,2)$\\ \hline
$G(3;p)$, $p=0,1$ & $\msl(2,\bR) \oplus G_{2,2p}$ & $ \bf (2,7)$\\ \hline
$D(2,1;\lambda,p)$, $p=0,1,2$ & $\msl(2,\bR) \oplus \so(p,4-p)$ & $ \bf (2,2,2)$\\ \hline
\end{tabular}
\caption{Real forms of the basic Lie superalgebras relevant for SCFTs. Note that \cite{DHoker:2008wvd, Frappat:1996pb} both incorrectly classify the maximal bosonic subalgebra of $F(4;p)$. The correct real forms can all be found in \cite{Parker:1980af}.}
\label{table:realSCA}
\end{center}
\end{table}

\subsubsection*{Superconformal Algebras for Unitary QFTs}

We now restrict our attention to unitary supersymmetric theories with conformal invariance. All known unitarity interacting SCFTs in $d \ge 3$ dimensions are in general described by a direct sum superalgebra $\mf{G}_s \oplus \mf{G}_f$, which comprises of the superconformal symmetry $\mf{G}_s$ as well as any (bosonic) global symmetry $\mf{G}_f$.\footnote{We only focus on continuous symmetries in this paper.} The maximal bosonic subalgebra of $\mf{G}_s$ contains the $d$-dimensional conformal algebra $\so(2,d)_\text{conf}$, which we take to be generated by dilatations $D$, translations $P_\mu$, Lorentz transformations $M_{\mu\nu}$, and special conformal transformations $K_\mu$. It also includes the R-symmetry algebra $\mf{R}$, which commutes with all of the (bosonic) conformal generators. The fermionic generators of $\mf{G}_s$ transform as an $\so(2,d)_\text{conf}$ spinor, which can be divided into $\so(1,d-1)$ spinors consisting of the Poincar\'e supercharges $Q$ and superconformal charges $S$, which transform in conjugate representations under $\mf{R}$.\footnote{Note that our definition of superconformal algebras requires nontrivial supersymmetry. In particular, we do not consider the bosonic subalgebra to be a  superconformal subalgebra.} Their anti-commutators take the schematic form
\ie{}
\{Q, Q\} \sim P, \quad \{S,S\} \sim K, \quad \{Q,S\} \sim D + M + R
\fe
where $R \in \mf{R}$ and we have suppressed Lorentz and R-symmetry indices for brevity. 

In addition to superconformal invariance, the SCFTs under study are assumed to obey unitarity, whose consistency with local interactions (i.e. the existence of a local stress-tensor operator) places further kinematical restrictions on the theory. In particular, the spacetime dimension is constrained to $d\leq 6$, and the maximum amount of supersymmetry $\cN$ is limited for interacting SCFTs in each dimension \cite{Cordova:2016emh}.\footnote{As is conventional, we let $\cN$ denote the number of minimal spinors. For $d=1$ algebras, which admit no spinor representations, we take $\cN_1$ to be the number of real Poincar\'e supercharges.} For the theories under consideration with $3\leq d\leq 6$, these constraints are sufficiently restrictive so as to uniquely assign a superconformal algebra to each allowed value of $d$ and $\cN$, namely\footnote{It can be shown that all 3d $\cN=7$ SCFTs automatically preserve $\cN=8$ SUSY \cite{Bashkirov:2011fr}.}
\ie{}\label{eq:bulkSCA}
&d = 3\,,  &&\osp(\cN|4;\bR) &&\supset \so(2,3)_{\text{conf}} \oplus \so(\cN)_R \,, &&\cN = 1, \ldots,6, 8 \,, \\
&d = 4\,,  &&\su(2,2|\cN) &&\supset \so(2,4)_{\text{conf}} \oplus \su(\cN)_R \oplus \uu(1)_R \,, &&\cN = 1, 2, 3 \,, \\
&\:   &&\psu(2,2|4) &&\supset \so(2,4)_{\text{conf}} \oplus \su(4)_R  \,, && \cN = 4 \,,  \\
&d = 5\,,  &&F(4;2) &&\supset \so(2,5)_{\text{conf}} \oplus \su(2)_R \,, && \: \cN =1 \,, \\
&d = 6\,,  &&\osp(8^*|2\cN) &&\supset \so(2,6)_{\text{conf}} \oplus \usp(2\cN)_R \,, && \cN =(1,0), (2,0) \,.
\fe
SCFTs described by superconformal algebras in $d>6$ dimensions or with sufficiently large $\cN$ are necessarily free, inconsistent with unitarity, or both. 

\subsubsection*{Superconformal Defect Algebras}

We now consider the addition of (planar) conformal defects extended along a subspace in the bulk. On a geometric level, a $p$-dimensional conformal defect is simply a subspace $\Sigma \subset \bR^{1,d-1}$ which could be timelike, lightlike, or spacelike depending on the signature of $\Sigma$. While all three cases can be realized as unitary defects in CFTs, we find that in practice, superconformal defects are either timelike or lightlike.\footnote{One way to see this is that to go from the superconformal symmetry preserved by a timelike defect to that of a spacelike defect requires an imaginary Lorentzian rotation which violates the reality conditions on the fermionic generators of the bulk superconformal algebra.} We further restrict our attention to timelike superconformal defects and leave the study of lightlike superconformal defects to future work.

For a timelike conformal defect, the worldvolume $\Sigma = \bR^{1,p-1}$ preserves
\ie{}\label{eq:timelikeDefectAlg}
\so(2,p)_{\text{conf}} \oplus \so(d-p)_{\text{rot}} \subset \so(2,d)_{\text{conf}} \,,
\fe
where the commutant $\so(d-p)_{\text{rot}}$ of $\so(2,p)_{\text{conf}} \subset \so(2,d)_{\text{conf}}$ describes rotations transverse to the defect. In superconformal theories, it is possible for such defects to preserve some of the bulk supersymmetry and global R-symmetry, which together with the $\so(2,p)_{\text{conf}}$ conformal symmetry form a superconformal algebra $\mf{g}_s$ contained inside the bulk superconformal algebra $\mf{G}_s$. 

A natural first step in classifying unitary superconformal defects is to determine the allowed superconformal subalgebras of the bulk superconformal algebras listed in \eqref{eq:bulkSCA}, with the additional restriction that such algebras describe timelike defects \`{a} la \eqref{eq:timelikeDefectAlg}.\footnote{All of the direct sum superconformal subalgebras $\mf{g}_s \oplus \mf{g}'_s$ in Table~\ref{table:subalg} that appear as subalgebras of the bulk superconformal algebras in \eqref{eq:bulkSCA} describe the symmetries of timelike 2d defects. They admit two types of maximal 1d defect subalgebras; assuming $\mf{g}_s \subset \mf{g}'_s$, these consist of the diagonal subalgebra $\mf{g}_s$ (associated to timelike line defects) and the chiral subalgebras $\mf{g}_s$ and $\mf{g}'_s$ (associated to lightlike line defects).} Using Table~\ref{table:realSCA}, it is a straightforward exercise to enumerate all 1d superconformal algebras:
\ie{}
\osp(4^*|2n) &\supset \so(2,1)_{\text{conf}} \oplus \so(3) \oplus \usp(2n) \,, \\
\su(1,1|n) &\supset \so(2,1)_{\text{conf}} \oplus \su(n)\oplus \uu(1)  \:\: (n \neq 2) \,, \\
\psu(1,1|2) &\supset  \so(2,1)_{\text{conf}} \oplus \su(2) \,, \\
\osp(n|2;\bR)  &\supset \so(2,1)_{\text{conf}} \oplus \so(n) \,, \\
F(4;2) &\supset \so(2,1)_{\text{conf}} \oplus \so(7) \,, \\
G(3;0) &\supset \so(2,1)_{\text{conf}} \oplus G_{2} \,, \\
D(2,1;\lambda,0) &\supset \so(2,1)_{\text{conf}} \oplus \so(4) \,.
\fe
\begin{table}[htbp]
\renewcommand{\arraystretch}{1.5}
\begin{center}
\begin{tabular}{|c|c|}
\hline
\textbf{Superalgebra} $\mf{G}^\phi$ & \textbf{Maximal Subalgebra} \\  \hline\hline
$\osp(m|2n;\bR)$ & $\osp(p|2q;\bR) \oplus \osp(m-p|2n-2q;\bR)$ \\ \hline
$\osp(2m|2n;\bR)$ & $\su(p,n-p|m) \oplus \uu(1)$ \\ \hline \hline
$\osp(2m^*|2n)$  & $\osp(2p^*|2q) \oplus \osp(2{m}-2p^*|2n-2q)$ \\ 
  & $\su(p,m-p|n)\oplus \uu(1)$ \\ \hline \hline
$\su(p,q|m)$  & $\su(p',q'|m') \oplus \su(p-p',q-q'|m-m') \oplus \uu(1)$  \\ 
$\psu(p,q|p+q)$  & $\su(p',q'|m') \oplus \su(p-p',q-q'|p+q-m')$   \\ 
  & $\uu(1) \rtimes [\psu(p',q'|p'+q') \oplus \psu(p-p',q-q'|p+q-p'-q')] \rtimes \uu(1)$  \\ \hline 
$\su(m,m|2n)$ & $\osp(2m^*|2n)$ \\ \hline
$\su(m,m|n)$ & $\osp(n|2m;\bR)$ \\ \hline \hline
$F(4;2)$  & $\su(2|1) \oplus \su(1,2)$ \\ 
& $D(2,1;2;0)\oplus\su(2)$ \\
& $\osp(2|4;\bR)\oplus\uu(1)$ \\ 
& $\su(2,2|1)$ \\ \hline
\end{tabular}
\caption{Subalgebras of the real forms relevant for classifying defects. The rules apply for all non-negative entries given that we identify  $\osp(m|0;\bR) = \so(m)$, $\osp(0|2n;\bR) = \msp(2n,\bR)$, $\osp(2m^*|0) = \so(2m^*)$, $\osp(0^*|2n) = \usp(2n)$, and $\su(p,q|0) = \su(p,q)$. 
Here in $\uu(1)_1 \rtimes \psu(p,q|p+q) \rtimes \uu(1)_2$,  $\psu(p,q|p+q)$ is centrally extended by $\uu(1)_1$ and further extended by $\uu(1)_2$ as an outer-automorphism. Instances of the superalgebras $\su(m,m|2m)$ (in the left column) should be replaced with $\psu(m,m|2m)$. Similarly, those of the maximal subalgebras $\su(p,q|m)\oplus\uu(1)$ with $p+q = m$ should be replaced by $\uu(1)\rtimes  \psu(p,q|p+q) \rtimes\uu(1)$.  }
\label{table:subalg}
\end{center}
\end{table}
Among these, we are only interested the ones which appear as subalgebras of \eqref{eq:bulkSCA}. Luckily, the maximal subalgebras of the $\su$ and $\osp$ series have already been classified \cite{DHoker:2008wvd}, as well as those of $F(4;2)$ \cite{Gutperle:2017nwo}, as listed in Table~\ref{table:subalg}.\footnote{Our results refine the classification in \cite{DHoker:2008wvd} for $\psu(p,q|m)$ maximal subalgebras, which properly show up as the centrally extended $\uu(1) \rtimes \psu(p,q|m) \rtimes \uu(1)$. We also correct a mistake in \cite{Gutperle:2017nwo}, which incorrectly identifies the real form of $\msl(2|1,\bC) \oplus \msl(3,\bC)$ inherited from that of $F(4;2)$ as having a maximal bosonic subalgebra given by $\so(2,1)_{\text{conf}} \oplus \su(3)\oplus\uu(1)$, when in reality it is $\su(2)\oplus\su(1,2)\oplus\uu(1)$.} Using these results, we find that the \textit{maximal} 1d superconformal subalgebras in the form of $\mf{g}_s\oplus \mf{g}_b$ are\footnote{A maximal 1d subalgebra $\mf{g} \subset \mf{G}$ by definition is not a proper subalgebra of another proper 1d subalgebra $\mf{g}' \subset \mf{G}$. Note that a maximal 1d subalgebra is not necessarily a maximal subalgebra.} 
\ie{}\label{eq:1dSCAs}
&d=3\,, &&\cN =2,3 \,, &&\su(1,1|1) \oplus \uu(1) \,, && \cN_1 = 2 \,, \\
& &&\cN = 4,5\,, \quad &&\uu(1) \rtimes \psu(1,1|2) \rtimes \uu(1)\,, &&  \cN_1 = 4 \,, \\
& &&\cN = 6\,, \quad &&\su(1,1|3) \oplus \uu(1)\,, &&  \cN_1 = 6 \,, \\
& &&\cN = 8\,, \quad &&\su(1,1|4) \oplus \uu(1)\,, &&  \cN_1 = 8 \,, \\
&d=4\,,  &&\cN = 2\,, \quad &&\osp(4^*|2)\,, && \cN_1 = 4 \,, \\
& &&\cN = 3 \,, \quad &&\osp(4^*|2)\oplus\uu(1)\,, && \cN_1 = 4 \,,  \\
& &&\cN = 4 \quad &&\osp(4^*|4)\,, && \cN_1 = 8  \,,\\
&d=5\,, &&\cN = 1\,, \quad &&D(2,1;2;0)\oplus \su(2)\,, && \cN_1 = 4 \,, \\
&d=6\,, &&\cN = (2,0)\,, \quad &&\osp(4^*|2)\oplus \uu(1)\,, && \cN_1 = 4 \,.
\fe

All possible superconformal line defects preserve one of the maximal 1d subalgebras $\mf{g}_s\oplus \mf{g}_b$ listed above in \eqref{eq:1dSCAs}, or a subalgebra thereof, which can be determined by applying the rules of Table~\ref{table:subalg}. It is noteworthy that none of the bulk $\cN=1$ superconformal algebras in $d=3,4,6$ dimensions admit timelike 1d superconformal subalgebras.

\section{Defect Superconformal Multiplets}
\label{sec:ureps}

In this section, we analyze the unitary multiplets of the 1d superconformal subalgebras of known superconformal algebras in $d=3,4,5,6$ dimensions, as discussed in Section~\ref{sec:subalg}. In particular, we outline the features essential for our classification program and include additional details relevant for the conformal bootstrap. We begin with a brief discussion on the implications of superconformal symmetry for unitary line defects, and outline our conventions for unitary superconformal multiplets and their decomposition into conformal multiplets.

\subsubsection*{Defect Conformal Symmetry}
The 1d conformal algebra $\so(2,1)_{\text{conf}}$ of the timelike line defect $\cD$ is generated by dilatations $D_L$, time translations $P_L$, and special conformal transformations $K_L$, which satisfy
\ie{}
[D_L, P_L] = -iP_L \,, \quad [D_L, K_L] = iK_L \,, \quad[P_L, K_L] = 2iD_L \,.
\fe
Conformal symmetry arranges local operators (both in the bulk and on the defect) into infinite families known as conformal multiplets \cite{Rychkov:2016iqz,Simmons-Duffin:2016gjk}. Such multiplets transform in positive-energy (lowest weight) unitary irreducible representations (UIRs) of the conformal algebra, where the energy $\Delta \ge 0$ is taken to be the eigenvalue of the conformal Hamiltonian $H_L = - \frac12 (P_L + K_L)$ following the convention in \cite{Minwalla:1997ka,Cordova:2016emh,Luscher:1974ez}.\footnote{In a general $d$-dimensional Lorentzian CFT, the physical conformal group is the universal cover $\widetilde{{SO}(2,d)}$ whose UIRs are in one-to-one correspondence with those of the conformal algebra $\mf{so}(2,d)$ \cite{Luscher:1974ez}. Note that the conformal Hamiltonian $H_L$ generates the compact $SO(2)$ subgroup in $SO(2,d)$ which is decompactified in the universal cover and allows for an unquantized spectrum in $\Delta$. 
This is akin to decompactifying the time direction in global $AdS_{d+1}$.
For $d=1$, the UIRs of $\widetilde{SL(2,\bR)}$ can be found in \cite{Kitaev:2017hnr}. The lowest weight UIRs correspond to the discrete series. Similar statements apply to the supergroups relevant for the superconformal line defects we study in this paper.
}

While the  defect CFT is defined in Lorentzian spacetime, it is often convenient to work with the Wick-rotated theory in Euclidean signature, where the line defect preserves the Euclidean conformal algebra $\so(1,2)_{\text{conf}}$ generated by
\ie{}
D_E = i H_L \,, \quad P_E = D_L + \tfrac{i}{2}(K_L - P_L) \,, \quad K_E = D_L -  \tfrac{i}{2} (K_L - P_L)  \,,
\label{ltoe}
\fe
with commutation relations
\ie{}
[D_E, P_E] = i P_E \,, \quad [D_E, K_E] = -i K_E \,, \quad [P_E, K_E] = 2i D_E \,,
\fe
and Hermiticity properties
\ie{}
D_E^\dagger = - D_E \,, \quad P_E^\dagger = K_E \,, \quad K_E^\dagger = P_E \,.
\label{rhc}
\fe
These all follow from the relation \eqref{ltoe} to the Hermitian Lorentzian generators. 
Here, $P_E$ and $K_E$ are related to the translations and special conformal transformations  of the Wick-rotated line by conjugation in $SO(1,2)$, while $D_E$ is the generator of dilatations of the Euclidean spacetime $\mR^d$. We will be mostly working with the Euclidean generators obeying \eqref{rhc} and henceforth drop the Lorentzian/Euclidean subscripts whenever they are clear from context.

\begin{figure}
    \centering
    \begin{tikzpicture}
\def\R{2}
\def\a{1}
\def\z{1}
\def\h{3.5*\R}

\usetikzlibrary{decorations.markings}
  \fill[gray,opacity=0.2]
    (0,0) circle (\R cm and \a cm);
  \draw (-\R,\h) -- (-\R,0) arc (180:360:\R cm and \a cm)
               -- (\R,\h) ++ (-\R,0) circle (\R cm and \a cm);
  \draw[densely dashed] (-\R,0) arc (180:0:\R cm and \a cm);
  
\coordinate (B) at (0,0);
\coordinate (A) at ($(B)+(0,\h)$);
\coordinate (E) at (-\R,\h);
\coordinate (F) at (\R,\h);
\coordinate (G) at (-\R,0);
\coordinate (H) at (\R,0);
\coordinate (C) at ($(B) + (225:{\R} and {\a})$);
\coordinate (D) at ($(C)+(0,\h)$);
\coordinate (Y) at ($(B) -(225:{\R} and {\a})$);
\coordinate (Z) at ($(Y)+(0,\h)$);



\draw[densely dotted] (-\R,2.2) arc (180:360:\R cm and \a cm);
\draw[densely dotted] (-\R,3.7) arc (180:360:\R cm and \a cm);
\draw[densely dotted] (-\R,5.2) arc (180:360:\R cm and \a cm);
\draw[densely dotted] (-\R,2.2) arc (180:0:\R cm and \a cm);
\draw[densely dotted] (-\R,3.7) arc (180:0:\R cm and \a cm);
\draw[densely dotted] (-\R,5.2) arc (180:0:\R cm and \a cm);
\coordinate (bl) at (2*\R,1);
\coordinate (tl) at (2*\R,\h-1);
\coordinate (tr) at (6*\R,\h-1);
\coordinate (center) at (4*\R, 0.5*\h);

\path [fill=white] (bl) rectangle (tr);

\path (bl) rectangle (tr);
\draw[densely dotted] (center) circle (0.75cm);
\draw[densely dotted] (center) circle (1.5cm);
\draw[densely dotted] (center) circle (2.25cm);

\coordinate (lineStart) at ($(center)+(0,-2*\R)$);
\coordinate (lineStop) at ($(center)+(0,2*\R)$);
\coordinate (midPoint) at ($(tr)+(bl)$);
\draw[red!70!gray,line width=1.2pt,->] (lineStart) -- (lineStop)  node at ($(center)+(0,1.5*\R)$) {\AxisRotator[black,rotate=-90]};
\node [scale=1] at ($(center)+(\R,1.5*\R)$) {$SO(d-1)$};

\tkzDrawPoints[fill = black,size = 10 pt](center);

\node [scale=1.25] at ($(center)-(0.5,0)$) {$\cO$};
\node [scale=1.25] at ($(0,0)$) {$|\cO\ra$};
\node [scale=1.25] at ($(center)-(-1,1.5*\R)$) {$\cD$};
\node [scale=1.25] at ($(-\R+1,\h-5)$) {$\cD$};
\node [scale=1.25] at ($(\R-1,\h-3)$) {$\cD$};
\node [scale=1.25] at ($(E)+(3.8*\R,0)$) {$\bR^d$};
\node [scale=1.25] at ($(E)-(1.5,0)$) {$\bR \times S^{d-1}$};
 
\draw ($(center)-(1.5*\R,2.1*\R)$) -- ($(center)+ (-1.5*\R,2.1*\R)$) -- ($(center)+ (1.5*\R,2.1*\R)$) -- ($(center)+ (1.5*\R,-2.1*\R)$) -- ($(center)-(1.5*\R,2.1*\R)$);
 
\begin{scope}[very thick,decoration={
    markings,
    mark=at position 0.5 with {\arrow{>}}}
    ] 
    \draw[red,line width=1pt,postaction={decorate}] (C)--(D);
        \draw[red,line width=1pt,postaction={decorate}] (Z)--(Y);
  \end{scope}
   


\end{tikzpicture}
    \caption{Schematic of a radially quantized CFT endowed with a line defect $\cD$ and a defect local operator $\cO$.  }
    \label{fig:defecth}
\end{figure}
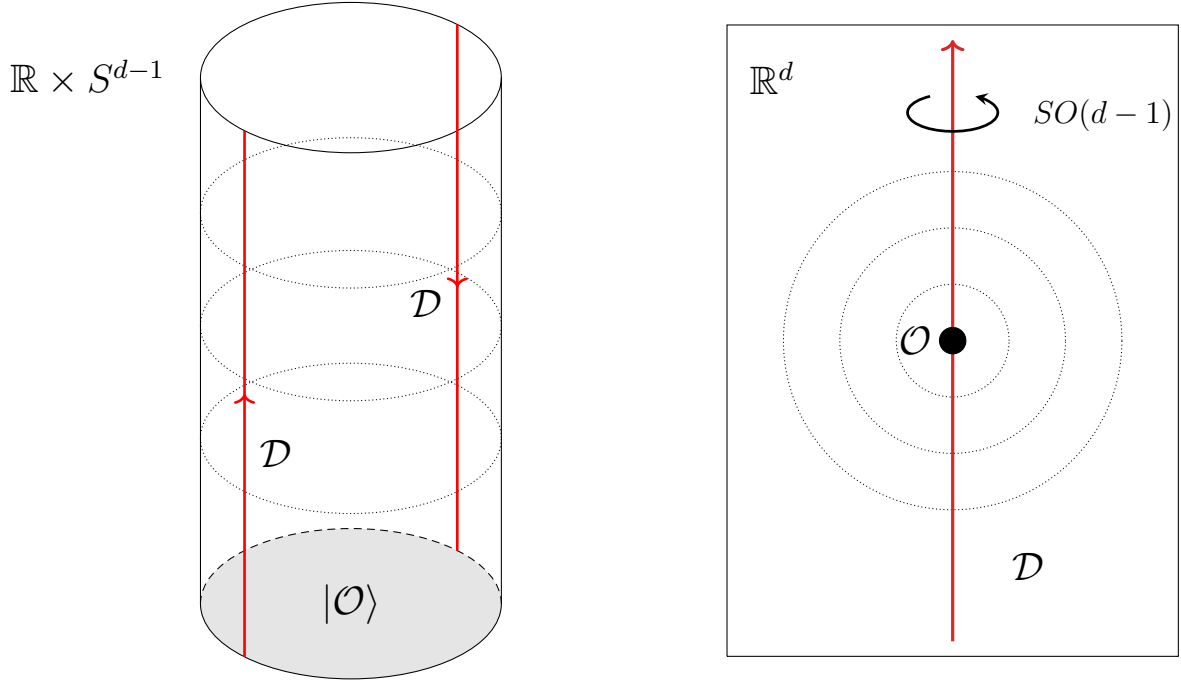

It is natural to quantize the theory by foliating the spacetime with spheres $S^{d-1}$ centered at the origin, treating the (Euclidean) dilatation $D$ as the Hamiltonian --- a procedure referred to as radial quantization.\footnote{This is equivalent to the N-S quantization on a constant time slice \cite{Rychkov:2016iqz}.} In the absence of defects, this gives rise to the usual state-operator correspondence between local operator insertions at the origin and states in the Hilbert space $\cH$ on $S^{d-1}$. With a line defect $\cD$ that passes through the origin, radial quantization establishes the equivalence between defect local operators $\cO$ and states $|\cO \ra$ in the Hilbert space $\cH_{\cD\bar\cD}$ on $S^{d-1}$ with two anti-podal defect points (see Figure~\ref{fig:defecth}). In particular, the identity operator corresponds to the defect vacuum state $|\cD \ra$, which is invariant under the full $\mf{so}(1,2)_{\text{conf}}$ conformal algebra. More generally, the positive-energy UIRs of defect local operators with respect to $\mf{so}(1,2)_{\text{conf}}$ correspond to conformal multiplets of states in $\cH_{\cD\bar\cD}$ of the form\footnote{Unitarity in the Lorentzian theory translates to reflection positivity of correlation functions in the Euclidean theory, which  implies in particular that all states in $\cH_{\cD\bar\cD}$ are required to have positive norm via the state/operator correspondence.}
 \ie{}
\cC = \text{Span} \left\{ \cV ~,~  P \cV ~,~ P^2 \cV  ~,~ \dots \right\} \,.
\fe
Here, $\cV$ is the conformal primary (CP), which is the lowest weight state in the multiplet and annihilated by $K$, while the rest are conformal descendants (CDs) arising from the action of $P$ on $\cV$.

There are several notable differences between defect CFTs in 1d and those in $d \ge 3$. In higher dimensional theories, unitarity imposes nontrivial constraints on the scaling dimension of the CP as a function of its $\so(d)$ representation. Operators that saturate this bound then reside in shortened multiplets, where states of zero norm are necessarily removed to ensure positivity and hence preserve unitarity. Examples of threshold CPs include free fields, such as a scalar field $\Phi$  with $\Delta_\Phi = \frac{d-2}{2}$ that obeys $\partial_\mu \partial^\mu \Phi = 0$, or spin 1 conserved currents $J^\mu$ with $\Delta_J = d-1$ and $\partial_\mu J^\mu = 0$. This should be contrasted with the 1d case, where unitarity only guarantees that $\Delta \ge 0$, and there are no nontrivial short multiplets.\footnote{It can happen that the defect Hilbert space $\cH_{\cD\bar\cD}$ contains degenerate ground states other than the defect vacuum $|\cD\ra$, which would comprise a topological sector among the defect local operators, and is relevant for a notion of conserved currents in 1d.}

\subsubsection*{Defect Superconformal Symmetry} 

We are interested in superconformal defects that preserve a subset of the bulk superconformal symmetries $\mf{g}_s\subset \mf{G}_s$. These symmetries generally include some of the bulk R-symmetry as well transverse rotations, which together form the \textit{defect} R-symmetry,\footnote{The defect can also preserve additional combinations of bulk R-symmetry and transverse rotations that commute with $\mf{g}_s$ which we denote as $\mf{g}_b$.} along with the Poincar\`e supercharges $\cQ$ and the superconformal charges $\cS$.\footnote{For certain algebras, the supercharges transform in a reducible representation of the R-symmetry, which can be decomposed into two conjugate irreducible representations (irreps). In this case, we will label the conjugate charges by $\cQ$ and $\bar{\cQ}$.} Local operators on the defect transform under both the defect conformal symmetry as well as the defect R-symmetry, and so can be characterized by their UIRs under the associated algebras. While there is no notion of Lorentz symmetry on the line defect, it is still useful to arrange our notation with an eye towards the bulk. One logical approach is to separate the global symmetries of the defect into preserved bulk R-symmetries and transverse rotations, as well as any additional flavor symmetries $\mf{g}_f\subset \mf{G}_f$. Given the $d$-dimensional bulk superconformal algebra $\mf{G}_s$, it can be shown that the superconformal subalgebra $\mf{g}_s \subset \mf{G}_s$ of a codimension $d-p$ defect is one of three types.\footnote{There is a similar classification of defect subalgebras for higher-dimensional defects whose proof we defer to future work.}
\begin{itemize}
    \item{\bf Type I ($d-p < 2$)} For these defect algebras, there is no transverse rotation symmetry, and   the R-symmetry of the defect coincides with a subalgebra $\mf{g}_\cR \subset \mf{R}$. We label UIRs by their quantum numbers $(R)$ under $\mf{g}_\cR$. A defect CP is represented as
    \ie{}
    [-]^{(R)}_{\Delta} \,.  
    \fe
    
    \item{\bf Type II ($d-p > 2$)} In this case, the R-symmetry of the defect admits a unique decomposition into $\mf{g}_{\text{rot}} \oplus \mf{g}_\cR$, where $\mf{g}_{\text{rot}} \subset \so(d-p)$ and $\mf{g}_\cR \subset \mf{R}$. That is, it is possible to categorize all defect R-symmetries as bulk R-symmetries or as transverse rotations. We consequently label UIRs by their quantum numbers $[j]$ and $(R)$ under  $\mf{g}_{\text{rot}}$ and $\mf{g}_\cR$, respectively. A defect CP is represented as
    \ie{}
    [j]^{(R)}_\Delta \,.
    \fe
    
    \item{\bf Type III ($d-p = 2$)} For codimension 2 defects, the defect R-symmetry is inexorably linked to a mixture of the R-symmetry of the bulk as well as $\so(2) \simeq \uu(1)_{\text{rot}}$ transverse rotations. The best we can do in this case is to decompose the R-symmetry of the defect as $\mf{g}_\cR \oplus \mf{g}_r$, where $\mf{g}_R \subset \mf{R}$, while $\mf{g}_r$ is in general a mixture of R-symmetries and transverse rotations. For these defect algebras, we label UIRs by their quantum numbers $[r]$ and $(R)$ under $\mf{g}_r$ and $\mf{g}_\cR$, respectively. A defect CP is represented as
    \ie{}
    [r]^{(R)}_\Delta \,. 
    \fe
\end{itemize}
In the discussion that follows, we only consider superconformal multiplets for Type II defect algebras, but the results extend in a straightforward manner to the other types.

With the classification in hand, we now consider the (positive-energy) UIRs of the superconformal algebra $\mf{g}_s$. Due to the fermionic nature of the supercharges, every superconformal multiplet $\cU$ can be decomposed into a \textit{finite} number of conformal multiplets, i.e.
\ie{}
\cU = \text{Span} \left\{ \bigoplus_{n=0}^{N} \cC_{n}[j_n]_{\Delta_n}^{(R_n)} \right\}\,, \quad \Delta_n = \Delta_0 +  \tfrac12 n \,,  
\label{SCtoC}
\fe
where each conformal multiplet $\cC_{n}$ transforms in some definite representation $[j_n]^{(R_n)}$ under $\mf{g}_\cR$ and $\mf{g}_{\text{rot}}$. The conformal multiplet of the lowest scaling dimension contains the superconformal primary (SCP), and is annihilated by the action of $\cS$ (in the same way the CP is annihilated by the action of $K$). The other conformal multiplets are generated by superconformal descendants (SCDs) and can be obtained by repeated application of the $\cQ$ supercharges. We refer to the conformal multiplet and its CP obtained from acting $\cQ^n$ on the SCP as level $n$, corresponding to  $\cC_{n}$  in \eqref{SCtoC}. It follows that the quantum numbers of the SCDs are fixed in terms of those of the SCP. 

In contrast to ordinary conformal multiplets, unitarity for 1d defects places stringent restrictions on the allowed quantum numbers of any superconformal multiplet \cite{Minwalla:1997ka}.\footnote{This is akin to the higher-dimensional story, where here $\mf{g}_{\text{rot}}$ transverse rotations play the role of the  $\so(1,d-1)$ Lorentz symmetry.} In the radially quantized theory, $\cQ^\dagger \sim \cS$, and so we have schematically
\ie{}\label{eq:1dQS}
\{\cQ, \cS\} \sim D + \cM + \cR \ge 0 \,,
\fe
where $\cM$ are $\mf{g}_{\text{rot}}$ transverse rotation generators and $\cR$ are the $\mf{g}_\cR$ generators. 
Here, the generators should be viewed as operators acting on states in the multiplet via the state-operator map. 
Given that the SCP of a multiplet $\cU$ is annihilated by $\cS$, it follows that positivity of the inner product along with \eqref{eq:1dQS} implies that the dimension of any SCP (and hence any state in the multiplet $\cU$) is bounded below by universal functions of the defect R-symmetry quantum numbers 
\ie{}\label{eq:1dunitaryBounds}
\Delta \ge \Delta_\star \,, \quad \Delta_\star = f(R_\cU) + g(j_\cU) \,.
\fe
Whenever the SCP saturates these bounds, some of the states in $\cU$ acquire zero norm, which themselves form a multiplet $\cU_{\text{null}}$ under $\mf{g}_s$. To preserve unitarity, we must instead consider the Verma module of physical states given by the quotient $\cU/\cU_{\text{null}}$. 

This analysis of level 1 SCDs, as discussed above, is generally insufficient to exhaust the list of consistent unitary multiplets. Indeed, the constraints derived from an analysis of level 2 descendants and beyond generically lead to unitary  multiplets \textit{below} the level 1 bounds for degenerate $\mf{g}_{\rm rot}$ and $\mf{g}_R$ quantum numbers, with dimensions given by  
\ie{}\label{eq:1disolatedBounds}
\Delta_B = \Delta_\star - \delta_B \,,
\fe
where $\delta_B$ is a constant shift.\footnote{The C and D-type unitarity bounds for superconformal algebras in dimensions $d \ge 5$, as presented in \cite{Cordova:2016emh}, do not arise for the 1d superconformal algebras under consideration. That is, 1d algebras only admit multiplets at threshold (A-type) and isolated multiplets with a single shift parameter $\delta_B$ (B-type).} The associated unitary multiplets are therefore isolated from the continuum of long multiplets. Following the conventions of \cite{Cordova:2016emh}, we denote long (generic) multiplets by $L[j]^{(R)}_\Delta$, short multiplets at threshold by $A[j]^{(R)}_\Delta$, and isolated short multiplets by $B[j]^{(R)}_\Delta$. Here, $A$ and $B$ label the specific shortening conditions of the multiplets. 

\subsubsection*{Defect Multiplets and the RS Algorithm}

Determining the conformal multiplets within a given Verma module  is straightforward in principle, but difficult in practice. This is true both in higher dimensions and in the case of 1d superconformal algebras. For both cases, the RS algorithm provides an efficient recipe to decompose superconformal multiplets into their constituent conformal multiplets, as was worked out systematically in \cite{Cordova:2016emh}. We outline the basic philosophy in what follows, but refer the reader to the seminal paper for further details, including the  intricacies of applying the algorithm to short multiplets. 

Consider first the fundamental problem in the representation theory of Lie algebras to determine the irreducible representations (irreps) that enter into the RHS of the tensor product
\ie{}
(R_1) \otimes (R_2) = \bigoplus_i  m_i (R_i) ,
\fe
where $m_i \in \bZ_{\ge 0}$ denotes the multiplicity of irrep $(R_i)$. The RS algorithm provides a method to extract the $m_i$ via the quantum numbers of the highest weight states, without having to explicitly construct any states in $(R_i)$. In short, the algorithm instructs us to simply add the weights (quantum numbers) of the highest weight state in ${ R}_1$ to all of those in ${R}_2$. We then perform Weyl reflections on the weights outside the Weyl chamber to obtain highest weight states of some irrep $(\widetilde{R}_r)$, and assign it a multiplicity $\widetilde{m}_r \in \bZ$ depending on the number of reflections. Some of the multiplicities may be negative, and thus remove trial representations from the decomposition, which handles the naive over-counting. It follows that \cite{Fuchs:1997jv}
\ie{}
\bigoplus_r \widetilde{m}_r (\widetilde{R}_r) = \bigoplus_{i} m_i (R_i) \,.
\fe

The algorithm outlined above can be adapted to the decomposition of superconformal multiplets, where the CPs of a generic superconformal multiplet can be obtained by adding the quantum numbers of the supercharges to those of the highest weight state in the SCP in accordance with the Grassmannality of the supercharges (i.e. each supercharge can be applied once and only once) \cite{Cordova:2016emh}.\footnote{For short multiplets, the algorithm must be adapted to carefully handle the decoupling of null states, and can outright fail in certain instances.} In what follows, we apply the RS algorithm to extract the CP content of candidate distinguished multiplets of superconformal lines (as introduced in  Section~\ref{sec:importantMult}), i.e. those whose top components have (half-)integer dimensions $\Delta \leq 2$.

\subsection{$\su(1,1| \frac12 \cN_1)$}
	
In this section, we study the UIRs of the 1d superconformal algebra
\ie{}
\su(1,1|\tfrac12\cN_1) \supset \su(1,1) \oplus \su(N)_\cR \oplus \uu(1)_{\cR_{N}} \,,
\fe
where $N \equiv \frac12 \cN_1$ and $\su(1,1) \simeq \so(2,1)_{\text{conf}}$. For $\cN_2 = 4$ the algebra contains a nontrivial central ideal $\uu(1)_{\cR_{2}}$ and also admits a nontrival $\mf{u}(1)_b$ outer-automorphism, 
\ie{}
\uu(1)_{\cR_2} \rtimes \psu(1,1|2) \rtimes \uu(1)_b \supset \su(1,1) \oplus \su(2)_\cR \oplus \uu(1)_{\cR_2} \oplus \uu(1)_b \,.
\fe
(Refer to Appendix~\ref{sec:su11NAlgebra} for more details). These all appear as maximal subalgebras of the 3d superconformal algebra
\ie{}
\osp(\cN_1|4;\bR) &\supset \su(1,1|\tfrac12\cN_1) \oplus \uu(1)_b  &&(\cN_1 \neq 4)  \,, \\
  &\supset \uu(1)_{\cR_2} \rtimes \psu(1,1|2) \rtimes \uu(1)_b  &&(\cN_1 = 4)   \,,
\fe
where $\uu(1)_{\cR_N}$ and $\uu(1)_b$ are a combination of the transverse rotation symmetry $\so(2) \simeq \uu(1)_{\text{rot}}$ and $\uu(1)_R$ inside the maximal subalgebra $\uu(1)_R \oplus \su(\tfrac12\cN_1)_R \subset \so(\cN_1)$, and so can be classified as type III defect algebras. Throughout the section, we label CPs by $[r]_\Delta^{(R)}$, where $r \in \bR$ is the $\uu(1)_r$ charge and $(R)$ is a UIR labeled by $\su(\tfrac12\cN_1)_R$ quantum numbers.

Note that there are two independent sets of supercharges, $\cQ$ and $\bar{\cQ}$, which are interchanged under complex conjugation. Each set leads to its own collection of shortening conditions that can be imposed independently from the other. To distinguish the two, we denote shortening conditions arising from $\cQ$ by $S$ and the conjugate conditions coming from $\bar{\cQ}$ by $\bar{S}$.
	
\subsubsection{$\cN_1=2$}

The $\cN_1=2$ superconformal algebra $\su(1,1|1)$ has an R-symmetry $\uu(1)_{\cR_1}$. There are two independent supercharges
\ie{}
&\cQ \: [\tfrac12]_{\frac12}, &&\bar{\cQ} \: [-\tfrac12]_{\frac12} \,.
\fe
The unitarity bounds and shortening conditions are presented in Table~\ref{table:su(1,1|1)}. 
\begin{table}[htbp]
\small
\renewcommand{\arraystretch}{1.5} 
\begin{center}
\begin{tabular}{|c|l|l|l|l|l|l|}
\hline
\textbf{Name} & \textbf{Primary} & \textbf{Unitarity Bound} & \textbf{BPS} & \textbf{Null State} & $\cQ\ket{\text{h.w.}} = 0$\\  \hline\hline
$L\bar{L}$ & $[r]_\Delta$ & $\Delta > |r|$ & $-$ & $-$ & $-$\\ \hline \hline
$A_1 \bar{L}$ & $[r]_\Delta$ & $\Delta = r$ & $(\frac12,0)$ & $[r+\tfrac12]_{\Delta+\frac12}$ & $\cQ$ \\ \hline 
$L \bar{A}_1 $ & $[r]_\Delta$ & $\Delta = - r$ & $(0,\frac12)$ & $[r-\tfrac12]_{\Delta+\frac12}$ & $\bar{\cQ}$ \\ \hline 
\end{tabular}
\caption{Unitary multiplets of $\su(1,1|1)$.}
\label{table:su(1,1|1)}
\end{center}
\end{table}
The long multiplets decompose at threshold $\Delta_\star = r$ ($\bar{\Delta}_\star = -r)$ as
\ie{}
L\bar{L}[r]_{\Delta\to\Delta_\star} &= A_1 \bar{L}[r]_{\Delta_\star} \oplus A_1 \bar{L}[r +\tfrac12]_{\Delta_\star+\frac12} \,, \\
L\bar{L}[r]_{\Delta\to\bar{\Delta}_\star} &= L\bar{A}_1[r]_{\bar{\Delta}_\star} \oplus L\bar{A}_1[r -\tfrac12]_{\bar{\Delta}_\star+\frac12} \,.
\fe
The multiplets that contain top components with (half-)integer scaling dimensions $\Delta \leq 2$ are given by
\ie{}
&\xymatrix{
*++[F-,][F**:Generic]{{\boldsymbol{ L \bar L}}}}  \qquad \xymatrix @C=7pc @R=7pc @!0 @dr {
*++[F][F**:Generic]{[r]_\Delta \,, \: |r| < \Delta = \frac12, 1 } \ar[r]|--{{~\bar \cQ~}} \ar[d]|--{~\cQ~}
& *++[F][F**:Generic]{[r-\frac12]_{\Delta+\frac12}} \ar[d]|--{~\cQ~}\\
*++[F][F**:Generic]{[r+\frac12]_{\Delta+\frac12}}\ar[r]|--{{~\bar \cQ~}}
& *++[F][F**:Generic]{ [r]_{\Delta+1}}
} \\ \\
\fe
\ie{}
&\xymatrix{
*++[F-,][F**:Generic]{{\boldsymbol{ L \bar{A}_1}}}}  \qquad \xymatrix @C=7pc @R=7pc @!0 @dr {
*++[F][F**:Generic]{[r]_{-r} \,, \: r= -\frac12, -1, -\frac32 }  \ar[d]|--{~\cQ~} \\
*++[F][F**:Generic]{[r+\tfrac12]_{-r+\frac12}}
} \\ \\
\fe
as well as their conjugate multiplets.

\subsubsection{$\cN_1=4$}

The $\cN_1=4$ superconformal algebra $\uu(1)_{\cR_2} \rtimes \psu(1,1|2) \rtimes \uu(1)_b$ has an $\su(2)_\cR \oplus \uu(1)_{\cR_2} \oplus \uu(1)_b$ subalgebra.\footnote{Note that the centrally extended $\uu(1) \rtimes \psu(1,1|2) \rtimes \uu(1)$ algebra admits a more intricate multiplet structure as compared to the small 2d $\cN=4$ superconformal algebra $\psu(1,1|2)$. This can be seen by comparing our results in Table~\ref{table:psu(1,1|2)} with, for instance, those of Table 3 in \cite{Lee:2019uen} (see also \cite{Dorey:2019kaf} where these multiplets were studied in the context of superconformal quantum mechanics with $\uu(1)\rtimes\psu(1,1|2)$ symmetry). Our $A_1 \bar{A}_1$ multiplet is their $A$ multiplet. In particular, $A_1 \bar{L}_1$ and its conjugate multiplet are absent for the small $\cN =4$ algebra.} We label the $\su(2)_\cR$ UIRs by $(R)$, where $R \in \bZ_{\ge 0}$ is an $\su(2)_\cR$ Dynkin label (contrasted with the usual $\su(2)_\cR$ isospin $\frac12 R \in \frac12\bZ$). The two sets of supercharges both transform as $\su(2)_\cR$ doublets,
\ie{}
&\cQ_{a} \in [0]_{\frac12}^{(1)} \,, \quad \bar{\cQ}^a \in [0]_{\frac12}^{(1)} \,,
\fe
where $a$ is an $\su(2)_\cR$ index. We assign weights to the individual supercharges as
\ie{}
&\cQ_1[0]_{\frac12}^{(+1)}, &&\cQ_2[0]_{\frac12}^{(-1)} \,, \\
&\bar{\cQ}^1 [0]_{\frac12}^{(-1)}, &&\bar{\cQ}^2 [0]_{\frac12}^{(+1)} \,,
\fe
where we have suppressed their $\uu(1)_b$ charges, as they only serve to distinguish the two sets of supercharges; in particular, $\uu(1)_b$ does not affect the unitarity bounds or multiplet structure. The unitarity bounds and shortening conditions are presented in Table~\ref{table:psu(1,1|2)}. 
\begin{table}[htbp]
\small
\renewcommand{\arraystretch}{1.5}
\begin{center}
\begin{tabular}{|c|l|l|l|l|l|l|}
\hline
\textbf{Name} & \textbf{Primary} & \textbf{Unitarity Bound} & \textbf{BPS} & \textbf{Null State} & $Q\ket{\text{h.w.}} = 0$\\  \hline\hline
$L\bar{L}$ & $[r]_\Delta^{(R)}$ & $\Delta > |r| + \frac12 R$ & $-$ & $-$ & $-$  \\ \hline \hline
$A_1 \bar{L}$ & $[r]_\Delta^{(R)}$ $(r<0)$ & $\Delta = -r + \frac12 R$ & $(\tfrac14,0)$ & $[r]_{\Delta+\frac12}^{(R+1)}$ & $\cQ_1$ \\ \hline 
\: & $[r]_\Delta^{(0)}$ & $\Delta = -r$ & $(\tfrac12,0)$ & $[r]_{\Delta+\frac12}^{(1)}$  & $\cQ_a$ \\ \hline  \hline
$L \bar{A}_1$ & $[r]_\Delta^{(R)}$ $(r>0)$ & $\Delta =r + \frac12 R$ &  $(0,\tfrac14)$ & $[r]_{\Delta+\frac12}^{(R+1)}$ & $\bar{\cQ}^2$ \\ \hline 
\: & $[r]_\Delta^{(0)}$ & $\Delta = r$ & $(0,\tfrac12)$ & $[r]_{\Delta+\frac12}^{(1)}$ & $\bar{\cQ}^a$ \\ \hline \hline
$A_1 \bar{A}_1$ & $[0]_\Delta^{(R)}$ & $\Delta = \frac12 R$ & $(\tfrac14,\tfrac14)$ & $[0]^{(R+1)}_{\Delta+\frac12}\oplus [0]^{(R+1)}_{\Delta+\frac12}$ & $\cQ_1$, $\bar{\cQ}^2$  \\ \hline
\end{tabular}
\caption{Unitary multiplets of $\uu(1)_{\cR_2} \rtimes \psu(1,1|2) \rtimes \uu(1)_b$, where we have suppressed the $\uu(1)_b$ charge.}
\label{table:psu(1,1|2)}
\end{center}
\end{table}
The long multiplets decompose at threshold $\Delta_\star = -r+\frac12 R$ ($\bar{\Delta}_\star = r+\frac12 R)$ as
\ie{}
L\bar{L}[r]_{\Delta \to \bar{\Delta}_\star}^{(R)} &= L\bar{A}_1[r]_{\bar{\Delta}_\star}^{(R)} \oplus L\bar{A}_1[r]_{\bar{\Delta}_\star+\frac12}^{(R+1)} ,\\
L\bar{L}[r]_{\Delta \to \Delta_\star}^{(R)} &= A_1\bar{L}[r]_{\Delta_\star}^{(R)} \oplus A_1 \bar{L}[r]_{\Delta_\star+\frac12}^{(R+1)} ,\\
L\bar{L}[0]_{\Delta \to \Delta_\star}^{(R)} &= A_1 \bar{A}_1[0]_{\Delta_\star}^{(R)} \oplus 2  A_1 \bar{A}_1[0]_{\Delta_\star+\frac12}^{(R+1)} \oplus  A_1 \bar{A} _1[0]_{\Delta_\star+1}^{(R+2)}.
\fe
Note that $\Delta_\star = \bar{\Delta}_\star = \frac12 R$ for $r=0$. The multiplets whose top component scales with (half-)integer $\Delta \leq 2$ are given by
\ie{}
&\xymatrix{
*++[F-,][F**:Generic]{{\boldsymbol{ A_1 \bar{A}_1}}}} \qquad \xymatrix @C=7pc @R=7pc @!0 @dr {
*++[F][F**:Generic]{[0]^{(1)}_{\frac12}} \ar[r]|--{{~\bar{\cQ}~}} \ar[d]|--{~\cQ~}
& *++[F][F**:Generic]{[0]^{(0)}_{1}}\\
*++[F][F**:Generic]{[0]^{(0)}_{1}}
} \\ \\
\fe
\ie{}
&\xymatrix{
*++[F-,][F**:Generic]{{\boldsymbol{ A_1 \bar{A}_1}}}} \qquad \xymatrix @C=7pc @R=7pc @!0 @dr {
*++[F][F**:Generic]{[0]^{(2)}_{1}} \ar[r]|--{{~\bar{\cQ}~}} \ar[d]|--{~\cQ~}
& *++[F][F**:Generic]{[0]^{(1)}_{\frac32}} \ar[d]|--{~\cQ~}\\
*++[F][F**:Generic]{[0]^{(1)}_{\frac32}}\ar[r]|--{{~\bar{\cQ}~}}
& *++[F][F**:Generic]{[0]^{(0)}_{2}}
} \\ \\
\fe
\ie{}
&\xymatrix{
*++[F-,][F**:Generic]{{\boldsymbol{ L \bar{A}_1}}}} \qquad \xymatrix @C=7pc @R=7pc @!0 @dr {
*++[F][F**:Generic]{[\frac12]^{(0)}_{\frac12}}  \ar[d]|--{~\cQ~} \\
*++[F][F**:Generic]{[\frac12]^{(1)}_{1}}  \ar[d]|--{~\cQ~} \\
*++[F][F**:Generic]{[\frac12]^{(0)}_{\frac32}} 
} \\ \\
\fe
\ie{}
&\xymatrix{
*++[F-,][F**:Generic]{{\boldsymbol{ L \bar{A}_1}}}} \qquad \xymatrix @C=7pc @R=7pc @!0 @dr {
*++[F][F**:Generic]{[1]^{(0)}_{1} }  \ar[d]|--{~\cQ~} \\
*++[F][F**:Generic]{[1]^{(1)}_{\frac32}}  \ar[d]|--{~\cQ~} \\
*++[F][F**:Generic]{[1]^{(0)}_{2}} 
} \\ \\
\fe
as well as their conjugate multiplets.

\subsubsection{$\cN_1=6$}

The $\cN_1=6$ superconformal algebra $\su(1,1|3)$ has an $\su(3)_\cR \oplus \uu(1)_{\cR_3}$ R-symmetry. We label the $\su(3)_\cR$ UIRs by $(R_1,R_2)$, where $R_1,R_2 \in \bZ_{\ge 0}$ are $\su(3)_\cR$ Dynkin labels. The fundamental is given by $\mathbf{3} = (1,0)$ and the anti-fundamental by $\bf{\bar{3}} = (0,1)$, while the adjoint is $\mathbf{8} = (1,1)$. The two sets transform as the (anti-)fundamental of $\su(3)_\cR$,
\ie{}
&\cQ_{a} \in [\tfrac12]_{\frac12}^{(1,0)} \,, \quad \bar{\cQ}^a \in [-\tfrac12]_{\frac12}^{(0,1)} \,,
\fe
where $a$ is an $\su(3)_\cR$ index. We assign weights to the individual supercharges as
\ie{}
&\cQ_1[+\tfrac12]_{\frac12}^{(+1,0)}\,, &&\cQ_2[+\tfrac12]_{\frac12}^{(-1,+1)}\,, &&\cQ_3[+\tfrac12]_{\frac12}^{(0,-1)}\,, \\
&\bar{\cQ}^1[-\tfrac12]_{\frac12}^{(-1,0)}\,, &&\bar{\cQ}^2[-\tfrac12]_{\frac12}^{(+1,-1)}\,, &&\bar{\cQ}^3[-\tfrac12]_{\frac12}^{(0,+1)} \,.
\fe
The unitarity bounds and shortening conditions are presented in Table~\ref{table:su(1,1|3)}. 
\begin{table}[htbp]
\footnotesize
\renewcommand{\arraystretch}{1.5}
\begin{center}
\begin{tabular}{|c|l|l|l|l|l|l|}
\hline
\textbf{Name} & \textbf{Primary} & \textbf{Unitarity Bound} & \textbf{BPS} & \textbf{Null State} & $Q\ket{\text{h.w.}} = 0$\\  \hline\hline
$L\bar{L}$ & $[r]_\Delta^{(R_1,R_2)}$ & $\Delta > \frac12(R_1+R_2)+\frac13\left|r - \frac12R_{12}\right|$ & $(0,0)$ & $-$ & $-$ \\ \hline \hline
$A_1 \bar{L}$ & $[r]_\Delta^{(R_1,R_2)}$ $(r < \tfrac{R_{12}}{2})$ & $\Delta = \frac13 (2R_1+R_2 - r)$ & $(\frac16,0)$ & $[r+\frac12]_{\Delta+\frac12}^{(R_1+1,R_2)}$ &   $\cQ_1$ \\ \hline 
\: & $[r]_\Delta^{(0,R_2)}$ & $\Delta = \frac13 (R_2 - r)$ & $(\frac13,0)$ & $[r+\frac12]_{\Delta+\frac12}^{(1,R_2)}$ &  $\cQ_1$, $\cQ_2$\\ \hline 
\: & $[r]_\Delta^{(0,0)}$ & $\Delta = -\frac13 r$ & $(\frac12,0)$ & $[r+\frac12]_{\Delta+\frac12}^{(1,0)}$  & $\cQ_a$ \\ \hline  \hline
$L \bar{A}_1 $ & $[r]_\Delta^{(R_1,R_2)}$ $(r > \tfrac{R_{12}}{2})$ & $\Delta = \frac13 (R_1+2R_2 + r)$ & $(0,\frac16)$ & $[r-\frac12]_{\Delta+\frac12}^{(R_1,R_2+1)}$  & $\bar{\cQ}^3$ \\ \hline 
\: & $[r]_\Delta^{(R_1,0)}$ & $\Delta = \frac13 (R_1+r)$  & $(0,\frac13)$ & $[r-\frac12]_{\Delta+\frac12}^{(R_1,1)}$& $\bar{\cQ}^2$, $\bar{\cQ}^3$ \\ \hline 
\: & $[r]_\Delta^{(0,0)}$ & $\Delta = \frac13 r$ & $(0,\frac12)$ & $[r-\frac12]_{\Delta+\frac12}^{(0,1)}$  & $\bar{\cQ}^a$ \\ \hline \hline
$A_1 \bar{A}_1$ & $[r]_\Delta^{(R_1,R_2)}$ & $\Delta = \frac12 (R_1+R_2)$ & $(\frac16,\frac16)$ & $[r+\tfrac12]_{\Delta+\frac12}^{(R_1+1,R_2)}$ & $\cQ_1$,   \\ 
\: & \: & $r=\frac12(R_1-R_2)$ & \: & $\oplus [r-\tfrac12]_{\Delta+\frac12}^{(R_1,R_2+1)}$ & $\bar{\cQ}^3$ \: \\ \hline
\: & $[r]_\Delta^{(0,R_2)}$ & $\Delta = \frac12 R_2$ & $(\frac13,\frac16)$ & $[r+\tfrac12]_{\Delta+\frac12}^{(1,R_2)} $ & $\cQ_1$, $\cQ_2$,     \\ 
\: & \: & $r = -\frac12 R_2$ & \: & $\oplus [r-\tfrac12]_{\Delta+\frac12}^{(0,R_2+1)}$ & $\bar{\cQ}^3$ \\ \hline
\: & $[r]_\Delta^{(R_1,0)}$ & $\Delta = \frac12 R_1$ & $(\frac16,\frac13)$ & $[r+\tfrac12]_{\Delta+\frac12}^{(R_1+1,0)}$ & $\cQ_1$,   \\ 
\: & \: & $r = \frac12 R_1$ & \: & $\oplus [r-\tfrac12]_{\Delta+\frac12}^{(R_1,1)}$ & $\bar{\cQ}^2$, $\bar{\cQ}^3$ \\ \hline
\end{tabular}
\caption{Unitary multiplets of $\su(1,1|3)$.}
\label{table:su(1,1|3)}
\end{center}
\end{table}
The long multiplets decompose at threshold $\Delta_\star = \frac13(2R_1+R_2-r)$ ($\bar{\Delta}_\star = \frac13(R_1+2R_2+r) )$ as
\ie{}
L \bar{L}[r]^{(R_1,R_2)}_{\Delta\to \bar{\Delta}_\star} &= L\bar{A}_1[r]^{(R_1,R_2)}_{\bar{\Delta}_\star} \oplus L\bar{A}_1[r-\tfrac12]^{(R_1,R_2+1)}_{\bar{\Delta}_\star+\frac12} \,, \\
L \bar{L}[r]^{(R_1,R_2)}_{\Delta\to \Delta_\star} &= A_1\bar{L}[r]^{(R_1,R_2)}_{\Delta_\star} \oplus A_1 \bar{L}[r+\tfrac12]^{(R_1+1,R_2)}_{\Delta_\star+\frac12} \,, \\
L \bar{L}[r=\tfrac12 (R_1-R_2)]_{\Delta\to\Delta_\star}^{(R_1,R_2)} &= A_1 \bar{A}_1[r]_{\Delta_\star}^{(R_1,R_2)} \oplus A_1 \bar{A}_1[r+\tfrac12]_{\Delta_\star+\frac12}^{(R_1+1,R_2)} \\&\quad \oplus A_1 \bar{A}_1[r-\tfrac12]_{\Delta_\star+\frac12}^{(R_1,R_2+1)} \oplus A_1 \bar{A}_1[r]_{\Delta_\star+1}^{(R_1+1,R_2+1)}  \,.
\fe
Note that $\Delta_\star = \bar{\Delta}_\star = \frac12(R_1+R_2)$ for $r=\frac12(R_1-R_2)$. The multiplets whose top component scales with (half-)integer $\Delta \leq 2$ are given by
\ie{}
&\xymatrix{
*++[F-,][F**:Generic]{{\boldsymbol{ A_1 \bar{A}_1}}}} \qquad \xymatrix @C=7pc @R=7pc @!0 @dr {
*++[F][F**:Generic]{[\tfrac12]^{(1,0)}_{\frac12}} \ar[r]|--{{~\bar{\cQ}~}} \ar[d]|--{~\cQ~}
& *++[F][F**:Generic]{[0]^{(0,0)}_{1}} \\ 
*++[F][F**:Generic]{[1]^{(0,1)}_{1}} \ar[d]|--{~\cQ~} \\
*++[F][F**:Generic]{[\tfrac32]^{(0,0)}_{\frac32}}
} \\ \\
\fe
\ie{}
&\xymatrix{
*++[F-,][F**:Generic]{{\boldsymbol{ L \bar{A}_1}}}} \qquad \xymatrix @C=7pc @R=7pc @!0 @dr {
*++[F][F**:Generic]{[\frac32]^{(0,0)}_{\frac12} }  \ar[d]|--{~\cQ~} \\
*++[F][F**:Generic]{[2]^{(1,0)}_{1}}  \ar[d]|--{~\cQ~} \\
*++[F][F**:Generic]{[\frac52]^{(0,1)}_{\frac32}} \ar[d]|--{~\cQ~} \\
*++[F][F**:Generic]{[3]^{(0)}_{2}}
} \\ \\
\fe
as well as their conjugate multiplets.
	
\subsubsection{$\cN_1=8$}

The $\cN_1=8$ superconformal algebra $\su(1,1|4)$ has an $\su(4)_\cR \oplus \uu(1)_{\cR_4}$ R-symmetry. We label the $\su(4)_\cR$ UIRs by $(R_1,R_2,R_3)$, where $R_1,R_2,R_3 \in \bZ_{\ge 0}$ are $\su(4)_\cR$ Dynkin labels. The fundamental is given by $\mathbf{4} = (1,0,0)$ and the anti-fundamental by $\mathbf{\bar{4}} = (0,0,1)$, while the adjoint is $\mathbf{15} = (1,0,1)$. The two sets transform as the (anti-)fundamental of $\su(4)_\cR$,
\ie{}
&\cQ_{a} \in [\tfrac12]_{\frac12}^{(1,0,0)} \,, \quad \bar{\cQ}^a \in [-\tfrac12]_{\frac12}^{(0,0,1)} 
\fe
where $a$ is an $\su(4)_\cR$ index. We assign weights to the individual supercharges as
\ie{}
&\cQ_1[+\tfrac12]_{\frac12}^{(+1,0,0)}, &&\cQ_2[+\tfrac12]_{\frac12}^{(-1,+1,0)}\,,  &&\cQ_3[+\tfrac12]_{\frac12}^{(0,-1,+1)}, &&\cQ_4[+\tfrac12]_{\frac12}^{(0,0,-1)} \,, \\
&\bar{\cQ}^1[-\tfrac12]_{\frac12}^{(-1,0,0)}, &&\bar{\cQ}^2[-\tfrac12]_{\frac12}^{(+1,-1,0)}\,,  &&\bar{\cQ}^3[-\tfrac12]_{\frac12}^{(0,+1,-1)}, &&\bar{\cQ}^4 [-\tfrac12]_{\frac12}^{(0,0,+1)} \,.
\fe
The unitarity bounds and shortening conditions are presented in Table~\ref{table:su(1,1|4)}. 
\begin{table}[htbp]
\footnotesize
\renewcommand{\arraystretch}{1.5}
\begin{center}
\begin{tabular}{|c|l|l|l|l|l|l|}
\hline
\textbf{Name} & \textbf{Primary} & \textbf{Unitarity Bound} & \textbf{BPS} & \textbf{Null State} & $\cQ_*\ket{\text{h.w.}} = 0$\\  
\hline\hline$L\bar{L}$ & $[r]_\Delta^{(R_1,R_2,R_3)}$ & $\Delta > \frac12(R_1+R_2+R_3)+\frac12\left|r - \frac12R_{13}\right|$ & $-$  & $-$  & $-$ \\ \hline \hline
$A_1 \bar{L}$ & $[r]_\Delta^{(R_1,R_2,R_3)}$ & $\Delta = \frac14 (3R_1+2R_2+R_3-2r)$ & $(\frac18,0)$ & $[r+\tfrac12]_{\Delta+\frac12}^{(R_1+1,R_2,R_3)}$  & $\cQ_1$ \\ \hline 
\: & $[r]_\Delta^{(0,R_2,R_3)}$ & $\Delta = \frac14 (2R_2+R_3-2r)$ & $(\frac14,0)$ & $[r+\tfrac12]_{\Delta+\frac12}^{(1,R_2,R_3)}$   & $\cQ_1$, $\cQ_2$\\ \hline 
\: & $[r]_\Delta^{(0,0,R_3)}$ & $\Delta = \frac14 (R_3-2r)$ & $(\frac38,0)$  & $[r+\tfrac12]_{\Delta+\frac12}^{(1,0,R_3)}$  & $\cQ_1$, $\cQ_2$, $\cQ_3$ \\ \hline  
\: & $[r]_\Delta^{(0,0,0)}$ & $\Delta = -\frac12 r$ & $(\frac12,0)$  & $[r+\tfrac12]_{\Delta+\frac12}^{(1,0,0)}$  & $\cQ_a$ \\ \hline  \hline
$L \bar{A}_1$ & $[r]_\Delta^{(R_1,R_2,R_3)}$ & $\Delta = \frac14 (R_1+2R_2+3R_3+2r)$ & $(0,\frac18)$  & $[r-\tfrac12]_{\Delta+\frac12}^{(R_1,R_2,R_3+1)}$  & $\bar{\cQ}^4$ \\ \hline 
\: & $[r]_\Delta^{(R_1,R_2,0)}$ & $\Delta = \frac14 (R_1+2R_2+2r)$ & $(0,\frac14)$ & $[r-\tfrac12]_{\Delta+\frac12}^{(R_1,R_2,1)}$   & $\bar{\cQ}^3$, $\bar{\cQ}^4$ \\ \hline 
\:  & $[r]_\Delta^{(R_1,0,0)}$ & $\Delta = \frac14 (R_1+2r)$ & $(0,\frac38)$  & $[r-\tfrac12]_{\Delta+\frac12}^{(R_1,0,1)}$   & $\bar{\cQ}^2$, $\bar{\cQ}^3$, $\bar{\cQ}^4$ \\ \hline 
\: & $[r]_\Delta^{(0,0,0)}$ & $\Delta = \frac12 r$ &  $(0,\frac12)$ & $[r-\tfrac12]_{\Delta+\frac12}^{(0,0,1)}$  & $\bar{\cQ}^a$ \\ \hline \hline
$A_1 \bar{A}_1$ & $[r]_\Delta^{(R_1,R_2,R_3)}$ & $\Delta = \frac12 (R_1+R_2+R_3)$ & $(\frac18,\frac18)$  & $[r+\tfrac12]_{\Delta+\frac12}^{(R_1+1,R_2,R_3)}$  & $\cQ_1$,  \\ 
\: & \: & $r = \frac12(R_1-R_3)$ & \: & $\oplus[r-\tfrac12]_{\Delta+\frac12}^{(R_1,R_2,R_3+1)}$ & $\bar{\cQ}^4$ \: \\ \hline
\: & $[r]_\Delta^{(0,R_2,R_3)}$ & $\Delta = \frac12 (R_2+R_3)$ & $(\frac14,\frac18)$  & $[r+\tfrac12]_{\Delta+\frac12}^{(1,R_2,R_3)}$  & $\cQ_1$, $\cQ_2$,   \\ 
\: & \: & $r=-\frac12 R_3$ & \: & $\oplus[r-\tfrac12]_{\Delta+\frac12}^{(0,R_2,R_3+1)}$ & $\bar{\cQ}^4$  \\ \hline
\: & $[r]_\Delta^{(R_1,R_2,0)}$ & $\Delta = \frac12 (R_1+R_2)$ & $(\frac18,\frac14)$ & $[r+\tfrac12]_{\Delta+\frac12}^{(R_1+1,R_2,0)}$  & $\cQ_1$,   \\ 
\: & \: & $r = \frac12R_1$ & \: & $\oplus[r-\tfrac12]_{\Delta+\frac12}^{(R_1,R_2,1)}$ & $\bar{\cQ}^3$,  $\bar{\cQ}^4$ \: \\ \hline
\:  & $[r]_\Delta^{(0,0,R_3)}$ & $\Delta = \frac12 R_3$ & $(\frac38,\frac18)$  & $[r+\tfrac12]_{\Delta+\frac12}^{(1,0,R_3)}$ & $\cQ_1$, $\cQ_2$, $\cQ_3$,   \\ 
\: & \: & $r=-\frac12 R_3$ & \: & $\oplus[r-\tfrac12]_{\Delta+\frac12}^{(0,0,R_3+1)}$ & $\bar{\cQ}^4$ \: \\ \hline
\: & $[r]_\Delta^{(R_1,0,0)}$ & $\Delta = \frac12 R_1$ & $(\frac18,\frac38)$  & $[r+\tfrac12]_{\Delta+\frac12}^{(R_1+1,0,0)}$  & $\cQ_1$,   \\ 
\: & \: & $r = \frac12R_1$ & \: & $\oplus[r-\tfrac12]_{\Delta+\frac12}^{(R_1,0,1)}$ & $\bar{\cQ}^2$, $\bar{\cQ}^3$, $\bar{\cQ}^4$ \: \\ \hline
\end{tabular}
\caption{Unitary multiplets of $\su(1,1|4)$.}
\label{table:su(1,1|4)}
\end{center}
\end{table}
The long multiplets decompose at threshold $\Delta_\star = \frac14(3R_1+2R_2+R_3-2r)$ ($\bar{\Delta}_\star = \frac14(R_1+2R_2+3R_3+2r ))$ as
\ie{}
L \bar{L}[r]_{\Delta \to \bar{\Delta}_\star}^{(R_1,R_2,R_3)} &= L\bar{A}_1[r]^{(R_1,R_2,R_3)}_{\bar{\Delta}_\star} \oplus L\bar{A}_1[r-\tfrac12]^{(R_1,R_2,R_3+1)}_{\bar{\Delta}_\star+\frac12} \,, \\
L \bar{L}[r]_{\Delta\to\Delta_\star}^{(R_1,R_2,R_3)} &= A_1\bar{L}[r]^{(R_1,R_2,R_3)}_{\Delta_\star} \oplus A_1\bar{L}[r+\tfrac12]^{(R_1+1,R_2,R_3)}_{\Delta_\star+\frac12} \,, \\
L \bar{L}[r=\tfrac12(R_1-R_3)]_{\Delta\to\Delta_\star}^{(R_1,R_2,R_3)} &= A_1 \bar{A}_1[r]^{(R_1,R_2,R_3)}_{\Delta_\star} \oplus A_1 \bar{A}_1[r+\tfrac12]^{(R_1+1,R_2,R_3)}_{\Delta_\star+\frac12} \\&\quad\oplus A_1 \bar{A}_1[r-\tfrac12]^{(R_1,R_2,R_3+1)}_{\Delta_\star+\frac12} \oplus A_1 \bar{A}_1[r]^{(R_1+1,R_2,R_3+1)}_{\Delta_\star+1} \,.
\fe
Note that $\Delta_\star = \bar{\Delta}_\star = \frac12(R_1+R_2+R_3)$ for $r=\frac12(R_1-R_3)$. The multiplets whose top component scales with (half-)integer $\Delta \leq 2$ are given by
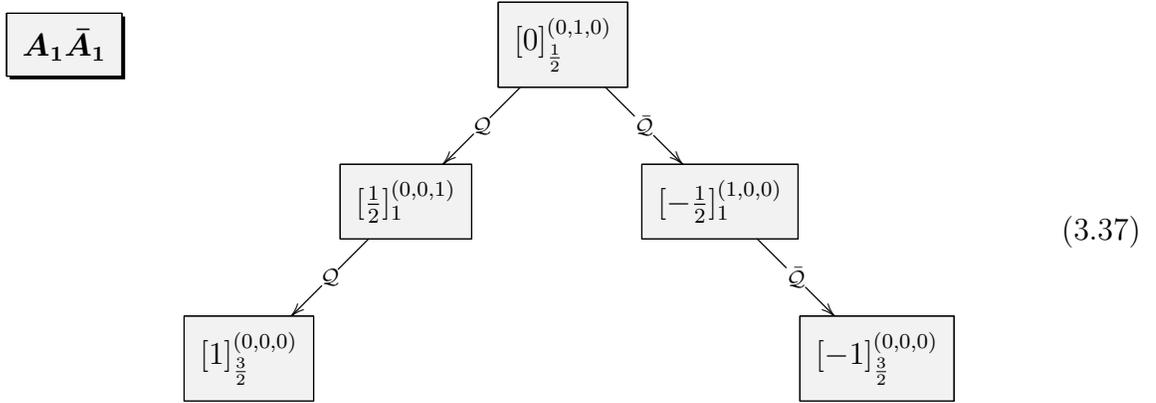
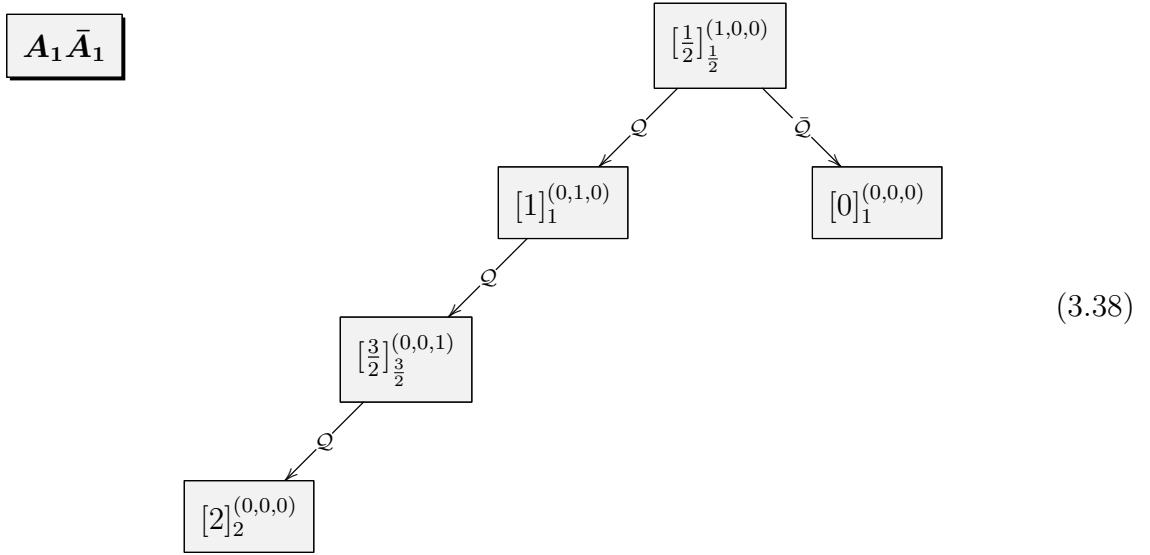
\ie{}
&\xymatrix{
*++[F-,][F**:Generic]{{\boldsymbol{ A_1 \bar{A}_1}}}} \qquad \xymatrix @C=7pc @R=7pc @!0 @dr {
*++[F][F**:Generic]{[0]^{(0,1,0)}_{\frac12}} \ar[r]|--{{~\bar{\cQ}~}} \ar[d]|--{~\cQ~}
& *++[F][F**:Generic]{[-\frac12]^{(1,0,0)}_{1}} \ar[r]|--{{~\bar{\cQ}~}} 
& *++[F][F**:Generic]{[-1]^{(0,0,0)}_{\frac32}} \\
*++[F][F**:Generic]{[\frac12]^{(0,0,1)}_{1}} \ar[d]|--{~\cQ~} \\
*++[F][F**:Generic]{[1]^{(0,0,0)}_{\frac32}}
} \\ \\
\fe
\ie{}
&\xymatrix{
*++[F-,][F**:Generic]{{\boldsymbol{ A_1 \bar{A}_1}}}} \qquad \xymatrix @C=7pc @R=7pc @!0 @dr {
*++[F][F**:Generic]{[\frac12]^{(1,0,0)}_{\frac12}} \ar[r]|--{{~\bar{\cQ}~}} \ar[d]|--{~\cQ~}
& *++[F][F**:Generic]{[0]^{(0,0,0)}_{1}} \\ 
*++[F][F**:Generic]{[1]^{(0,1,0)}_{1}} \ar[d]|--{~\cQ~} \\
*++[F][F**:Generic]{[\frac32]^{(0,0,1)}_{\frac32}} \ar[d]|--{~\cQ~} \\
*++[F][F**:Generic]{[2]^{(0,0,0)}_{2}}
} \\ \\
\fe
as well as their conjugate multiplets.

\subsection{$\osp(4^*|\frac12\cN_1)$}

In this section, we study the unitary multiplets of the 1d superconformal algebra
\ie{}
\osp(4^*|\tfrac12\cN_1) \supset \so(4^*) \oplus \usp(\tfrac12\cN_1)_\cR \,,
\fe
where $\so(4^*) \simeq \so(2,1)_{\text{conf}} \oplus \so(3)_{\text{rot}}$ (consult Appendix~\ref{sec:osp4star2NAlgebra} for more details).\footnote{For rules to determine the UIRs of $\osp(2m^*|2n)$, we refer the reader to \cite{Gunaydin:1990ag}.} It appears as a maximal subalgebra of the 4d superconformal algebra
\ie{}
\su(2,2|\tfrac12\cN_1) \supset \osp(4^*|\tfrac12\cN_1)  \,,
\fe
where the transverse rotation symmetry is $\so(3)_{\text{rot}} \simeq \su(2)_{\text{rot}}$, and so can be classified as a type II defect algebra. Through the section, we label UIRs by $[j]_\Delta^{(R)}$, where $(R)$ are $\usp(\tfrac12\cN_1)_\cR$ quantum numbers.
	
\subsubsection{$\cN_1=4$}

The $\cN_1=4$ superconformal algebra $\osp(4^*|2)$ has an $\su(2)_{\text{rot}} \oplus \su(2)_\cR$ R-symmetry. We label the UIRs of $\su(2)_\cR$ by $(R)$, where $R \in \bZ_{\ge 0}$ is an $\su(2)_\cR$ Dynkin label. The supercharges transform as $\su(2)_{\text{rot}} \oplus \su(2)_\cR$ doublets,
\ie{}
\cQ_{a\alpha} \in [1]_{\frac12}^{(1)} \,,
\fe
where $a$ is an $\su(2)_\cR$ index and $\alpha$ is an $\su(2)_{\text{rot}}$ index. We assign weights to the individual supercharges as
\ie{}
&\cQ_{11}[+1]^{(+1)}_{\frac12}, &&\cQ_{21}[+1]^{(-1)}_{\frac12} \,, \\
&\cQ_{21}[-1]^{(+1)}_{\frac12}, &&\cQ_{22}[-1]^{(-1)}_{\frac12} \,.
\fe
The unitarity bounds and shortening conditions are presented in Table~\ref{table:osp4star2}. 
\begin{table}[bhtp]
\small
\begin{center}
\renewcommand{\arraystretch}{1.5}
\begin{tabular}{|c|l|l|l|l|l|}
\hline
\textbf{Name} & \textbf{Primary} & \textbf{Unitarity Bound} & \textbf{BPS} & \textbf{Null State} & $\cQ_*^n\ket{\text{h.w.}} = 0$\\  \hline\hline
$L$ & $[j]_\Delta^{(R)}$ & $\Delta > R + \tfrac12 j+1 $ & $-$ & $-$ & $-$ \\ \hline \hline
$A_1$ & $[j]_\Delta^{(R)}$ $(j>0)$ & $\Delta =R + \tfrac12 j + 1$ & $\frac14$ & $[j-1]^{(R+1)}_{\Delta+\frac12}$ & $\cQ_{12}$ \\ \hline 
\: & $[j]_\Delta^{(0)}$ & $\Delta = \tfrac12 j +1$ & $\frac12$ & $[j-1]^{(1)}_{\Delta+\frac12}$ & $\cQ_{a 2}$ \\ \hline 
$A_2$ & $[0]_\Delta^{(R)}$ & $\Delta = R+1$ & $\frac14$ & $[0]^{(R+2)}_{\Delta+1}$ & $\cQ_{12} \cQ_{11}$  \\ \hline 
\: & $[0]_\Delta^{(0)}$ & $\Delta = 1$ & $\frac12$ & $[0]^{(2)}_{\Delta+1}$ & $\cQ_{a2}\cQ_{11} +\cQ_{12}\cQ_{a1}$  \\ \hline  \hline
$B_1$ & $[0]_\Delta^{(R)}$ & $\Delta = R$ & $\frac12$ & $[0]^{(R+1)}_{\Delta+\frac12}$ & $\cQ_{1\alpha}$ \\ \hline 
\end{tabular}
\caption{Unitary multiplets of $\osp(4^*|2)$.}
\label{table:osp4star2}
\end{center}
\end{table}
The long multiplets decompose at threshold $\Delta_\star = R+\frac12 j +1$ as
\ie{}
L[j]^{(R)}_{\Delta \to \Delta_\star} &= A_1[j]^{(R)}_{\Delta_\star} \oplus A_1[j-1]_{\Delta_\star+\frac12}^{(R+1)} \,, \\
L[0]^{(R)}_{\Delta \to \Delta_\star} &= A_2[0]_{\Delta_\star}^{(R)} \oplus B_1[0]^{(R+2)}_{\Delta_\star+1} \,.
\fe
It follows that $B_1[0]^{(R)}_R$ with $R=1$ is an absolutely protected multiplet. The multiplets whose top components scale with (half-)integer $\Delta \leq 2$ are given by
\ie{}
&\xymatrix{
*++[F-,][F**:Generic]{{\boldsymbol{ B_1}}}} \qquad \xymatrix @C=3pc @R=3pc @r {
*++[F][F**:Generic]{[0]_{1}^{(1)}} \ar[r]|--{{~\cQ~}} 
&*++[F][F**:Generic]{[1]^{(0)}_{\frac32}}
&*++[]{\qquad \:\:}
} \\ \\
\fe
\ie{}
&\xymatrix{
*++[F-,][F**:Generic]{{\boldsymbol{ A_2}}}} \qquad \xymatrix @C=3pc @R=3pc @r {
*++[F][F**:Generic]{[0]_{1}^{(0)}} \ar[r]|--{{~\cQ~}} 
&*++[F][F**:Generic]{[1]^{(1)}_{\frac32}}\ar[r]|--{{~\cQ~}} 
&*++[F][F**:Generic]{[2]^{(0)}_{2}}
}
\fe

\subsubsection{$\cN_1=8$}

The $\cN_1=8$ superconformal algebra $\osp(4^*|4)$ has an $\su(2)_{\text{rot}} \oplus \usp(4)_\cR$ R-symmetry. We label the UIRs of $\usp(4)_\cR$ by $(R_1,R_2)$, where $R_1,R_2 \in \bZ_{\ge 0}$ are $\usp(4)_\cR$ Dynkin labels. In our conventions, $(1,0) = \mathbf{4}$ is the $\usp(4)_\cR$ fundamental and $(0,1) = \mathbf{5}$.\footnote{There is an exceptional isomorphism $\usp(4)_\cR \simeq \so(5)$, under which the Dynkin labels are related by $(R_1,R_2)_{\usp(4)_\cR} \simeq (R_2,R_1)_{\so(5)}$.} The supercharges transform in the fundamental UIR of $\su(2)_{\text{rot}} \oplus \usp(4)_\cR$, with
\ie{}
\cQ_{a\alpha} \in [1]_{\frac12}^{(1,0)} \,,
\fe
where $a$ is an $\usp(4)_\cR$ fundamental index and $\alpha$ is an $\su(2)_{\text{rot}}$ index. We assign weights to the individual supercharges as
\ie{}
&\cQ_{11}[+1]^{(1,0)}_{\frac12}, &&\cQ_{21}[+1]^{(-1,1)}_{\frac12}, &&\cQ_{31}[+1]^{(1,-1)}_{\frac12}, &&\cQ_{41}[+1]^{(-1,0)}_{\frac12} \,, \\
&\cQ_{12}[-1]^{(1,0)}_{\frac12}, &&\cQ_{22}[-1]^{(-1,1)}_{\frac12}, &&\cQ_{32}[-1]^{(1,-1)}_{\frac12}, &&\cQ_{42}[-1]^{(-1,0)}_{\frac12} \,.
\fe
The unitarity bounds and shortening conditions are presented in Table~\ref{table:osp4star4}. 
\begin{table}[htbp]
\small
\begin{center}
\renewcommand{\arraystretch}{1.5}
\begin{tabular}{|c|l|l|l|l|l|}
\hline
\textbf{Name} & \textbf{Primary} & \textbf{Unitarity Bound} & \textbf{BPS} & \textbf{Null State} & $\cQ_*^n\ket{\text{h.w.}} = 0$\\  \hline\hline
$L$ & $[j]_\Delta^{(R_1,R_2)}$ & $\Delta > R_1 +R_2 + \tfrac12 j+1 $ & $-$ & $-$ & $-$ \\ \hline \hline
$A_1$ & $[j]_\Delta^{(R_1,R_2)}$ $(j>0)$ & $\Delta =R_1 + R_2 + \tfrac12 j + 1$ & $\frac18$ & $[j-1]^{(R_1+1,R_2)}_{\Delta+\frac12}$ & $\cQ_{12}$ \\ \hline 
\: & $[j]_\Delta^{(0,R_2)}$  & $\Delta = \tfrac12 j + R_2 +1$ & $\frac14$ & $[j-1]^{(1,R_2)}_{\Delta+\frac12}$ & $\cQ_{12}$, $\cQ_{22}$ \\ \hline 
\: & $[j]_\Delta^{(0,0)}$  & $\Delta = \tfrac12 j +1$ & $\frac12$ & $[j-1]^{(1,0)}_{\Delta+\frac12}$ & $\cQ_{a2}$ \\ \hline 
$A_2$ & $[0]_\Delta^{(R_1,R_2)}$ & $\Delta = R_1+R_2+1$ & $\frac18$ & $[0]^{(R_1+2,R_2)}_{\Delta+1}$ & $\cQ_{12} \cQ_{11}$  \\ \hline 
\: & $[0]_\Delta^{(0,R_2)}$ & $\Delta = R_2+1$ & $\frac14$ & $[0]^{(2,R_2)}_{\Delta+1}$ & $\cQ_{12} \cQ_{11}$, $\cQ_{22}\cQ_{11} +\cQ_{12}\cQ_{21}$  \\ \hline 
\: & $[0]_\Delta^{(0)}$ & $\Delta = 1$ & $\frac12$ & $[0]^{(2,0)}_{\Delta+1}$ & $\cQ_{a2}\cQ_{11} +\cQ_{12}\cQ_{a1}$ \\ \hline  \hline
$B_1$ & $[0]_\Delta^{(R_1,R_2)}$ & $\Delta = R_1+R_2$ & $\frac14$ & $[0]^{(R_1+1,R_2)}_{\Delta+\frac12}$ & $\cQ_{1\alpha}$ \\ \hline 
\: & $[0]_\Delta^{(0,R_2)}$ & $\Delta = R_2$ & $\frac12$ & $[0]^{(1,R_2)}_{\Delta+\frac12}$ & $\cQ_{1\alpha}$, $\cQ_{2\alpha}$ \\ \hline 
\end{tabular}
\caption{Unitary multiplets of $\osp(4^*|4)$.}
\label{table:osp4star4}
\end{center}
\end{table}
The long multiplets decompose at threshold $\Delta_\star = R_1+R_2+\frac12 j +1$ as
\ie{}
L[j]^{(R_1,R_2)}_{\Delta \to \Delta_\star} &= A_1[j]^{(R_1,R_2)}_{\Delta_\star} \oplus A_1[j-1]_{\Delta_\star+\frac12}^{(R_1+1,R_2)} \,, \\
L[0]^{(R_1,R_2)}_{\Delta\to \Delta_\star} &= A_2[0]_{\Delta_\star}^{(R_1,R_2)} \oplus B_1[0]^{(R_1+2,R_2)}_{\Delta_\star+1} \,.
\fe
It follows that $B_1[0]^{(R_1,R_2)}_{R_1+R_2}$ is absolutely protected for $R_1 \leq 1$ and $R_2 \in \bZ_{\ge 0}$. The only multiplet whose top components scales with (half-)integer $\Delta \leq 2$ is
\ie{}
\xymatrix{
*++[F-,][F**:Generic]{{\boldsymbol{ B_1}}}}  \qquad  \xymatrix @C=3pc @R=3pc @r {
*++[F][F**:Generic]{[0]_{1}^{(0,1)}} \ar[r]|--{{~\cQ~}} 
&*++[F][F**:Generic]{[1]^{(1,0)}_{\frac32}}\ar[r]|--{{~\cQ~}} 
&*++[F][F**:Generic]{[2]^{(0)}_{2}}
}
\fe

\subsection{$\osp(\cN_1|2;\bR)$}

In this section, we study the unitary multiplets of the 1d superconformal algebra
\ie{}
\osp(\cN_1|2;\bR) \supset \msp(2,\bR) \oplus \so(\cN_1)_\cR \,,
\fe
where $\msp(2,\bR) \simeq \so(2,1)_{\text{conf}}$. It so happens that this is the only instance of a superconformal algebra $\mf{g}_s$ (with $\cN_1 = 1,3,4$) that never appears as a maximal 1d  subalgebra of the bulk superconformal symmetry $\mf{G}_s$, but rather through a sequence of subalgebras given by\footnote{On the other hand, $\osp(\cN_1|2;\bR)$ appears as a maximal 1d diagonal superconformal subalgebra of the 2d $(\cN_1,\cN_1)$ global superconformal algebra $\osp(\cN_1|2;\bR) \oplus \osp(\cN_1|2;\bR)$. }
\ie{}
\osp\left(\cN_1|2;\bR\right) \subset \su\left(1,1|\cN_1\right) \subset \cdots \subset \mf{G}_s \,,
\fe	
for $\cN_1 = 1,3,4$, where $\so(\cN_1)$ is fully contained within the bulk R-symmetry algebra. It follows that the algebra describes a type I defect, and so its UIRs are labeled by $[-]_\Delta^{(R)}$, where $(R)$ are $\so(\cN_1)_\cR$ quantum numbers. For $\cN_1 = 2$ there is an exceptional isomorphism $\osp(2|2;\bR) \simeq \su(1,1|1)$, where the bosonic algebras are identified as $\msp(2,\bR) \simeq \su(1,1)$ and $\so(2)_\cR \simeq \uu(1)_{\cR_1}$.

\subsubsection{$\cN_1=1$}

The $\cN_1=1$ superconformal algebra $\osp(1|2;\bR)$ has no R-symmetry. It has a single real supercharge 
\ie{}
\cQ \in [-]_{\frac12} \,.
\fe
Consequently, there are no shortening conditions no additional unitarity bounds besides that of the conformal symmetry. Every multiplet consists of the minimal conformal substructure
\ie{}
\xymatrix{
*++[F-,][F**:Generic]{{\boldsymbol{L}}}}  \qquad  \xymatrix @C=3pc @R=3pc @r {
*++[F][F**:Generic]{[-]_{\Delta}} \ar[r]|--{{~\cQ~}} 
&*++[F][F**:Generic]{[-]_{\Delta+\frac12}}
}
\fe
	
\subsubsection{$\cN_1=3$}

The $\cN_1=3$ superconformal algebra $\osp(3|2;\bR)$ has an $\so(3)_\cR \simeq \su(2)_\cR$ R-symmetry, whose UIRs we label by $(R)$, where $R \in \bZ$ is an $\su(2)_\cR$ Dynkin label. The supercharges transform as an $\so(3)_\cR$ vector,
\ie{}
\cQ_{ab} \in [-]_{\frac12}^{(2)} \,,
\fe
where $a,b$ are $\su(2)_\cR$ indices. We assign weights to the individual supercharges as
\ie{}
&\cQ_{11}[-]^{(+2)}_{\frac12}\,, &&\cQ_{12}[-]^{(0)}_{\frac12}\,, &&\cQ_{22}[-]^{(-2)}_{\frac12} \,.
\fe
The unitarity bounds and shortening conditions are presented in Table~\ref{table:osp32R}. 
\begin{table}[htbp]
\small
\begin{center}
\renewcommand{\arraystretch}{1.5}
\begin{tabular}{|c|l|l|l|l|l|}
\hline
\textbf{Name} & \textbf{Primary} & \textbf{Unitarity Bound} & \textbf{BPS} & \textbf{Null State} & $\cQ_*^n\ket{\text{h.w.}} = 0$\\  \hline\hline
$L$ & $[-]_\Delta^{(R)}$ & $\Delta > \frac14 R$ & $-$ & $-$ & $-$ \\ \hline \hline
$A_1$ & $[-]_\Delta^{(R)}$ & $\Delta = \frac14 R $ & $\frac13$ & $[-]^{(R+2)}_{\Delta+\frac12}$ & $\cQ_{11}$ \\ \hline 
\end{tabular}
\caption{Unitary multiplets of $\osp(3|2;\bR)$.}
\label{table:osp32R}
\end{center}
\end{table}
The long multiplets decompose at threshold $\Delta_\star = \frac14 R$ as
\ie{}
L[-]^{(R)}_{\Delta \to \Delta_\star} = A_1[-]_{\Delta_\star}^{(R)}  \oplus A_1[-]_{\Delta_\star+\frac12}^{(R+2)} \,.
\fe
The multiplets whose top component scales with (half-)integer $\Delta \leq 2$ are given by
\ie{}
&\xymatrix{
*++[F-,][F**:Generic]{{\boldsymbol{L}}}}  \qquad  \xymatrix @C=3pc @R=3pc @r {
*++[F][F**:Generic]{[-]^{(0)}_{\frac12}} \ar[r]|--{{~\cQ~}} 
&*++[F][F**:Generic]{[-]_{1}^{(2)}}\ar[r]|--{{~\cQ~}} 
&*++[F][F**:Generic]{[-]_{\frac32}^{(2)}}\ar[r]|--{{~\cQ~}} 
&*++[F][F**:Generic]{[-]_{2}^{(0)}}
} \\ \\
\fe
\ie{}
&\xymatrix{
*++[F-,][F**:Generic]{{\boldsymbol{A}_1}}}  \quad \:\: \xymatrix @C=3pc @R=3pc @r {
*++[F][F**:Generic]{[-]^{(2)}_{\frac12}} \ar[r]|--{{~\cQ~}} 
&*++[F][F**:Generic]{[-]^{(0) \oplus (2)}_1}\ar[r]|--{{~\cQ~}} 
&*++[F][F**:Generic]{[-]^{(0)}_{\frac32}}
} \\ \\
\fe
\ie{}
&\xymatrix{
*++[F-,][F**:Generic]{{\boldsymbol{A}_1}}} \quad \:\:  \xymatrix @C=3pc @R=3pc @r {
*++[F][F**:Generic]{[-]^{(4)}_{1}} \ar[r]|--{{~\cQ~}} 
&*++[F][F**:Generic]{[-]^{(2) \oplus (4)}_{\frac32}}\ar[r]|--{{~\cQ~}} 
&*++[F][F**:Generic]{[-]^{(2)}_{2}}
} \\ \\
\fe
Here, $[-]^{(0)}_1 \in A_1[-]^{(2)}_{\frac12}$ and $[-]^{(2)}_{\frac32} \in A_1[-]^{(4)}_{1}$ are sporadic (but manifest) top components. For later convenience, we note that there is no multiplet with a top component transforming as $[-]_1^{(4)}$.
	
\subsubsection{$\cN_1=4$}

The $\cN_1=4$ superconformal algebra $\osp(4|2;\bR)$ has an $\so(4)_\cR \simeq \su(2)_{R_1} \oplus \su(2)_{R_2}$ R-symmetry, whose UIRs we label by $(R_1,R_2)$, where $R_1,R_2 \in \bZ_{\ge 0}$ are $\su(2)_{R_1}$ and $\su(2)_{R_2}$ Dynkin labels, respectively. The supercharges transform as an $\so(4)_\cR$ vector, 
\ie{}
\cQ_{a \dot{a}} \in [-]^{(1,1)}_{\frac12} \,,
\fe
where $a,\dot{a}$ are $\su(2)_{R_1}$ and $\su(2)_{R_2}$ indices, respectively. We assign the individual supercharges weights according to
\ie{}
&\cQ_{1\dot{1}}[-]^{(+1,+1)}_{\frac12}, &&\cQ_{1\dot{2}}[-]^{(+1,-1)}_{\frac12} \,, \\
&\cQ_{2\dot{1}}[-]^{(-1,+1)}_{\frac12}, &&\cQ_{2\dot{2}}[-]^{(-1,-1)}_{\frac12} \,.
\fe
The unitarity bounds and shortening conditions are presented in Table~\ref{table:osp42R}.
\begin{table}[htbp]
\small
\begin{center}
\renewcommand{\arraystretch}{1.5}
\begin{tabular}{|c|l|l|l|l|l|}
\hline
\textbf{Name} & \textbf{Primary} & \textbf{Unitarity Bound} & \textbf{BPS} & \textbf{Null State} & $\cQ_*^n\ket{\text{h.w.}} = 0$\\  \hline\hline
$L$ & $[-]_\Delta^{(R_1,R_2)}$ & $\Delta > \frac14(R_1+R_2) $ & $-$ & $-$ & $-$ \\ \hline \hline
$A_1$ & $[-]_\Delta^{(R_1,R_2)}$ & $\Delta = \frac14(R_1+R_2) $ & $\frac14$ & $[-]^{(R_1+1,R_2+1)}_{\Delta+\frac12}$ & $\cQ_{1 \dot 1}$ \\ \hline 
\: & $[-]_\Delta^{(0,R_2)}$ & $\Delta = \tfrac14 R_2$ & $\frac12$ & $[-]^{(1,R_2+1)}_{\Delta+\frac12}$ & $\cQ_{1 \dot 1}$, $\cQ_{2 \dot 1}$\\ \hline
\: & $[-]_\Delta^{(R_1,0)}$ & $\Delta = \tfrac14 R_1$ & $\frac12$ & $[-]^{(R_1+1,1)}_{\Delta+\frac12}$ & $\cQ_{1 \dot 1}$, $\cQ_{1 \dot 2}$  \\ \hline
\end{tabular}
\caption{Unitary multiplets of $\osp(4|2;\bR)$.}
\label{table:osp42R}
\end{center}
\end{table}
The long multiplets decompose at threshold $\Delta_\star = \frac14(R_1+R_2)$ as
\ie{}
L[-]^{(R_1,R_2)}_{\Delta \to \Delta_\star} = A_1[-]_{\Delta_\star}^{(R_1,R_2)}  \oplus  A_1[-]_{\Delta_\star+\frac12}^{(R_1+1,R_2+1)}  \,.
\fe
The multiplets whose top component scales with (half-)integer $\Delta \leq 2$ are given by
\ie{}
&\xymatrix{
*++[F-,][F**:Generic]{{\boldsymbol{A}_1}}}  \qquad \xymatrix @C=3pc @R=3pc @r {
*++[F][F**:Generic]{[-]^{(1,1)}_{\frac12}} \ar[r]|--{{~\cQ~}} 
&*++[F][F**:Generic]{[-]_1^{(0,0)\oplus(2,0)\oplus(0,2)}}\ar[r]|--{{~\cQ~}} 
&*++[F][F**:Generic]{[-]_{\frac32}^{(1,1)}}\ar[r]|--{{~\cQ~}} 
&*++[F][F**:Generic]{[-]_{2}^{(0,0)}}
} \\ \\
\fe
\ie{}
&\xymatrix{
*++[F-,][F**:Generic]{{\boldsymbol{A}_1}}} \qquad \xymatrix @C=3pc @R=3pc @r {
*++[F][F**:Generic]{[-]^{(2,0)}_{\frac12}} \ar[r]|--{{~\cQ~}} 
&*++[F][F**:Generic]{[-]^{(1,1)}_{1}}\ar[r]|--{{~\cQ~}} 
&*++[F][F**:Generic]{[-]^{(0,0)}_{\frac32}}
} \\ \\
\fe
\ie{}
&\xymatrix{
*++[F-,][F**:Generic]{{\boldsymbol{A}_1}}} \qquad \xymatrix @C=3pc @R=3pc @r {
*++[F][F**:Generic]{[-]^{(0,2)}_{\frac12}} \ar[r]|--{{~\cQ~}} 
&*++[F][F**:Generic]{[-]^{(1,1)}_{1}}\ar[r]|--{{~\cQ~}} 
&*++[F][F**:Generic]{[-]^{(0,0)}_{\frac32}}
} \\ \\
\fe
\ie{}
&\xymatrix{
*++[F-,][F**:Generic]{{\boldsymbol{A}_1}}} \qquad \xymatrix @C=3pc @R=3pc @r {
*++[F][F**:Generic]{[-]^{(4,0)}_{1}} \ar[r]|--{{~\cQ~}} 
&*++[F][F**:Generic]{[-]^{(3,1)}_{\frac32}}\ar[r]|--{{~\cQ~}} 
&*++[F][F**:Generic]{[-]^{(2,0)}_{2}}
} \\ \\
\fe
\ie{}
&\xymatrix{
*++[F-,][F**:Generic]{{\boldsymbol{A}_1}}} \qquad \xymatrix @C=3pc @R=3pc @r {
*++[F][F**:Generic]{[-]^{(0,4)}_{1}} \ar[r]|--{{~\cQ~}} 
&*++[F][F**:Generic]{[-]^{(1,3)}_{\frac32}}\ar[r]|--{{~\cQ~}} 
&*++[F][F**:Generic]{[-]^{(0,2)}_{2}}
} \\ \\
\fe
where $[-]_1^{(0,0)} \in A_1[-]_{\frac12}^{(1,1)}$ is a sporadic (but manifest) top component. For later convenience, we note that there is no multiplet with a top component transforming as $[-]_1^{(2,2)}$.

\subsection{$D(2,1;2;0)$}

In this section, we study the unitary multiplets of the $\cN_1=4$ superconformal algebra
\ie{}
D(2,1;2;0) \supset \msl(2,\bR) \oplus \so(4)\,,
\fe
where $\msl(2,\bR) \simeq \so(2,1)_{\text{conf}}$ and $\so(4) \simeq \su(2)_{\text{left}} \oplus \su(2)_\cR$ (consult Appendix~\ref{sec:D21lambdaAlgebra} for more details). It appears as a maximal subalgebra of the 5d superconformal algebra
\ie{}
F(4;2) \supset D(2,1;2;0) \oplus \su(2)_{\text{right}} \,,
\fe
where the transverse rotation symmetry is $\so(4)_{\text{rot}} \simeq \su(2)_{\text{left}} \oplus \su(2)_{\text{right}}$, and so can be classified as a type II defect algebra. We label UIRs by $([j]_\Delta^{(R)}, \bf{j'}+1)$, where $j,R,j' \in \bZ_{\ge 0}$ are $\su(2)_{\text{left}}$, $\su(2)_\cR$, and $\su(2)_{\text{right}}$ Dynkin labels, respectively. From this viewpoint, $([1],\bf{0})$ and $([0],\bf{1})$ are chiral and anti-chiral $\so(4)_{\text{rot}}$ spinors, respectively, while $([1],\bf{2})$ is the vector. In what follows we suppress the $\su(2)_{\text{right}}$ quantum numbers when discussing the unitary multiplets of $D(2,1;2;0)$, though it will return later when we classify broken symmetries of the defect (see Section~\ref{sec:classify}). The supercharges transform as $\su(2)_{\text{left}} \oplus \su(2)_\cR$ doublets, 
\ie{}
\cQ_{a \alpha} \in [1]_{\frac12}^{(1)} \,,
\fe
where $a$ is an $\su(2)_\cR$ index and $\alpha$ is an $\su(2)_{\text{left}}$ index. We assign weights to the individual supercharges as
\ie{}
&\cQ_{11}[+1]^{(+1)}_{\frac12}, &&\cQ_{21}[+1]^{(-1)}_{\frac12} \,, \\
&\cQ_{12}[-1]^{(+1)}_{\frac12}, &&\cQ_{22}[-1]^{(-1)}_{\frac12} \,.
\fe
The unitarity bounds and shortening conditions are presented in Table~\ref{table:D212}. The long multiplets decompose at threshold $\Delta=\frac32R + j + 2$ as
\begin{table}[htbp]
\small
\begin{center}
\renewcommand{\arraystretch}{1.5}
\begin{tabular}{|c|l|l|l|l|l|}
\hline
\textbf{Name} & \textbf{Primary} & \textbf{Unitarity Bound} & \textbf{BPS} & \textbf{Null State} & $\cQ_*^n\ket{\text{h.w.}} = 0$\\  \hline\hline
$L$ & $[j]_\Delta^{(R)}$ & $\Delta > \frac32R + j +2 $ & $-$ & $-$ & $-$ \\ \hline \hline
$A_1$ & $[j]_\Delta^{(R)}$ $(j>0)$ & $\Delta = \frac32 R + j + 2$ & $\frac14$ & $[j-1]^{(R+1)}_{\Delta+\frac12}$ & $\cQ_{12}$ \\ \hline 
\: & $[j]_\Delta^{(0)}$ & $\Delta = j + 2$ & $\frac12$ & $[j-1]^{(1)}_{\Delta+\frac12}$ & $\cQ_{a2}$ \\ \hline 
$A_2$ & $[0]_\Delta^{(R)}$ & $\Delta = \frac32 R + 2$ & $\frac14$ & $[0]^{(R+2)}_{\Delta+1}$ & $\cQ_{12} \cQ_{11}$  \\ \hline 
\: & $[0]_\Delta^{(0)}$ & $\Delta = 2$ & $\frac12$ & $[0]^{(2)}_{\Delta+1}$ & $\cQ_{a2}\cQ_{11} +\cQ_{12}\cQ_{a1}$  \\ \hline  \hline
$B_1$ & $[0]_\Delta^{(R)}$ & $\Delta = \frac32 R$ & $\frac12$ & $[0]^{(R+1)}_{\Delta+\frac12}$ & $\cQ_{1\alpha}$ \\ \hline 
\end{tabular}
\caption{Unitary multiplets of $D(2,1;2;0)$. }
\label{table:D212}
\end{center}
\end{table}
\ie{}
L[j]^{(R)}_{\Delta \to \Delta_\star} &= A_1[j]^{(R)}_{\Delta_\star} \oplus A_1[j-1]_{\Delta_\star+\frac12}^{(R+1)} \,, \\
L[0]^{(R)}_{\Delta \to \Delta_\star} &= A_2[0]_{\Delta_\star}^{(R)} \oplus B_1[0]^{(R+2)}_{\Delta_\star+1} \,.
\fe
It follows that $B_1[0]^{(R)}_\Delta$ with $\Delta=\frac32 R$ and $R=1$ is an absolutely protected multiplet. The only multiplet whose top component scales as a (half-)integer $\Delta \leq 2$ is
\ie{}
\xymatrix{
*++[F-,][F**:Generic]{{\boldsymbol{ B_1}}}} \qquad \xymatrix @C=3pc @R=3pc @r {
*++[F][F**:Generic]{[0]_{\frac32}^{(1)}} \ar[r]|--{{~\cQ~}} 
&*++[F][F**:Generic]{[1]^{(0)}_{2}}
}
\fe


\section{Displacement Multiplets and General Broken Current Multiplets}
\label{sec:importantMult}

Symmetry-breaking is an innate nature of defects. While it is well-known that QFTs in general are nontrivially constrained by the symmetries they obey, here we shall see that broken symmetries (together with the preserved ones) are also quite powerful in determining a part of the operator spectrum on a conformal defect. Let us briefly recall the (broken) symmetries of the setup. The $d$-dimensional bulk SCFT has the full symmetry algebra $\mf{G}=\mf{G}_s\oplus \mf{G}_f$, which consists of the superconformal symmetry $\mf{G}_s$ and a possibly continuous flavor symmetry $\mf{G}_f$. The $p$-dimensional superconformal defect $\cD$ only preserves a symmetry subalgebra $\mf{g}=\mf{g}_s\oplus \mf{g}_b\oplus \mf{g}_f$ that consists of the superconformal subalgebra $\mf{g}_s$ on the defect worldvolume $\Sigma$, a commuting bosonic factor $\mf{g}_b$ from combinations of bulk R-symmetry and transverse rotations $\mf{so}(d-p)$, and a residual flavor symmetry $\mf{g}_f\subset \mf{G}_f$. The broken symmetry charges are characterized by the quotient $\mf{G}/\mf{g}$.

As we explain in what follows, there is a distinguished set of defect local operators that keeps track of the symmetries in the parent SCFT broken by the defect insertion. They can be identified by studying modifications of the Ward identities for bulk conserved currents by contact terms localized on the defect worldvolume. Furthermore, by including these with the preserved superconformal symmetry $\mf{g}_s$ on the defect $\cD$, we deduce simple structural theorems for the candidate superconformal multiplets that accommodate these defect operators. Existence of the corresponding superconformal multiplets as UIRs of the superconformal algebra $\mf{g}_s$ then leads to nontrivial constraints on the defect $\cD$ itself, which we explore in more detail in the subsequent section.

We emphasize that the discussion in this section concerns a general $p$-dimensional superconformal defect $\cD$ in a general $d$-dimensional SCFT.

\subsection{Defect Operators from Broken Bulk Symmetries}
\label{sec:brokencurrents}
\subsubsection{Broken Conformal Symmetries and Displacement Operators}
The most fundamental defect operator is the displacement operator ${\mathsf D}^i$, which measures transverse deformations of the defect. It can be identified as the coefficient of the contact term present in the modified Ward identity for translation invariance in the normal directions \cite{Billo:2016cpy},\footnote{Throughout this work, we assume that the displacement operator and more general defect operators associated to broken bulk currents are \textit{normalizable}.}
\ie
&\pa_\m T^{\m i} (x) = -\D_\Sigma (x_\perp) {\mathsf D}^i (x_\parallel) \,.
\label{disward}
\fe
The Ward identities for translation invariance in the longitudinal directions as well as those for scale invariance are preserved,
\ie
&\pa_\m T^{\m a}(x) = T^\m {}_\m(x)=0\,.
\fe
If the defect further breaks the  transverse  $\so(d-p)_{\text{rot}}$ rotation symmetry, there is an additional defect operator $\lambda^{[ij]}$ from
\ie
& T^{[ij]}(x)=\D_\Sigma (x_\perp )\lambda^{[ij]}(x_\parallel)\,.
\label{rotward}
\fe
We emphasize that while these (modified) Ward identities in the presence of the defect are subject to ambiguities from redefinitions of the bulk stress-tensor $T_{\m\n}$ by defect contact terms involving $\D_\Sigma (x_\perp)$ and its normal derivatives, the operators ${\mathsf D}^i$ and $\lambda^{[ij]}$ are unambiguously defined up to conformal descendants along the defect.\footnote{In particular, our stress-tensor $T_{\m\n}$  is related to $T^{\mathsf tot}_{\m\n}$ in (5.21) of \cite{Billo:2016cpy} by 
	\ie
	T^{\m\n}(x)=T_{\mathsf tot}^{\m\n}(x)+\D^\m_i \D^\n_j \D_\Sigma (x_\perp )\lambda^{[ij]}(x_\parallel)\,.
	\fe
	Note that the analysis in \cite{Billo:2016cpy} focuses on defects that preserve the transverse rotation symmetry, in which case $\lambda^{[ij]}$ is a conformal descendant (total derivative on the defect). For the case of defects that break transverse rotations, $\lambda^{[ij]}$ is a nontrivial conformal primary. 
} 
The last ambiguity can be repaired by requiring the stress-tensor $T_{\m\n}$ (or rather its components $T_{ab},T_{ai}$ and $T_{ij}$) in the coincident limit with the defect to transform as a conformal primary with respect to the defect conformal algebra $\so(2,p)_{\text{conf}}$. This ensures that the operators ${\mathsf D}^i$ and $\lambda^{[ij]}$ in \eqref{disward} and \eqref{rotward} are automatically conformal primaries.\footnote{One can check this explicitly by acting with the special conformal generators $K_a$ in the defect conformal subalgebra on the left-hand-side of \eqref{disward} and \eqref{rotward}.}

The spacetime quantum numbers of the defect operators immediately follow from the Ward identities. As in \eqref{disward}, one can deduce that the displacement operator ${\mathsf D}^i$ is a scalar of dimension $\Delta_{{\mathsf D}}=p+1$, and transforms as a vector under $\so(d-p)_{\text{rot}}$ transverse rotations. Similarly from \eqref{rotward}, we have that $\lambda^{[ij]}$ is a scalar operator of dimension $\Delta_\lambda=p$ and transforms in the rank-two anti-symmetric representation of $\so(d-p)_{\text{rot}}$.\footnote{More generally, if the defect breaks a subgroup of $SO(d-p)_{\text{rot}}$, the primary components of $\lambda^{[ij]}$ transform in the adjoint representation of this subgroup.} 

In the radially quantized CFT (after Wick rotation), we find that these modified Ward identities, after integration over the full spacetime, lead to commutation relations involving the defect with the conserved charges
\ie
{} [P_i,  \cD]  =&\int_\Sigma      d^p x_\parallel \,    {\mathsf{D}}_i(x_\parallel)  \cD\,,
\\
{} [M_{ij},  \cD]  =&\int_\Sigma     d^p x_\parallel \,    {\lambda}_{[ij]}(x_\parallel)  \cD\,.
\label{PMcom}
\fe
Here, we take the defect to be spherical (after a conformal transformation) and sandwiched between two radial slices of the spacetime, $S^{d-1}_{\tau_1}$ and $S^{d-1}_{\tau_2}$, with $\tau_1<\tau_2$. The transverse translation and rotation charges are defined in the usual way as $P_i\equiv \int dS^\m T_{\m i}$ and $M_{ij}\equiv \int dS^\m T_{\m [i}x_{j]}$, where $S^\m$ denotes the area element on the radial slices. We emphasize that the equations above are operator identities that hold in the presence of other (defect) operator insertions in the CFT, as long as their radial locations satisfy $\tau<\tau_1$ or $\tau>\tau_2$. 
One notable feature of the extended defect as compared to local operators in the CFT is that the defect transforms \textit{non-linearly} under the conformal symmetries (and general zero-form symmetries).

\subsubsection{Broken Supersymmetries}
In an SCFT, the defect additionally breaks superconformal symmetries. Following the same reasoning as for the broken conformal symmetries, we find that the modified Ward identities  
\ie
&\pa_\m S^{\m }_\A(x) = \D_\Sigma (x_\perp)  {\mathsf S}_\A(x_\parallel),\quad S^\m_\A(x) (\C_\m)^{\A}{}_\B=0
\label{sward}
\fe
give rise to a fermionic defect primary ${\mathsf S}_\A$ of dimension $\Delta_{{\mathsf S}}=p+{1\over 2}$ that keeps track of the broken supersymmetries. The integrated Ward identity determines the action of the broken supercharge $Q_B$ on the defect $\cD$,
\ie
{} [Q_B^\A,  \cD]  =&\int_\Sigma     d^p x_\parallel \,    {\mathsf S}^\A(x_\parallel)  \cD\,.
\label{Scom}
\fe
The defect operator  ${\mathsf S}_\A$ transforms in the spinor representation of $\so(d-p)_{\text{rot}}$ and in certain representations of the residual R-symmetry preserved by the defect,  namely those of the broken supercurrents that appear in the modified Ward identities \eqref{sward}.

\subsubsection{Broken R-symmetries and Flavor Symmetries}
Similarly, for the broken  R-symmetry and flavor symmetry currents $J_R,J_F$ we have
\ie
&\pa_\m J_{R,F}^{\m }(x) = \D_\Sigma (x_\perp)  {\mathsf J}_{R,F}(x_\parallel)\,, \label{jward}
\fe
which lead to marginal scalar primaries ${\mathsf J}_{R,F}$ on the defect worldvolume. Correspondingly, the broken R-symmetry and flavor symmetry charges act on $\cD$ as
\ie
{} [R_B ,  \cD]  =&\int_\Sigma     d^p x_\parallel \,    {\mathsf J}_R (x_\parallel)  \cD\,,
\\
[F_B ,  \cD]  =&\int_\Sigma     d^p x_\parallel \,    {\mathsf J}_F (x_\parallel)  \cD\,.
\label{RFcom}
\fe
From the modified Ward identity \eqref{jward} for the broken currents, we see that the defect operators $J_{R,F}$ are $\so(d-p)_{\text{rot}}$ singlets, and transform under the residual R-symmetry and flavor symmetry accordingly.  

\subsubsection{Broken Higher-form Symmetries}

Before we end our discussion on broken symmetries, let us briefly comment on higher-form symmetries, which can also be broken by the (super)conformal defect $\cD$ and consequently give rise to local operators on its worldvolume $\Sigma$.\footnote{It follows from the general classification  of conserved current multiplets in \cite{Cordova:2016emh} that \textit{interacting} unitary SCFTs never admit higher-form symmetry currents. We thank Ken Intriligator for explaining this point to us.} Consider a $q$-form symmetry current $J^{\m \m_1\dots \m_q}$ with $q\geq 1$: its conservation equation can be modified in the presence of a $p$-dimensional conformal defect as
\ie
\pa_\m J^{\m \m_1\dots \m_q} =\D_\Sigma(x_\perp){\mathsf J}^{i_1\dots i_q}(x_\parallel)\D_{[i_1\dots i_q]}^{\m_1\dots \m_q}\,.
\fe
Since $J^{\m \m_1\dots \m_q}$ has scaling dimension $\Delta=d-q-1$, we see by unitarity that the RHS can be nontrivial only if $p\geq q$, with ${\mathsf J}^{i_1\dots i_q}$ the induced defect primary.\footnote{Recall that the operator which generates a $q$-form symmetry is a $d-q-1$-dimensional topological defect. Thus it can only act nontrivially on defect operators of dimension $p$ such that $d-q-1+p\geq d-1$, otherwise the topological defect can be shrunk at no cost.} In the case $p=q$, ${\mathsf J}^{i_1\dots i_q}$ is topological and measures the charge of the defect under the $q$-form symmetry.

\subsection{Structural Theorems for Defect Superconformal Multiplets}
\label{sec:struc}
We have seen how broken symmetries of the bulk CFT, due to the defect insertion, lead  to conformal primaries on the defect worldvolume. For defects that preserves a fraction of the bulk superconformal symmetries, the defect conformal primaries further organize into superconformal multiplets of $\mf{g}_s$. These multiplets must obey certain structure criteria, which are as follows.

\begin{theorem}
\label{thm1}
	The displacement operator ${\mathsf D}^i$ is always a top component in a defect superconformal multiplet (possibly reducible). The corresponding multiplet is referred to as the $\mf{g}_s$ displacement multiplet of the defect $\cD$.  
\end{theorem}

To show this, we act with one of the preserved supercharges $\cQ \in  \mf{g}_s$  on the commutation relation involving $P_i$ in \eqref{PMcom}
\ie
{}[\cQ, [P_i,  \cD]]  =&\int_\Sigma      d^p x_\parallel \,  [\cQ,  {\mathsf{D}}_i(x_\parallel)]  \cD\,.
\\
\fe
Using the fact that $[\cQ, \cD]=0$ and the Jacobi identities, we conclude $[\cQ, [P_i,  \cD]] =[\cQ,P_i],\cD]=0$, and consequently $ [\cQ,  {\mathsf{D}}_i(x_\parallel)]$ must be a total derivative (if non-vanishing). Thus we have that the displacement operator ${\mathsf D}_i$ is a top component in the corresponding superconformal multiplet of $\mf{g}_s$, namely the displacement multiplet. 

By a similar argument, we deduce 
\begin{theorem}A broken charge  $U_B$ in $\mf{G}/\mf{g}$ that commutes with the supercharges $\cQ \in  \mf{g}_s$ gives rise to a supermultiplet  of $\mf{g}_s$ whose top component  is the conformal primary  $\cO_{U_B}$ within
	\ie
	{}  [U_B,  \cD]   =&\int_\Sigma      d^p x_\parallel \,    \cO_{U_B}(x_\parallel)  \cD\,.
	\\
	\fe
	In particular, a broken flavor symmetry charge always induces a top component of a broken F-current multiplet.
\label{thm2}
\end{theorem}
Note that the top component ${\mathsf J}_F$ induced from a broken flavor symmetry charge $F_B$ is marginal on the defect worldvolume $\Sigma$. Its integral $\int_\Sigma     d^p x_\parallel \,    {\mathsf J}_F (x_\parallel)$ is a singlet under the defect superconformal algebra $\mf{g}_s$, and thus preserves  $\mf{g}_b$, but breaks (part of) the flavor symmetry $\mf{g}_f$. Consequently it gives rise to a marginal supersymmetric deformation of the defect $\cD$ (see Section~\ref{sec:defor} for a related discussion).  

One can also deduce structures deeper down in the defect superconformal multiplets using the (anti)commutation relations between $\mf{g}_s$ and the broken charges. In particular, a set of broken charges in $\mf{G}/\mf{g}$ that are related by  (anti)commutators with elements in $\mf{g}_s$ induce conformal primaries that sit in the same $\mf{g}_s$ supermultiplet.
For example, if a combination of anti-commutators of $U_B\in   \mf{G}/\mf{g}$ with $\cQ$ (multiple-times) lead to $P_i$,  the corresponding defect operator $\cO_{U_B}$ must be a part of the displacement multiplet. We emphasize that not all conformal primaries in a $\mf{g}_s$ multiplet whose top component is associated to a broken symmetry current arise this way. For example, the superconformal primary of the displacement multiplet can be {\it emergent} and so does not correspond to any bulk symmetry broken by the defect.\footnote{This is akin to the stress-tensor multiplets in SCFTs with extended supersymmetry, in which case the superconformal primaries are not conserved currents \cite{Cordova:2016emh}.}  Nevertheless, as we will explain in the next section, the above information together with the unitary representations of $\mf{g}_s$ (see Section~\ref{sec:classify}) are sufficient to pin down the entire multiplet structure.

The displacement multiplet for the defect $\cD$ is generally reducible under the defect superconformal algebra $\mf{g}_s$. 
\begin{theorem}
\label{thm3}
	If the transverse rotation symmetry is preserved, the displacement multiplet is irreducible under $\mf{g}_s$.
\end{theorem}
This is obvious if the transverse rotation symmetry $\mf{so}(d-p)_{\text{rot}}$ is a subalgebra of $\mf{g}_s$. More generally, it suffices to show that given two transverse translation generators $P_i$ and $P_j$, there exist $\cQ,\cS\in \mf{g}_s$ such that 
\ie
{}[\{\cQ,\cS\}],P_i]= \A  P_i+ \B P_j 
\fe
for $\B \neq 0$. Since $M_{ij} \in \mf{g}$ by assumption, it must map the preserved supercharges to themselves, and so we can assign definite $U(1)$ charges to $\cQ$ as $\cQ_\pm$ under the $M_{ij}$ rotation (similarly for the superconformal charges $\cS$). We take any one of these supercharges $\cQ_+$, and its Hermitian conjugate  (under radial quantization), which we refer to as $\cS_-$. Then from the structure of the bulk superconformal algebra, we have in general,
\ie
{}[\{\cQ_+,\cS_-\}],P_i]=\A_i  P_i+ \A_j P_j + \sum_{k\neq i,j} \A_k P_k
\fe
with $\A_i\neq 0$.
Performing a $\pi$-rotation by $M_{ij}$ on both sides, we find
\ie
{}[\{\cQ_+,\cS_-\}],P_i]=\A_i  P_i+ \A_j P_j - \sum_{k\neq i,j} \A_k P_k\,,
\fe
and thus 
\ie
{}[\{\cQ_+,\cS_-\}],P_i]=\A_i  P_i+ \A_j P_j  \,.
\fe
Instead, performing a $\pi/2$-rotation by $M_{ij}$ on the above equation gives
\ie 
{}[\{\cQ_+,\cS_-\}],P_j]= \A_i  P_j- \A_j P_i \,.
\fe
Together with 
\ie{}
[\{\cQ_+,\cS_-\}],P_i-i P_j]= 0\,, 
\fe
which follows from the Jacobi identity and $[P_i-i P_j,\cS_-]=0$, we conclude
\ie
{}[\{\cQ,\cS\}],P_i]=\A_i( P_i+ i P_j) 
\fe
as desired.

Note that the argument above only relies on the fact that $\mf{so}(p-d)_{\text{rot}}$ transforms the supercharges $\cQ \in\mf{g}_s$ among themselves. Thus, the displacement multiplet will stay irreducible as long as this feature remains, even if $\mf{so}(p-d)_{\text{rot}}$ is no longer a symmetry of the defect $\cD$.

Given the defect supermultiplets that accommodate broken bulk currents for a defect $\cD$ preserving some symmetry algebra $\mf{g}$, it is straightforward to deduce the corresponding defect supermultiplets for another defect $\cD'$ (of the same dimension) that preserves a further subalgebra $\mf{g}'\subset \mf{g}$. The broken current defect supermultiplets for $\cD'$ can come from two sources. In the first, the broken current multiplets for $\cD$ (including the displacement multiplet) naturally decompose into those of $\cD'$. In the second, there are additional broken supercharges in $\mf{g}/\mf{g}'$ that induce top components in $\mf{g}'_s$ multiplets of dimension $\Delta_{{\mathsf S}}=p+{1\over 2}$ as in \eqref{Scom}. Note that these must be top components because their (anti)commutators with the preserved supercharges in $ \mf{g}'_s$ can only give rise to translations along $\Sigma$, which are preserved by the defect.  If any bosonic symmetry generators in $\mf{g}_b\oplus \mf{g}_f$ are broken by $\cD'$, they generate additional $\mf{g}'_s$ multiplets with top components of dimension $\Delta_{{\mathsf J}}=p$ as in \eqref{RFcom}. Thus we have the following.
\begin{theorem}
\label{thm4}
	For a $p$-dimensional superconformal defect $\cD'$ that preserves a symmetry subalgebra $\mf{g}' \subset \mf{g}$ of the superconformal  defect $\cD$, the further broken charges in $\mf{g}/\mf{g}'$ on the defect $\cD'$ give rise to broken supercurrent multiplets and broken R/F-current multiplets, which contain top components with dimensions $\Delta_{{\mathsf S}}=p+{1\over 2}$ and $\Delta_{{\mathsf J}}=p$, respectively.
\end{theorem}

\section{Classification of Superconformal Line Defects}
\label{sec:classify}

\subsection{Classification Procedure and Summary of Results}

In this section, we provide a systematic classification of all possible superconformal line defects $(p=1)$ based on the allowed 1d superconformal subalgebras $\mf{g}_s$ and their multiplet structure. The logic is quite simple and follows immediately from the results of Section~\ref{sec:ureps}. Each 1d subalgebra $\mf{g}_s$ within a given bulk superconformal algebra $\mf{G}_s$ corresponds to the superconformal symmetries preserved by a putative line defect. As explained in Section~\ref{sec:struc}, the existence of a defect $\cD$ relies crucially on the admission of several distinguished multiplets induced by broken symmetries in the bulk. That is, every broken spacetime (super)current and R/F-symmetry current must be accounted for by a primary operator on the defect with the correct transformation properties, and all such operators must lie in \textit{some} multiplet of the 1d superconformal algebra $\mf{g}_s$.\footnote{We relegate details of specific 1d superconformal subalgebras and the broken bulk generators to Appendix~\ref{sec:1dSCA}.} This suggests a natural classification strategy, which we describe as follows:

\begin{enumerate}
    \item Consider the full symmetry algebra $\mf{G}$ of some SCFT in the bulk. For each \textit{maximal} 1d symmetry subalgebra $\mf{g} \subset \mf{G}$ associated with a putative defect $\cD$ containing the superconformal subalgebra $\mf{g}_s$, we determine the broken symmetries $\mf{G}/\mf{g}$ and their induced operators on $\cD$, and in particular how such operators fit within $\mf{g}_s$ multiplets. 
    
    \item We next consider all possible 1d symmetry subalgebras $\mf{g}' \subset \mf{g}$ associated to defects $\cD'$ preserving some fraction of the superconformal subalgebra $\mf{g}'_s\subset \mf{g}_s$ of $\cD$. The broken symmetries $\mf{G}/\mf{g}'$ are all from either $\mf{G}/\mf{g}' \cap \mf{G}/\mf{g}$, in which case the induced operator on $\cD'$ sits inside a multiplet of $\mf{g}$ and follows directly from the branching rules of $\mf{g} \subset \mf{g}'$, or from $\mf{g}/\mf{g}' \subset \mf{G}$, in which case the induced operator on $\cD'$ resides in a separate multiplet not associated to $\cD$. An important example of the former is the displacement multiplet of $\cD$, which always decomposes into the displacement multiplet of $\cD'$ and several broken R-current multiplets. 
    
    \item  If any induced defect operator from a broken symmetry cannot be accommodated by the multiplet structure of $\mf{g}$  ($\mf{g}'$), the defect $\cD$ is \textit{forbidden}.
\end{enumerate}

For convenience, we recall that the distinguished set of operators induced by broken symmetries consists of the displacement operator $\mathsf{D}_i$ (broken transverse translations), the broken transverse rotation (TR) operator $\lambda_{[ij]}$, the broken supersymmetry operators $\mathsf{S}_{\alpha}$, and the broken R/F-symmetry operators $\mathsf{J}_{R,F}$. Generic operators on the defect with no \textit{a priori} relation to the broken symmetries of the bulk are labeled as $\cO$. For clarity, each CP within a given multiplet is color-coded as follows:
    \ie{}
    \xymatrix @C=7pc @R=7pc @!0 @r {
    *++[F-,][F**:Displacement]{ \mathsf{D}_i  } 
    &*++[F-,][F**:BrokenRot]{ \mathsf{\lambda}_{[ij]} } 
    &*++[F-,][F**:BrokenSUSY]{ \mathsf{S}_\alpha } 
    &*++[F-,][F**:BrokenCurrent]{ \mathsf{J}_{R,F} } 
    &*++[F-,][F**:Generic]{ {\cal O} } 
    }
    \fe

We now present the results of our classification, categorizing the defects by superconformal symmetry in Table~\ref{table:defectClass}, and summarize the main points below. All allowed superconformal line defects in $d=3,4,5$ dimensions preserve a superconformal subalgebra $\mf{g}_s \subset \mf{G}_s$ within the same superalgebra family (e.g. $\su(1,1|N)$ or $\osp(4^*|2n)$) for $\cN_1>1$, according to the bulk dimension. Furthermore, specifying the ambient spacetime dimension and the amount of preserved supersymmetries $\cN_1$ \textit{uniquely} fixes the defect superconformal symmetry (if nontrivial). This is akin to situation in the bulk, where the superconformal symmetry is also fixed by the dimension and the number of bulk supersymmetries $\cN$. In particular, we find that that all superconformal lines with $\cN_1 > 1$ organize into a fairly short list consisting of
\ie{}\label{eq:lineClass}
&d=3: 
&&\begin{cases}
\su(1,1|1) \subset \osp(\cN|4;\bR)\,, &\cN = 2,3,4,5,6,8 \\
\uu(1)\rtimes\psu(1,1|2)\rtimes\uu(1) \subset \osp(\cN|4;\bR)\,, &\cN = 4,5,6,8 \\
\su(1,1|3) \subset \osp(\cN|4;\bR)\,, &\cN = 6,8 \\
\su(1,1|4) \subset \osp(8|4;\bR)\,, \\
\end{cases} \\
&d=4: 
&&\begin{cases}
\osp(4^*|2) \subset \su(2,2|\cN)\,, &\cN=2,3 \\
\osp(4^*|2) \subset \psu(2,2|4)\,, \\
\osp(4^*|4) \subset \psu(2,2|4) \,,
\end{cases} \\
&d=5: && D(2;1;2;0) \subset F(4;2) \,.
\fe
Note the absence of superconformal lines with $\cN_1>1$ in 6d SCFTs.

There is a further sub-classification of \eqref{eq:lineClass} into lines with reduced bosonic symmetry. In general, we find that all $d=3$ lines are allowed to break transverse rotations and any fraction of the bulk flavor symmetry, if any. However, superconformal lines in $d=4,5$ are too rigid and must preserve all bosonic symmetries commuting with their 1d superconformal algebras (see Section~\ref{sec:defor} for related discussions about deforming the superconformal lines).

The case $\cN_1 = 1$ behaves somewhat different from the classification above and so we handle it separately. As it turns out, all SCFTs with 1d superconformal subalgebras automatically admit the \textit{minimal line}, which we define as any defect that preserves the minimal superconformal algebra $\osp(1|2;\bR)$ with a single real supercharge $\cQ$. In all cases, it must completely break $\so(d-1)$ transverse rotations (otherwise the supersymmetry would be enhanced),
but can optionally preserve some of the R-symmetries, including a single $\uu(1)$ generated by a combination of transverse rotations and the bulk R-symmetry, so long as they commute with $\cQ$. Given that the algebra has only a single supercharge and no R-symmetry, it follows that the 1d operators arrange into boson-fermion pairs. Furthermore, since all broken symmetries and their induced operators on the defect can be recast into this form, our classification program automatically allows such defects. In other words, the superconformal kinematics is not powerful enough to rule out any of these lines on the basis of broken symmetries alone. 

For \textit{all} other 1d superconformal algebras, we find that the associated superconformal line defect is forbidden, either because the algebra is not a timelike sub-superconformal algebra of any bulk superconformal algebra, or because the broken symmetries and multiplet structure are incompatible (see Section~\ref{sec:forbiddenLines} relevant details). It is noteworthy that $d=3,4,6$ $\cN=1$ SCFTs cannot accommodate any superconformal lines due to the simple fact that their superconformal algebras do not admit 1d superconformal subalgebras. In addition, $d=6$ $\cN=(2,0)$ SCFTs only admit minimal lines.

\begin{table}[htbp]
\begin{center}
\renewcommand{\arraystretch}{1.5}
\begin{tabular}{|c|c|c|c|c|c|}
\hline
bulk $\backslash$ line & $\cN_1=1$   &  $\cN_1=2$   & $\cN_1=4$  & $\cN_1=6$ & $\cN_1=8$  \\ \hline
3d $\cN = 1$ & & & & & \\
3d $\cN = 2$ &\rejectCross & \acceptBoth & & & \\
3d $\cN = 3$ &\rejectCross & \acceptBoth & & & \\
3d $\cN = 4$ &\rejectCross & \acceptBoth  & \acceptBoth & & \\
3d $\cN = 5$ &\rejectCross & \acceptBoth & \acceptBoth & & \\
3d $\cN = 6$ &\rejectCross & \acceptBoth & \acceptBoth & \acceptBoth & \\
3d $\cN = 8$ & \rejectCross & \acceptBoth & \acceptBoth & \rejectBoth & \acceptBoth \\ \hline
4d $\cN = 1$ & & & & & \\
4d $\cN = 2$ &\rejectCross & & \acceptCirc &  & \\
4d $\cN = 3$ &\rejectCross & & \rejectCirc & & \\
4d $\cN = 4$ &\rejectCross & & \acceptCirc & & \acceptCirc \\ \hline
5d $\cN = 1$ &\rejectCross &  & \acceptCirc & & \\ \hline
6d $\cN = (1,0)$ & & & & & \\
6d $\cN = (2,0)$ &\rejectCross & & & & \\ \hline
\end{tabular}
\caption{Allowed superconformal line defects for interacting SCFTs in $3 \leq d \leq 6$ dimensions, classified by $\cN_1$, the number of preserved real supercharges. We denote SCFTs that admit rotation-preserving (breaking) lines by \protect\sampleCirc \: (\protect\sampleCross) and those that admit both by $\protect\sampleBoth$. Cases found in the literature are colored blue, while those currently absent are colored red. Descriptions and references for the known line defects can be found in Section~\ref{sec:examplelines}.}
\label{table:defectClass}
\end{center}
\end{table}

\subsection{Superconformal Lines in $d=3$}

\subsubsection{$d=3$, $\cN=2$}

\subsubsection*{$\frac12$-BPS Lines}

The $\frac12$-BPS line preserves a \textit{maximal} 1d subalgebra
\ie{}
\su(1,1|1) \oplus \uu(1)_b \oplus \mathfrak{g}_f \supset \so(2,1)_\text{conf} \oplus \uu(1)_{\cR_1} \oplus \uu(1)_b \oplus \mathfrak{g}_f \,,
\fe
of the flavored 3d $\cN=2$ algebra
\ie{}
\osp(2|4;\bR) \oplus \mathfrak{G}_f \supset \so(2,3)_\text{conf} \oplus \so(2)_R \oplus \mathfrak{G}_f \,.
\fe
where $\so(2)_R  \oplus \so(2)_{\text{rot}} \simeq \uu(1)_{\cR_1} \oplus \uu(1)_b$ with the transverse rotation symmetry  $\so(2)_{\text{rot}} \subset \so(1,2) \subset \so(2,3)_\text{conf}$. Analyzing the symmetries broken by the defect leads to the following required multiplets. (Note that we do not label the trivial representation under $\mf{g}_f$, leaving it implicit).

\begin{itemize}

\item{\it Displacement Multiplet:} The broken supercharges induce fermions $[\pm \tfrac32]_{\frac 32}$ on the defect. Together with the displacement operator $[\pm 1]_2$, they sit inside the displacement multiplet $L\bar{A}_1[-\frac32]_{\frac32}$ (and its conjugate multiplet $A_1\bar{L}[\frac32]_{\frac32}$):
\ie{}\label{eq:su(1,1|1)DisplaceMult}
&\xymatrix{
*++[F-,][F**:Displacement]{{\boldsymbol{ L \bar{A}_1}}}}  \qquad \xymatrix @C=7pc @R=7pc @!0 @dr {
*++[F][F**:BrokenSUSY]{[-\frac32]_{\frac32}}  \ar[d]|--{~\cQ~} \\
*++[F][F**:Displacement]{[-1]_{2}}
}
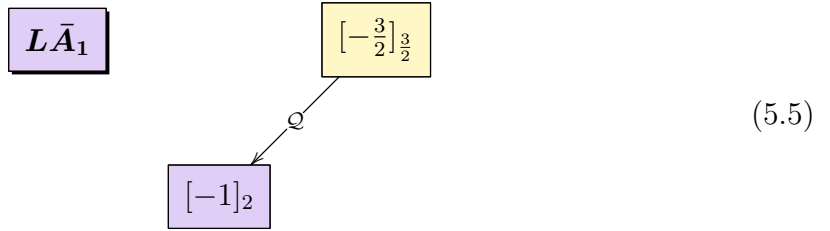
\fe

\item{\it Broken F-Current Multiplet:} The broken flavor symmetries induce marginal scalars $([0]_1,\mf{G}_f/\mf{g}_f)$ on the defect, each of which sits at the top of a broken F-current multiplet $L\bar{A}_1[-\frac12]_{\frac12}$ (or its conjugate multiplet $A_1\bar{L}[\frac12]_{\frac12}$):
\ie{}\label{eq:su(1,1|1)FCurrentMult}
&\xymatrix{
*++[F-,][F**:BrokenCurrent]{{\boldsymbol{ L \bar{A}_1}}}}  \qquad \xymatrix @C=7pc @R=7pc @!0 @dr {
*++[F][F**:Generic]{[-\frac12]_{\frac12}}  \ar[d]|--{~\cQ~} \\
*++[F][F**:BrokenCurrent]{[0]_{1}}
}
\fe
\item{\it Broken TR Multiplet:} It is possible for the defect to break $\uu(1)_b$ (a combination of transverse rotations and the 3d R-symmetry), which induces a marginal scalar $[0]_1$ at the top of the broken TR multiplet $L\bar{A}_1[-\frac12]_{\frac12}$ (or its conjugate multiplet $A_1\bar{L}[\frac12]_{\frac12}$):\footnote{Note that the defect operators induced by broken transverse rotations generally do not correspond to top components in multiplets of the preserved superconformal algebra $\mf{g}_s$. Here the broken $\uu(1)_b$ involves a particular combination of the 3d R-symmetry and $\so(2)_{\text{rot}}$ transverse rotations that commutes with $\mf{g}_s$, and thus the induced defect operator is a top component in a corresponding supermultiplet, namely the broken TR multiplet.}
\ie{}\label{eq:su(1,1|1)TRMult}
&\xymatrix{
*++[F-,][F**:BrokenRot]{{\boldsymbol{ L \bar{A}_1}}}}  \qquad \xymatrix @C=7pc @R=7pc @!0 @dr {
*++[F][F**:Generic]{[-\frac12]_{\frac12}}  \ar[d]|--{~\cQ~} \\
*++[F][F**:BrokenRot]{[0]_{1}}
}
\fe

\end{itemize}

\subsubsection{$d=3$, $\cN=3$}

\subsubsection*{$\frac13$-BPS Line}

The $\frac13$-BPS line preserves a \textit{maximal} 1d subalgebra
\ie{}
\su(1,1|1) \oplus \uu(1)_b \oplus \mathfrak{g}_f \supset \so(2,1)_\text{conf} \oplus \uu(1)_{\cR_1} \oplus \uu(1)_b \oplus \mathfrak{g}_f \,.
\fe
of the flavored 3d $\cN=3$ algebra
\ie{}
\osp(3|4;\bR) \oplus \mathfrak{G}_f \supset \so(2,3)_\text{conf} \oplus \so(3)_{R} \oplus \mathfrak{G}_f \,.
\fe
where $\so(2)_R  \oplus \so(2)_{\text{rot}} \simeq \uu(1)_{\cR_1} \oplus \uu(1)_b$ with $\so(2)_R \subset \so(3)_R$ and the transverse rotation symmetry  $\so(2)_{\text{rot}} \subset \so(1,2) \subset \so(2,3)_\text{conf}$. Analyzing the symmetries broken by the defect leads to the following required multiplets. (Note that we do not label the trivial representation under $\mf{g}_f$, leaving it implicit).

\begin{itemize}

\item{\it Displacement Multiplet:} The $\cN=2$ broken supercharges induce fermions $[\pm \tfrac32]_{\frac 32}$ on the defect. Together with the displacement operator $[\pm 1]_2$, they sit inside the displacement multiplet $L\bar{A}_1[-\frac32]_{\frac32}$ \eqref{eq:su(1,1|1)DisplaceMult} (and its conjugate multiplet $A_1\bar{L}[\frac32]_{\frac32}$).

\item{\it Broken Supercurrent Multiplet:} The broken $\so(3)_R/\so(2)_R$ currents induce marginal scalars $[\pm 1]_1$ on the defect, while the remaining $\cN=3$ broken supercharges induce fermions $[\pm \frac12]_{\frac32}$. Altogether, they sit inside a broken supercurrent multiplet $L\bar{A}_1[-1]_{1}$ (and its conjugate multiplet $A_1\bar{L}[1]_{1}$): 
\ie{}\label{eq:su(1,1|1)RCurrentMult}
&\xymatrix{
*++[F-,][F**:BrokenSUSY]{{\boldsymbol{ L \bar{A}_1}}}}  \qquad \xymatrix @C=7pc @R=7pc @!0 @dr {
*++[F][F**:BrokenCurrent]{[-1]_{1} }  \ar[d]|--{~\cQ~} \\
*++[F][F**:BrokenSUSY]{[-\frac12]_{\frac32}}
}
\fe

\item{\it Broken F-Current Multiplet:} The broken flavor symmetries induce marginal scalars $([0]_1,\mf{G}_f/\mf{g}_f)$ on the defect, each of which sits at the top of a broken current multiplet $L\bar{A}_1[-\frac12]_{\frac12}$ \eqref{eq:su(1,1|1)FCurrentMult} (or its conjugate multiplet $A_1\bar{L}[\frac12]_{\frac12}$).

\item{\it Broken TR Multiplet:} It is possible for the defect to break $\uu(1)_b$ (a combination of transverse rotations and the 3d R-symmetry), which induces a marginal scalar $[0]_1$ at the top of the broken TR multiplet $L\bar{A}_1[-\frac12]_{\frac12}$ \eqref{eq:su(1,1|1)TRMult} (or its conjugate multiplet $A_1\bar{L}[\frac12]_{\frac12}$).

\end{itemize}

\subsubsection{$d=3$, $\cN=4$}

\subsubsection*{$\frac12$-BPS Lines (maximal)}

The $\frac12$-BPS line preserves a \textit{maximal} 1d subalgebra
\ie{}
\uu(1)_{\cR_2} \rtimes \psu(1,1|2) \rtimes \uu(1)_b \oplus \mathfrak{g}_f \supset \so(2,1)_\text{conf}  \oplus \su(2)_\cR \oplus \uu(1)_{\cR_2} \oplus \uu(1)_b \oplus \mathfrak{g}_f  
\fe
of the flavored 3d $\cN=4$ algebra
\ie{}
\osp(4|4;\bR) \oplus \mathfrak{G}_f \supset \so(2,3)_\text{conf} \oplus \so(4)_R \oplus \mathfrak{G}_f \,,
\fe
where $\uu(1)_c  \oplus \so(2)_{\text{rot}} \simeq \uu(1)_{\cR_2} \oplus \uu(1)_b$. Here, $\uu(1)_c$ the abelian factor of $\su(2)_\cR \oplus \uu(1)_c \subset \so(4)_R$, and $\so(2)_{\text{rot}}$ is the transverse rotation symmetry inside $\so(1,2) \subset \so(2,3)_\text{conf}$. Analyzing the symmetries broken by the defect leads to the following required multiplets. (Note that we do not label the trivial representation under $\mf{g}_f$, leaving it implicit).

\begin{itemize}

\item{\it Displacement Multiplet:} The broken $\so(4)_R/(\su(2)_R \oplus \uu(1)_c)$ currents induce marginal scalars $[\pm 1]_1^{(0)}$ on the defect, while the broken supercharges induce fermions $[\pm 1]^{(1)}_{\frac 32}$. Together with the displacement operator $[\pm 1]^{(0)}_2$, they sit inside the displacement multiplet $L \bar{A}_1[1]^{(0)}_{1}$ (and its conjugate multiplet $A_1 \bar{L}[-1]_1^{(0)}$ with opposite $\uu(1)_b$ charge):
\ie{}\label{eq:su(1,1|2)DisplaceMult}
&\xymatrix{
*++[F-,][F**:Displacement]{{\boldsymbol{ L \bar{A}_1}}}} \qquad \xymatrix @C=7pc @R=7pc @!0 @dr {
*++[F][F**:BrokenCurrent]{[1]^{(0)}_{1}}  \ar[d]|--{~\cQ~} \\
*++[F][F**:BrokenSUSY]{[1]^{(1)}_{\frac32}}  \ar[d]|--{~\cQ~} \\
*++[F][F**:Displacement]{[1]^{(0)}_{2}} 
}
\fe

\item{\it Broken F-Current Multiplet:} The broken flavor symmetries induce marginal scalars $([0]^{(0)}_1,\mf{G}_f/\mf{g}_f)$ on the defect, each of which sits at the top of a broken F-current multiplet $A_1 \bar{A}_1[0]^{(1)}_{\frac12}$ (or its conjugate multiplet with opposite $\uu(1)_b$ charge):
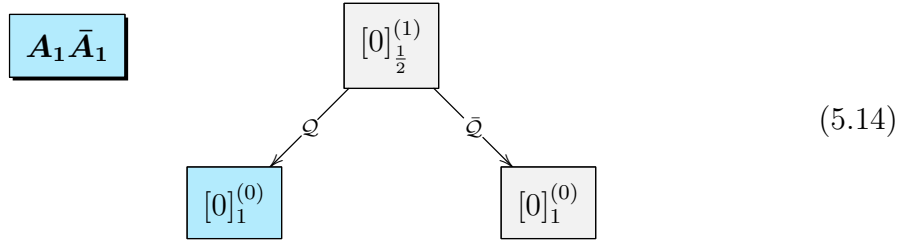
\ie{}\label{eq:su(1,1|2)FCurrentMult}
&\xymatrix{
*++[F-,][F**:BrokenCurrent]{{\boldsymbol{ A_1 \bar{A}_1}}}} \qquad \xymatrix @C=7pc @R=7pc @!0 @dr {
*++[F][F**:Generic]{[0]^{(1)}_{\frac12}} \ar[r]|--{{~\bar{\cQ}~}} \ar[d]|--{~\cQ~}
& *++[F][F**:Generic]{[0]^{(0)}_{1}}\\
*++[F][F**:BrokenCurrent]{[0]^{(0)}_{1}}
}
\fe

\item{\it Broken TR Multiplet:} It is possible for the defect to break $\uu(1)_b$ (a combination of transverse rotations and the 3d R-symmetry), which induces a marginal scalar $[0]^{(0)}_1$ at the top of the broken TR multiplet $A_1 \bar{A}_1[0]^{(1)}_{\frac12}$ (or its conjugate multiplet with opposite $\uu(1)_b$ charge):
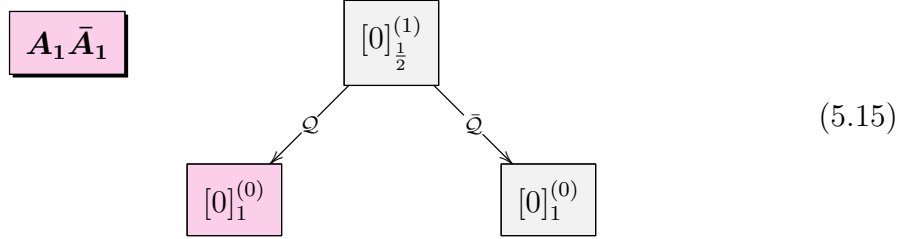
\ie{}\label{eq:su(1,1|2)TRMult}
&\xymatrix{
*++[F-,][F**:BrokenRot]{{\boldsymbol{ A_1 \bar{A}_1}}}} \qquad \xymatrix @C=7pc @R=7pc @!0 @dr {
*++[F][F**:Generic]{[0]^{(1)}_{\frac12}} \ar[r]|--{{~\bar{\cQ}~}} \ar[d]|--{~\cQ~}
& *++[F][F**:Generic]{[0]^{(0)}_{1}}\\
*++[F][F**:BrokenRot]{[0]^{(0)}_{1}}
}
\fe

\end{itemize}

\subsubsection*{$\frac14$-BPS Lines}

The $\frac14$-BPS line preserves a subalgebra
\ie{}
\su(1,1|1) \oplus \uu(1)_{b'} \oplus \uu(1)_{b''} \oplus \mf{g}_f \supset \so(2,1)_\text{conf} \oplus \uu(1)_{\cR_1} \oplus \uu(1)_{b'} \oplus \uu(1)_{b''} \oplus \mf{g}_f \,,
\fe
of the $\frac12$-BPS algebra $\uu(1)_{\cR_2} \rtimes \psu(1,1|2) \rtimes \uu(1)_b$, where $\uu(1)_{\cR_1} \oplus \uu(1)_{b'} \oplus \uu(1)_{b''}$ is a combination of $\uu(1)_{\cR_2} \oplus \uu(1)_b$ and $\uu(1)_R \subset \su(2)_\cR$. Analyzing the symmetries broken by the $\frac14$-BPS defect in the lens of the $\frac12$-BPS defect leads to the following required multiplets.

\begin{itemize}

\item{\it Displacement / Broken Supercurrent Multiplets:} The $\frac12$-BPS displacement multiplet decomposes into the $\frac14$-BPS displacement multiplet \eqref{eq:su(1,1|1)DisplaceMult} as well as a $\frac14$-BPS broken supercurrent multiplet \eqref{eq:su(1,1|1)RCurrentMult}, 
\ie{}
L \bar{A}_1[1]_1^{(0)} \to L \bar{A}_1[-\tfrac32]_{\frac32} \oplus L\bar{A}_1[-1]_1 \,.
\fe
 These multiplets account for the operators induced by the broken $\osp(4|4;\bR)/(\uu(1)_{\cR_2} \rtimes \psu(1,1|2) \rtimes \uu(1)_b)$ symmetries.

\item{\it Broken F-Current Multiplet:} The $\frac12$-BPS broken F-current multiplet decomposes into the $\frac14$-BPS broken F-current multiplet in \eqref{eq:su(1,1|1)FCurrentMult} (and its conjugate multiplet):
\ie{}
(A_1 A_1[0]_{\frac12}^{(1)},\mathfrak{G}_f / \mathfrak{g}_f) \to (L\bar{A}_1[-\tfrac12]_{\frac12}\oplus A_1 \bar{L}[\tfrac12]_{\frac12},\mathfrak{G}_f / \mathfrak{g}_f) \,.
\fe

\item{\it Broken Supercurrent Multiplet:} The broken $\su(2)_R/\uu(1)_R$ currents induce marginal scalars $[\pm 1]_{1}$ on the defect, while the broken $\frac12$-BPS supercharges induce fermions  $[\pm\tfrac12]_{\frac32}$. Altogether they form an additional $\frac14$-BPS broken supercurrent multiplet $L \bar{A}_1[-1]_1$ \eqref{eq:su(1,1|1)RCurrentMult} (and its conjugate multiplet $A_1 \bar{L}[1]_1$).  

\item{\it Broken TR Multiplet:} It is possible for the defect to break any combination of $\uu(1)_{b'}$ and $\uu(1)_{b''}$, both of which are a combination of transverse rotations and the 3d R-symmetry. Each broken current induces a marginal scalar $[0]^{(0)}_1$ at the top of a broken TR multiplet $L \bar{A}_1[-\frac12]_{\frac12}$ \eqref{eq:su(1,1|1)TRMult} (or its conjugate multiplet $A_1 \bar{L}[\frac12]_{\frac12}$). Note that the defect can also break a certain combination of $\uu(1)_{b'}$ and $\uu(1)_{b''}$ that is a pure R-symmetry, in which case the induced multiplet should properly be interpreted as a broken R-current multiplet.

\end{itemize}

\subsubsection{$d=3$, $\cN=5$}

\subsubsection*{$\frac25$-BPS Lines (maximal)}

The $\frac25$-BPS line preserves a \textit{maximal} 1d subalgebra
\ie{}
\uu(1)_{\cR_2} \rtimes \psu(1,1|2) \rtimes \uu(1)_b \oplus \mathfrak{g}_f \supset \so(2,1)_\text{conf}  \oplus \su(2)_\cR \oplus \uu(1)_{\cR_2} \oplus \uu(1)_b 
\fe
of the 3d $\cN=5$ algebra
\ie{}
\osp(5|4;\bR) \supset \so(2,3)_\text{conf} \oplus \so(5)_R \,,
\fe
where $\uu(1)_c  \oplus \so(2)_{\text{rot}} \simeq \uu(1)_{\cR_2} \oplus \uu(1)_b$. Here, $\uu(1)_c$ is contained in the chain of subalgebras
\ie{}
\uu(1)_c \oplus \su(2)_\cR \subset \su(2) \oplus \su(2)_\cR \simeq \so(4)_R \subset \so(5)_R \,,
\fe
while $\so(2)_{\text{rot}}$ is the transverse rotation symmetry inside $\so(1,2) \subset \so(2,3)_\text{conf}$. Analyzing the symmetries broken by the defect leads to the following required multiplets.

\begin{itemize}

\item{\it Displacement Multiplet:} The broken $\so(4)_R/(\su(2)_\cR \oplus \uu(1)_c)$ currents induce marginal scalars $[\pm 1]_1^{(0)}$ on the defect, while the broken supercharges induce fermions $[\pm 1]^{(1)}_{\frac 32}$. Together with the displacement operator $[\pm 1]^{(0)}_2$, they sit inside the displacement multiplet $L \bar{A}_1[1]^{(0)}_{1}$ \eqref{eq:su(1,1|2)DisplaceMult} (and its conjugate multiplet $A_1 \bar{L}[-1]_1^{(0)}$ with opposite $\uu(1)_b$ charge).

\item{\it Broken Supercurrent Multiplet:} The broken $\so(5)_R/\so(4)_R$ currents induce marginal scalars $[\pm \frac12]^{(1)}_1$ on the defect, while the remaining $\cN=5$ broken supercharges induce fermions $[\pm \frac12]^{(0)}_{\frac32}$. Altogether, they sit inside a broken supercurrent multiplet $L\bar{A}_1[\frac12]^{(0)}_{\frac12}$ (and its conjugate multiplet $A_1\bar{L}[\frac12]^{(0)}_{\frac12}$):  
\ie{}\label{eq:su(1,1|2)RCurrentMult}
&\xymatrix{
*++[F-,][F**:BrokenSUSY]{{\boldsymbol{ L \bar{A}_1}}}} \qquad \xymatrix @C=7pc @R=7pc @!0 @dr {
*++[F][F**:Generic]{[\frac12]^{(0)}_{\frac12}}  \ar[d]|--{~\cQ~} \\
*++[F][F**:BrokenCurrent]{[\frac12]^{(1)}_{1}}  \ar[d]|--{~\cQ~} \\
*++[F][F**:BrokenSUSY]{[\frac12]^{(0)}_{\frac32}} 
}
\fe

\item{\it Broken TR Multiplet:} It is possible for the defect to break $\uu(1)_b$ (a combination of transverse rotations and the 3d R-symmetry), which induces a marginal scalar $[0]^{(0)}_1$ at the top of the broken TR multiplet $A_1 \bar{A}_1[0]^{(1)}_{\frac12}$ \eqref{eq:su(1,1|2)TRMult} (or its conjugate multiplet with opposite $\uu(1)_b$ charge).

\end{itemize}

\subsubsection*{$\frac15$-BPS Lines}

The $\frac15$-BPS line preserves a subalgebra
\ie{}
\su(1,1|1) \oplus \uu(1)_{b'} \oplus \uu(1)_{b''} \supset \so(2,1)_\text{conf} \oplus \uu(1)_{\cR_1} \oplus \uu(1)_{b'} \oplus \uu(1)_{b''} \,,
\fe
of the $\frac12$-BPS algebra $\uu(1)_{\cR_2} \rtimes \psu(1,1|2) \rtimes \uu(1)_b$, where $\uu(1)_{\cR_1} \oplus \uu(1)_{b'} \oplus \uu(1)_{b''}$ is a combination of $\uu(1)_{\cR_2} \oplus \uu(1)_b$ and $\uu(1)_R \subset \su(2)_\cR$. Analyzing the symmetries broken by the $\frac15$-BPS defect in the lens of the $\frac25$-BPS defect leads to the following required multiplets.

\begin{itemize}

\item{\it Displacement / Broken Supercurrent Multiplets:} The $\frac25$-BPS displacement multiplet decomposes into the $\frac15$-BPS displacement multiplet \eqref{eq:su(1,1|1)DisplaceMult} as well as a $\frac15$-BPS broken supercurrent multiplet \eqref{eq:su(1,1|1)RCurrentMult},  
\ie{}
L\bar{A}_1[1]_1^{(0)} \to L\bar{A}_1[-\tfrac32]_{\frac32} \oplus L\bar{A}_1[-1]_1 \,.
\fe
These multiplets account for the operators induced by the broken $\osp(4|4;\bR)/(\uu(1)_{\cR_2} \rtimes \psu(1,1|2) \rtimes \uu(1)_b)$ symmetries with $\osp(4|4;\bR) \subset \osp(5|4;\bR)$.

\item{\it Broken Supercurrent / R-Current Multiplets:} The $\frac25$-BPS broken supercurrent multiplet containing operators induced by the $\so(5)_R/\so(4)_R$ broken currents decomposes into a $\frac15$-BPS broken R-current multiplet \eqref{eq:su(1,1|1)FCurrentMult} as well as a broken supercurrent multiplet \eqref{eq:su(1,1|1)RCurrentMult}, 
\ie{}
L\bar{A_1}[\tfrac12]_{\frac12}^{(0)} \to L\bar{A}_1[-\tfrac12]_{\frac12} \oplus  L\bar{A}_1[-1]_1 \,.
\fe

\item{\it Broken Supercurrent Multiplet:} The broken $\su(2)_\cR/\uu(1)_R$ currents induce marginal scalars $[\pm 1]_{1}$ on the defect, while the $\frac25$-BPS broken supercharges induce fermions  $[\pm\tfrac12]_{\frac32}$. Altogether they form an additional $\frac15$-BPS broken supercurrent multiplet $L \bar{A}_1[-1]_1$ \eqref{eq:su(1,1|1)RCurrentMult} (and its conjugate multiplet $A_1 \bar{L}[1]_1$).  

\item{\it Broken TR Multiplet:} It is possible for the defect to break any combination of $\uu(1)_{b'}$ and $\uu(1)_{b''}$, both of which are a combination of transverse rotations and the 3d R-symmetry. Each broken current induces a marginal scalar $[0]^{(0)}_1$ at the top of a broken TR multiplet $L \bar{A}_1[-\frac12]_{\frac12}$ \eqref{eq:su(1,1|1)TRMult} (or its conjugate multiplet $A_1 \bar{L}[\frac12]_{\frac12}$). Note that the defect can also break a certain combination of $\uu(1)_{b'}$ and $\uu(1)_{b''}$ that is a pure R-symmetry, in which case the induced multiplet should properly be interpreted as a broken R-current multiplet.

\end{itemize}

\subsubsection{$d=3$, $\cN=6$}

\subsubsection*{$\frac12$-BPS Lines (maximal)}

The $\frac12$-BPS line preserves a \textit{maximal} 1d subalgebra
\ie{}
\su(1,1|3) \oplus \uu(1)_b \subset \so(2,1)_\text{conf} \oplus \su(3)_\cR  \oplus \uu(1)_{\cR_3} \oplus \uu(1)_b
\fe
of the 3d $\cN=6$ algebra
\ie{}
\osp(6|4;\bR)\supset \so(2,3)_\text{conf} \oplus \so(6)_R \,.
\fe
where $\uu(1)_c  \oplus \so(2)_{\text{rot}} \simeq \uu(1)_{\cR_3} \oplus \uu(1)_b$. Here, $\uu(1)_c$ is the commutant of $\su(3)_\cR \subset \so(6)_R$ and $\so(2)_{\text{rot}}$ is the transverse rotation symmetry inside $\so(1,2) \subset \so(2,3)_\text{conf}$. Analyzing the symmetries broken by the defect leads to the following required multiplets.

\begin{itemize}

\item{\it Displacement Multiplet:} The broken $\so(6)_R/(\su(3)_\cR \oplus \uu(1)_c)$ currents induce marginal scalars $[2]_1^{(1,0)} \oplus [-2]_1^{(0,1)}$ on the defect, while the broken supercharges induce fermions $[\tfrac52]^{(0,1)}_{\frac 32} \oplus [-\tfrac52]^{(1,0)}_{\frac 32}$. Together with the displacement operator $[\pm 3]^{(0,0)}_2$, they sit inside the displacement multiplet $L\bar{A}_1[\frac32]^{(0,0)}_{\frac32}$ (and its conjugate multiplet $A_1\bar{L}[-\frac32]^{(0,0)}_{\frac32}$):
\ie{}\label{eq:su(1,1|3)DisplaceMult}
&\xymatrix{
*++[F-,][F**:Displacement]{{\boldsymbol{ L \bar{A}_1}}}} \qquad \xymatrix @C=7pc @R=7pc @!0 @dr {
*++[F][F**:Generic]{[\frac32]^{(0,0)}_{\frac12} }  \ar[d]|--{~\cQ~} \\
*++[F][F**:BrokenCurrent]{[2]^{(1,0)}_{1}}  \ar[d]|--{~\cQ~} \\
*++[F][F**:BrokenSUSY]{[\frac52]^{(0,1)}_{\frac32}} \ar[d]|--{~\cQ~} \\
*++[F][F**:Displacement]{[3]^{(0)}_{2}}
}
\fe

\item{\it Broken TR Multiplet:} It is possible for the defect to break $\uu(1)_b$ (a combination of transverse rotations and the 3d R-symmetry), which induces a marginal scalar $[0]_1$ at the top of the broken TR multiplet $A_1\bar{A}_1[\frac12]^{(1,0)}_{\frac12}$ (or its conjugate multiplet $A_1\bar{A}_1[-\frac12]^{(0,1)}_{\frac12}$):
\ie{}\label{eq:su(1,1|3)TRMult}
&\xymatrix{
*++[F-,][F**:BrokenRot]{{\boldsymbol{ A_1 \bar{A}_1}}}} \qquad \xymatrix @C=7pc @R=7pc @!0 @dr {
*++[F][F**:Generic]{[\tfrac12]^{(1,0)}_{\frac12}} \ar[r]|--{{~\bar{\cQ}~}} \ar[d]|--{~\cQ~}
& *++[F][F**:BrokenRot]{[0]^{(0,0)}_{1}} \\ 
*++[F][F**:Generic]{[1]^{(0,1)}_{1}} \ar[d]|--{~\cQ~} \\
*++[F][F**:Generic]{[\tfrac32]^{(0,0)}_{\frac32}}
}
\fe

\end{itemize}

\subsubsection*{$\frac13$-BPS Lines}
The $\frac13$-BPS line preserves a subalgebra
\ie{}
&\uu(1)_{\cR_2} \rtimes \psu(1,1|2) \rtimes \uu(1)_{b'} \oplus \uu(1)_{b}  \supset \so(2,1)_\text{conf} \oplus \su(2)_\cR \oplus \uu(1)_{\cR_2} \oplus \uu(1)_b \oplus \uu(1)_{b'}
\fe
of the $\frac12$-BPS algebra $\su(1,1|3) \oplus \uu(1)_b$, where $\uu(1)_{\cR_2} \oplus \uu(1)_{b'}$ is a combination of $\uu(1)_{\cR_3}$ and the $\uu(1)_R$ commutant of $\su(2)_\cR \subset \su(3)_\cR$.  Analyzing the symmetries broken by the $\frac13$-BPS defect in the lens of the $\frac12$-BPS defect leads to the following required multiplets.

\begin{itemize}

\item{\it Displacement / Broken Supercurrent Multiplets:} The $\frac12$-BPS displacement multiplet decomposes into the $\frac13$-BPS displacement multiplet \eqref{eq:su(1,1|2)DisplaceMult} and a $\frac13$-BPS broken supercurrent multiplet \eqref{eq:su(1,1|2)RCurrentMult}, 
\ie{}
L \bar{A}_1[\tfrac32]_{\frac12}^{(0,0)} \to L\bar{A}_1[1]_1^{(0)} \oplus L\bar{A}_1[\tfrac12]_{\frac12}^{(0)} \,.
\fe
These multiplets account for the operators induced by the broken $\osp(6|4;\bR)/(\su(1,1|3)\oplus\uu(1)_b)$ symmetries. 

\item{\it Broken Supercurrent Multiplet:} The broken $\su(3)_\cR/(\su(2)_\cR \oplus \uu(1)_R)$  currents induce marginal scalars $[\pm\tfrac12]^{(1)}_{1}$ on the defect, while the broken $\frac12$-BPS supercharges induce fermions $[\pm\tfrac12]^{(0)}_{\frac32}$. Altogether they form an additional $\frac13$-BPS broken supercurrent multiplet $L \bar{A}_1[\frac12]^{(0)}_{\frac12}$ \eqref{eq:su(1,1|2)RCurrentMult} (and its conjugate multiplet $A_1 \bar{L}[-\frac12]^{(0)}_{\frac12}$): 

\item{\it Broken TR Multiplet:} It is possible for the defect to break any combination of $\uu(1)_b$ and $\uu(1)_{b'}$, both of which are a combination of transverse rotations and the 3d R-symmetry. Each broken current induces a marginal scalar $[0]^{(0)}_1$ at the top of a broken TR multiplet $A_1 \bar{A}_1[0]^{(1)}_{\frac12}$ \eqref{eq:su(1,1|2)TRMult} (or its conjugate multiplet). Note that it is possible to break a certain combination of $\uu(1)_b$ and $\uu(1)_{b'}$ that is a pure R-symmetry, in which case the induced multiplet should properly be interpreted as a broken R-current multiplet.

\end{itemize}

\subsubsection*{$\frac16$-BPS Lines}

The $\frac16$-BPS line preserves a subalgebra
\ie{}
\su(1,1|1) \oplus \su(2)_\cR \oplus \uu(1)_{b} \oplus \uu(1)_{b'} \supset \so(2,1)_\text{conf} \oplus \uu(1)_{\cR_1} \oplus \su(2)_\cR \oplus \uu(1)_{b}  \oplus \uu(1)_{b'} \,.
\fe
of the $\frac12$-BPS algebra $\su(1,1|3) \oplus \uu(1)_b$, where $\uu(1)_{\cR_1} \oplus \uu(1)_{b'}$ is a combination of $\uu(1)_{\cR_3}$ and the $\uu(1)_R$ commutant of $\su(2)_\cR \subset \su(3)_\cR$.  Analyzing the symmetries broken by the $\frac16$-BPS defect in the lens of the $\frac12$-BPS defect leads to the following required multiplets, where we label UIRs of $\su(2)_\cR$ by their dimension (i.e. the $\su(2)_\cR$ fundamental is labeled as $\bf{2}$).

\begin{itemize}

\item{\it Displacement / Broken Current Multiplets:} The $\frac12$-BPS displacement multiplet decomposes into a $\frac16$-BPS displacement multiplet \eqref{eq:su(1,1|1)DisplaceMult} as well as several $\frac16$-BPS broken R/super-current multiplets \eqref{eq:su(1,1|1)FCurrentMult} and \eqref{eq:su(1,1|1)RCurrentMult} , 
\ie{}
L \bar{A}_1[\tfrac32]_{\frac12}^{(0,0)} \to (L\bar{A}_1[-\tfrac32]_{\frac32} \oplus L\bar{A}_1[-\tfrac12]_{\frac12}, \mathbf{1}) \oplus (L\bar{A}_1[-1]_1, \mathbf{2}) \,.
\fe
These multiplets account for the operators induced by the broken $\osp(6|4;\bR)/(\su(1,1|3)\oplus\uu(1)_b)$ symmetries. 

\item{\it Broken Supercurrent Multiplet:} The broken $\su(3)_\cR/(\su(2)_\cR \oplus \uu(1)_R)$ currents induce marginal scalars $([\pm 1]_{1},\mathbf{2}) $ on the defect, while the broken $\frac12$-BPS supercharges induce fermions $([\pm \tfrac12]_{\frac32},\mathbf{2})$. Altogether they form an additional $\frac16$-BPS broken supercurrent multiplet $(L \bar{A}_1[-1]_1, \bf{2})$ \eqref{eq:su(1,1|1)RCurrentMult} 
 (and its conjugate multiplet $(A_1 \bar{L}[1]_1, \mathbf{2})$). 

\item{\it Broken R-Current Multiplet:} It is possible for the defect to break $\su(2)_\cR$ to some (possibly trivial) subalgebra $\mf{g}_b$, which induces a set of marginal scalars $([0]_1, \su(2)_\cR/\mf{g}_b)$ on the defect within a broken R-current multiplet $(L\bar{A}_1[-\frac12]_{\frac12},\su(2)_R/\mf{g}_b)$ \eqref{eq:su(1,1|1)FCurrentMult} (or its conjugate multiplet $(A_1\bar{L}[\frac12]_{\frac12},\su(2)_R/\mf{g}_b)$).

\item{\it Broken TR Multiplet:} It is possible for the defect to break any combination of $\uu(1)_b$ and $\uu(1)_{b'}$, both of which are a combination of transverse rotations and the 3d R-symmetry. Each broken current induces a marginal scalar $[0]^{(0)}_1$ at the top of a broken TR multiplet $L \bar{A}_1[-\frac12]_{\frac12}$ \eqref{eq:su(1,1|2)TRMult} (or its conjugate multiplet $A_1 \bar{L}[\frac12]_{\frac12}$). Note that the defect can also break a certain combination of $\uu(1)_b$ and $\uu(1)_{b'}$ that is a pure R-symmetry, in which case the induced multiplet should properly be interpreted as a broken R-current multiplet.

\end{itemize}

\subsubsection{$d=3$, $\cN=8$}

\subsubsection*{$\frac12$-BPS Lines (maximal)}

The $\frac12$-BPS line preserves a \textit{maximal} 1d subalgebra
\ie{}
\su(1,1|4) \oplus \uu(1)_b \supset \so(2,1)_\text{conf} \oplus \su(4)_\cR  \oplus \uu(1)_{\cR_4} \oplus \uu(1)_b
\fe
of the 3d $\cN=8$ algebra
\ie{}
\osp(8|4;\bR)\supset \so(2,3)_{\text{conf}} \oplus \so(8)_R \,,
\fe
where $\uu(1)_c  \oplus \so(2)_{\text{rot}} \simeq \uu(1)_{\cR_4} \oplus \uu(1)_b$. Here, $\uu(1)_c$ is the commutant of $\su(4)_\cR \subset \so(8)_R$ and $\so(2)_{\text{rot}}$ is the transverse rotation symmetry inside $\so(1,2) \subset \so(2,3)_\text{conf}$. Analyzing the symmetries broken by the defect leads to the following required multiplets.

\begin{itemize}

\item{\it Displacement Multiplet:} The broken $\so(8)_R/(\su(4)_\cR \oplus \uu(1)_c)$ currents induce marginal scalars $[\pm 1]_1^{(0,1,0)}$ on the defect, while the broken supercharges induce fermions $[\tfrac32]^{(0,0,1)}_{\frac 32} \oplus [-\tfrac32]^{(1,0,0)}_{\frac 32}$. Together with the displacement operator $[\pm 2]^{(0,0,0)}_2$, they sit inside the displacement multiplet $A\bar{A}_1[\frac12]^{(1,0,0)}_{\frac12}$ (and its conjugate multiplet $A_1\bar{A}_1[-\frac12]^{(0,0,1)}_{\frac12}$):
\ie{}\label{eq:3dN8halfBPSdisplacementMult}
&\xymatrix{
*++[F-,][F**:Displacement]{{\boldsymbol{ A_1 \bar{A}_1}}}} \qquad \xymatrix @C=7pc @R=7pc @!0 @dr {
*++[F][F**:Generic]{[\frac12]^{(1,0,0)}_{\frac12}} \ar[r]|--{{~\bar{\cQ}~}} \ar[d]|--{~\cQ~}
& *++[F][F**:Generic]{[0]^{(0,0,0)}_{1}} \\ 
*++[F][F**:BrokenCurrent]{[1]^{(0,1,0)}_{1}} \ar[d]|--{~\cQ~} \\
*++[F][F**:BrokenSUSY]{[\frac32]^{(0,0,1)}_{\frac32}} \ar[d]|--{~\cQ~} \\
*++[F][F**:Displacement]{[2]^{(0,0,0)}_{2}}
}
\fe

\item{\it Broken TR Multiplet:} It is possible for the defect to break $\uu(1)_b$ (a combination of transverse rotations and the 3d R-symmetry), which induces a marginal scalar $[0]^{(0,0,0)}_1$ \textit{either} in the displacement multiplet \eqref{eq:3dN8halfBPSdisplacementMult} or at the top of a broken TR multiplet $A_1\bar{A}_1[\frac12]^{(1,0,0)}_{\frac12}$ (or its conjugate multiplet $A_1\bar{A}_1[-\frac12]^{(0,0,1)}_{\frac12}$):\footnote{Note that this is a unique feature of the $\frac12$-BPS line defect in 3d $\cN=8$ SCFTs. That is, symmetry considerations alone are insufficient to conclude whether the induced scalar resides in the displacement multiplet or in a separate multiplet.}
\ie{}
&\xymatrix{
*++[F-,][F**:BrokenRot]{{\boldsymbol{ A_1 \bar{A}_1}}}} \qquad \xymatrix @C=7pc @R=7pc @!0 @dr {
*++[F][F**:Generic]{[\frac12]^{(1,0,0)}_{\frac12}} \ar[r]|--{{~\bar{\cQ}~}} \ar[d]|--{~\cQ~}
& *++[F][F**:BrokenRot]{[0]^{(0,0,0)}_{1}} \\ 
*++[F][F**:Generic]{[1]^{(0,1,0)}_{1}} \ar[d]|--{~\cQ~} \\
*++[F][F**:Generic]{[\frac32]^{(0,0,1)}_{\frac32}} \ar[d]|--{~\cQ~} \\
*++[F][F**:Generic]{[2]^{(0,0,0)}_{2}}
}
\fe

\end{itemize}

\subsubsection*{$\frac38$-BPS Lines}

The $\frac38$-BPS line preserves a subalgebra
\ie{}
\su(1,1|3) \oplus \uu(1)_b \oplus \uu(1)_{b'} \supset \so(2,1)_\text{conf} \oplus \su(3)_\cR \oplus \uu(1)_{\cR_3} \oplus \uu(1)_b \oplus \uu(1)_{b'} \,.
\fe
of the $\frac12$-BPS algebra $\su(1,1|4) \oplus \uu(1)_b$, where $\uu(1)_{\cR_3} \oplus \uu(1)_{b'}$ is a combination of $\uu(1)_{\cR_4}$ and the $\uu(1)_R$ commutant of $\su(3)_\cR \subset \su(4)_\cR$.  Analyzing the symmetries broken by the $\frac38$-BPS defect in the lens of the $\frac12$-BPS defect leads to the following required multiplets.

\begin{itemize}

\item{\it Displacement / Broken Supercurrent Multiplets:} The $\frac12$-BPS displacement multiplet decomposes into a $\frac38$-BPS displacement multiplet \eqref{eq:su(1,1|3)DisplaceMult} as well as a $\frac38$-BPS broken supercurrent multiplet \eqref{eq:su(1,1|3)RCurrentMult},
\ie{}
A_1 \bar{A}_1[\tfrac12]_{\frac12}^{(1,0,0)}  \to L\bar{A}_1[\tfrac32]^{(0,0)}_{\frac32} \oplus A_1\bar{A}_1[\tfrac12]^{(1,0)}_{\frac12} \,,
\fe
where the broken supercurrent multiplet is given by 
\ie{}\label{eq:su(1,1|3)RCurrentMult}
&\xymatrix{
*++[F-,][F**:BrokenSUSY]{{\boldsymbol{ A_1 \bar{A}_1}}}} \qquad \xymatrix @C=7pc @R=7pc @!0 @dr {
*++[F][F**:Generic]{[\tfrac12]^{(1,0)}_{\frac12}} \ar[r]|--{{~\bar{\cQ}~}} \ar[d]|--{~\cQ~}
& *++[F][F**:Generic]{[0]^{(0,0)}_{1}} \\ 
*++[F][F**:BrokenCurrent]{[1]^{(0,1)}_{1}} \ar[d]|--{~\cQ~} \\
*++[F][F**:BrokenSUSY]{[\tfrac32]^{(0,0)}_{\frac32}}
}
\fe
These multiplets account for the operators induced by the broken $\osp(8|4;\bR)/(\su(1,1|4)\oplus\uu(1)_b)$ symmetries. 

\item{\it Broken Supercurrent Multiplet:} The broken $\su(4)_\cR/(\su(3)_\cR \oplus \uu(1)_R)$ currents induce marginal scalars $[1]^{(0,1)}_{1} \oplus [-1]^{(1,0)}_{1}$ on the defect, while the broken $\frac12$-BPS supercharges induce fermions $[\pm  \frac32]^{(0,0)}_{\frac32}$. Altogether they form an additional $\frac38$-BPS broken supercurrent multiplet $ A_1 \bar{A}_1[\tfrac12]^{(1,0)}_{\frac12}$  \eqref{eq:su(1,1|3)RCurrentMult} (and its conjugate multiplet $ A_1 \bar{A}_1[-\tfrac12]^{(0,1)}_{\frac12}$). 

\item{\it Broken TR Multiplet:} It is possible for the defect to break any combination of $\uu(1)_b$ and $\uu(1)_{b'}$, both of which are a combination of transverse rotations and the 3d R-symmetry. Each broken current induces a marginal scalar $[0]^{(0,0)}_1$ at the top of a broken TR multiplet $A_1 \bar{A}_1[\frac12]^{(1,0)}_{\frac12}$  \eqref{eq:su(1,1|3)TRMult} (or its conjugate multiplet $A_1 \bar{A}_1[-\frac12]^{(0,1)}_{\frac12}$).

\end{itemize}

\subsubsection*{$\frac14$-BPS Lines}
The $\frac14$-BPS line preserves a subalgebra
\ie{}
&\uu(1)_{\cR_2} \rtimes \psu(1,1|2) \rtimes \uu(1)_{b'} \oplus \su(2)_{R'} \oplus \uu(1)_{b} \\ &\quad\quad \supset \so(2,1)_\text{conf} \oplus \su(2)_\cR \oplus \su(2)_{R'} \oplus \uu(1)_{\cR_2} \oplus \uu(1)_b \oplus \uu(1)_{b'}
\fe
of the $\frac12$-BPS algebra $\su(1,1|4) \oplus \uu(1)_b$, where $\uu(1)_{\cR_2} \oplus \uu(1)_{b'}$ is a combination of $\uu(1)_{\cR_4}$ and the $\uu(1)_R$ commutant of $\su(2)_\cR \oplus \su(2)_{R'} \subset \su(4)_\cR$.  Analyzing the symmetries broken by the $\frac18$-BPS defect in the lens of the $\frac12$-BPS defect leads to the following required multiplets, where we label UIRs of $\su(2)_{R'}$ by their dimension (i.e. the $\su(2)_{R'}$ fundamental is labeled as $\bf{2}$).

\begin{itemize}

\item{\it Displacement / Broken Current Multiplets:} The $\frac12$-BPS displacement multiplet decomposes into the $\frac14$-BPS displacement multiplet \eqref{eq:su(1,1|2)DisplaceMult} as well as several $\frac14$-BPS broken R/super-current multiplets \eqref{eq:su(1,1|2)FCurrentMult} and \eqref{eq:su(1,1|2)RCurrentMult}, 
\ie{}
A_1 \bar{A}_1[\tfrac12]_{\frac12}^{(1,0,0)} \to \left(L\bar{A}_1[1]_1^{(0)} \oplus A_1\bar{A}_1[0]_{\frac12}^{(1)},\mathbf{1}\right) \oplus \left(L\bar{A}_1[\tfrac12]_{\frac12}^{(0)}, \mathbf{2}\right) \,.
\fe
These multiplets account for the operators induced by the broken $\osp(8|4;\bR)/(\su(1,1|4)\oplus\uu(1)_b)$ symmetries. 

\item{\it Broken Supercurrent Multiplet:} The broken $\su(4)_\cR/(\su(2)_\cR \oplus \su(2)_{R'} \oplus \uu(1)_R)$ currents induce marginal scalars $([\pm \tfrac12]^{(1)}_{1}, \mathbf{2})$ on the defect, while the broken $\frac12$-BPS supercharges induce fermions $([\pm \tfrac12]^{(0)}_{\frac52},\mathbf{2})$. Altogether they form an additional $\frac14$-BPS broken supercurrent multiplet $(L \bar{A}_1[\frac12]^{(0)}_{\frac12}, \bf{2})$ \eqref{eq:su(1,1|2)RCurrentMult} (and its conjugate multiplet $(A_1 \bar{L}[-\frac12]^{(0)}_{\frac12}, \bf{2})$). 

\item{\it Broken R-Current Multiplet:} It is possible for the defect to break $\su(2)_{R'}$ to some (possibly trivial) subalgebra $\mf{g}_b$, which induces a set of marginal scalars $([0]^{(0)}_1, \su(2)_{R'}/\mf{g}_b)$ on the defect within a broken R-current multiplet $(A_1 \bar{A}_1[0]_{\frac12}^{(1)},\su(2)_{R'}/\mf{g}_b)$ \eqref{eq:su(1,1|2)FCurrentMult} (or its conjugate multiplet).

\item{\it Broken TR Multiplet:} It is possible for the defect to break any combination of $\uu(1)_b$ and $\uu(1)_{b'}$, both of which are a combination of transverse rotations and the 3d R-symmetry. Each broken current induces a marginal scalar $[0]^{(0)}_1$ at the top of a broken TR multiplet $A_1 \bar{A}_1[0]^{(1)}_{\frac12}$ \eqref{eq:su(1,1|2)TRMult}  (or its conjugate multiplet). Note that it is possible to break a certain combination of $\uu(1)_b$ and $\uu(1)_{b'}$ that is a pure R-symmetry, in which case the induced multiplet should properly be interpreted as a broken R-current multiplet.

\end{itemize}

\subsubsection*{$\frac18$-BPS Lines}

The $\frac18$-BPS line preserves a subalgebra
\ie{}
\su(1,1|1) \oplus \su(3)_\cR \oplus \uu(1)_{b} \oplus \uu(1)_{b'} \supset \so(2,1)_\text{conf} \oplus \uu(1)_{\cR_1} \oplus \su(3)_\cR \oplus \uu(1)_{b}  \oplus \uu(1)_{b'} \,.
\fe
of the $\frac12$-BPS algebra $\su(1,1|4) \oplus \uu(1)_b$, where $\uu(1)_{\cR_1} \oplus \uu(1)_{b'}$ is a combination of $\uu(1)_{\cR_4}$ and the $\uu(1)_R$ commutant of $\su(3)_\cR \subset \su(4)_\cR$.  Analyzing the symmetries broken by the $\frac18$-BPS defect in the lens of the $\frac12$-BPS defect leads to the following required multiplets, where we label UIRs of $\su(3)_\cR$ by their dimension (i.e. the $\su(3)_\cR$ fundamental is labeled as $\bf{3}$), excluding the trivial representation (which we leave implicit).

\begin{itemize}

\item{\it Displacement / Broken Current Multiplets:} The $\frac12$-BPS displacement multiplet decomposes into the $\frac18$-BPS displacement multiplet \eqref{eq:su(1,1|1)DisplaceMult} as well as several $\frac18$-BPS broken super/R-current multiplets \eqref{eq:su(1,1|1)RCurrentMult} and \eqref{eq:su(1,1|1)FCurrentMult} , 
\ie{}
A_1 \bar{A}_1[\frac12]_{\frac12}^{(1,0,0)} \to \left(L\bar{A}_1[-\tfrac32]_{\frac32} \oplus A_1 \bar{L}[\tfrac12]_{\frac12}, \mathbf{1} \right) \oplus \left( L\bar{A}_1[-\tfrac12]_{\frac12}, \mathbf{3} \right) \oplus \left( L\bar{A}_1[-1]_{1}, \mathbf{\bar{3}} \right) \,.
\fe
These multiplets account for the operators induced by the broken $\osp(8|4;\bR)/(\su(1,1|4)\oplus\uu(1)_b)$ symmetries. 

\item{\it Broken Supercurrent Multiplet:} The broken $\su(4)_\cR/(\su(3)_\cR \oplus \uu(1)_R)$ currents induce marginal scalars $([-1]_{1}, \mathbf{3}) \oplus ([1]_{1}, \mathbf{\bar{3}})$ on the defect, while the broken $\frac12$-BPS supercharges induce fermions $([-\frac12]_{\frac52},\mathbf{3}) \oplus [\frac12]_{\frac52},\mathbf{\bar{3}})$. Altogether they form an additional $\frac18$-BPS broken supercurrent multiplet $(L \bar{A}_1[-1]_{1}, \bf{3})$ \eqref{eq:su(1,1|1)RCurrentMult}  (and its conjugate multiplet $(A_1 \bar{L}[1]_1, \mathbf{\bar{3}})$). 

\item{\it Broken R-Current Multiplet:} It is possible for the defect to break $\su(3)_\cR$ to some (possibly trivial) subalgebra $\mf{g}_b$, which induces a set of marginal scalars $([0]_1, \su(3)_\cR/\mf{g}_b)$ on the defect within a broken R-current multiplet $(L\bar{A}_1[-\frac12]_{\frac12},\su(3)_\cR/\mf{g}_b)$ \eqref{eq:su(1,1|1)FCurrentMult} (or its conjugate multiplet $(A_1\bar{L}[\frac12]_{\frac12},\su(3)_\cR/\mf{g}_b)$).

\item{\it Broken TR Multiplet:} It is possible for the defect to break any combination of $\uu(1)_b$ and $\uu(1)_{b'}$, both of which are a combination of transverse rotations and the 3d R-symmetry. Each broken current induces a marginal scalar $[0]_1$ at the top of a broken TR multiplet $L \bar{A}_1[-\frac12]_{\frac12}$ \eqref{eq:su(1,1|1)TRMult} (or its conjugate multiplet $A_1 \bar{L}[\frac12]_{\frac12}$). Note that it is possible to break a certain combination of $\uu(1)_b$ and $\uu(1)_{b'}$ that is a pure R-symmetry, in which case the induced multiplet should properly be interpreted as a broken R-current multiplet.

\end{itemize}

\subsection{Superconformal Lines in $d=4$}

\subsubsection{$d=4$, $\cN=2$}

\subsubsection*{$\frac12$-BPS Lines (maximal)}

The $\frac12$-BPS line preserves a \textit{maximal} 1d subalgebra
\ie{}
\osp(4^*|2) \oplus \mathfrak{G}_f \supset \so(2,1)_\text{conf} \oplus \su(2)_{\text{rot}} \oplus \su(2)_\cR \oplus \mathfrak{G}_f 
\fe
of the flavored 4d $\cN=2$ algebra
\ie{}
\su(2,2|2) \oplus \mathfrak{G}_f \supset \so(2,4)_\text{conf} \oplus \su(2)_R \oplus \uu(1)_R \oplus \mathfrak{G}_f \,,
\fe
where the transverse rotation symmetry $\su(2)_{\text{rot}} \simeq \so(3)_{\text{rot}}$ sits inside $\so(1,3) \subset \so(2,4)_{\text{conf}}$ and $\su(2)_\cR \simeq \su(2)_R$. Analyzing the symmetries broken by the defect leads to the following required multiplets. Note that everything transforms trivially under $\mf{G}_f$.

\begin{itemize}

\item{\it Displacement Multiplet:} The broken $\uu(1)_R$ current induces a marginal scalar  $[0]_1^{(0)}$ on the defect, while the broken supercharges induce fermions $[1]^{(1)}_{\frac 32}$. Together with the displacement operator $[2]^{(0)}_2$, they sit inside the displacement multiplet $A_2[0]_1^{(0)}$:
\ie{}
\label{4dn2dis}
\xymatrix{
*++[F-,][F**:Displacement]{{\boldsymbol{ A_2}}}} \qquad \xymatrix @C=3pc @R=3pc @r {
*++[F][F**:BrokenCurrent]{[0]_{1}^{(0)}} \ar[r]|--{{~\cQ~}} 
&*++[F][F**:BrokenSUSY]{[1]^{(1)}_{\frac32}}\ar[r]|--{{~\cQ~}} 
&*++[F][F**:Displacement]{[2]^{(0)}_{2}}
}
\fe 

\item The defect must preserve the full 4d flavor symmetry $\mf{G}_f$ because $\osp(4^*|2)$ does not admit any F-current multiplets.
 
\item Note that because the $\su(2)_{\text{rot}}$ is fully contained within $\osp(4^*|2)$, it is not possible for the defect to break transverse rotation symmetry without breaking any further supersymmetry.

\end{itemize}

\subsubsection{$d=4$, $\cN=3$}

\subsubsection*{$\frac13$-BPS Lines (maximal)}

The $\frac13$-BPS line preserves a \textit{maximal} 1d subalgebra
\ie{}
\osp(4^*|2) \oplus \uu(1)_b \supset \so(2,1)_\text{conf} \oplus \su(2)_{\text{rot}} \oplus \su(2)_\cR \oplus \uu(1)_b
\fe
of the 4d $\cN=3$ algebra
\ie{}
\su(2,2|3) \supset \so(2,4)_\text{conf} \oplus \su(3)_R \oplus \uu(1)_R \,,
\fe
where the transverse rotation symmetry $\su(2)_{\text{rot}} \simeq \so(3)_{\text{rot}}$ sits inside $\so(1,3) \subset \so(2,4)_{\text{conf}}$ and $\uu(1)_b$ is the commutant of $\osp(4^*|2) \subset \su(2,2|3)$. Analyzing the symmetries broken by the defect leads to the following required multiplets.

\begin{itemize}

\item{\it Displacement Multiplet:} The broken $\uu(1)_R$ current induces a marginal scalar $[0]_1^{(0)}$ on the defect, while the broken $\cN=2$ supercharges induce fermions $[1]^{(1)}_{\frac 32}$. Together with the displacement operator $[2]^{(0)}_2$, they sit inside the displacement multiplet $A_2[0]_1^{(0)}$ \eqref{4dn2dis}.

\item{\it Broken Supercurrent Multiplet:} The broken $\su(3)_R/(\su(2)_\cR \oplus \uu(1)_R)$ currents induce marginal scalars $[0]^{(1)}_1$ on the defect, while the remaining $\cN=3$ broken supercharges induce fermions $[1]^{(0)}_{\frac32}$. Altogether, they sit inside the broken supercurrent multiplet $B_1[0]_1^{(1)}$: 
\ie{}\label{4dn3R}
\xymatrix{
*++[F-,][F**:BrokenSUSY]{{\boldsymbol{ B_1}}}} \qquad \xymatrix @C=3pc @R=3pc @r {
*++[F][F**:BrokenCurrent]{[0]_{1}^{(1)}} \ar[r]|--{{~\cQ~}} 
&*++[F][F**:BrokenSUSY]{[1]^{(0)}_{\frac32}}
&*++[]{\qquad \:\:}
}
\fe
 
\item Note that because $\su(2)_{\text{rot}}$ is fully contained within $\osp(4^*|2)$, it is not possible for the defect to break transverse rotations without breaking any further supersymmetry.

\end{itemize}

\subsubsection{$d=4$, $\cN=4$}

\subsubsection*{$\frac12$-BPS Lines (maximal)}

The $\frac12$-BPS line preserves a \textit{maximal} 1d subalgebra
\ie{}
\osp(4^*|4) \supset \so(2,1)_\text{conf} \oplus \su(2)_{\text{rot}} \oplus \usp(4)_\cR
\fe
of the 4d $\cN=4$ algebra
\ie{}
\psu(2,2|4) \supset \so(2,4)_\text{conf} \oplus \su(4)_R \,,
\fe
where the transverse rotation symmetry $\su(2)_{\text{rot}} \simeq \so(3)_{\text{rot}}$ sits inside $\so(1,3) \subset \so(2,4)_{\text{conf}}$. Analyzing the symmetries broken by the defect leads to the following required multiplets.

\begin{itemize}

\item{\it Displacement Multiplet:} The broken $\su(4)_R/\usp(4)_\cR$ currents induce marginal scalars $[0]_1^{(0,1)}$ on the defect, while the broken supercharges induce fermions $[1]^{(1,0)}_{\frac 32}$. Together with the displacement operator $[2]^{(0,0)}_2$, they sit inside the displacement multiplet $B_1[0]_1^{(0,1)}$:
\ie{}
\label{4dn4dis}
\xymatrix{
*++[F-,][F**:Displacement]{{\boldsymbol{ B_1}}}}  \qquad  \xymatrix @C=3pc @R=3pc @r {
*++[F][F**:BrokenCurrent]{[0]_{1}^{(0,1)}} \ar[r]|--{{~\cQ~}} 
&*++[F][F**:BrokenSUSY]{[1]^{(1,0)}_{\frac32}}\ar[r]|--{{~\cQ~}} 
&*++[F][F**:Displacement]{[2]^{(0,0)}_{2}}
}
\fe
 
\item Note that because $\su(2)_{\text{rot}}$ is fully contained within $\osp(4^*|4)$, it is not possible for the defect to break transverse rotations without breaking any further supersymmetry.

\end{itemize}

\subsubsection*{$\frac14$-BPS Lines}

The $\frac14$-BPS line preserves a subalgebra
\ie{}
\osp(4^*|2) \oplus \su(2)_F \supset \so(2,1)_\text{conf} \oplus \su(2)_{\text{rot}} \oplus \su(2)_\cR \oplus \su(2)_F
\fe
of the $\frac12$-BPS algebra $\osp(4^*|4)$, where $\su(2)_\cR \oplus \su(2)_F \subset \usp(4)_\cR$. Analyzing the symmetries broken by the $\frac14$-BPS defect in the lens of the $\frac12$-BPS defect leads to the following required multiplets, where we label UIRs of $\su(2)_F$ by their dimension (i.e. the $\su(2)_F$ doublet is labeled as $\bf{2}$).

\begin{itemize}

\item{\it Displacement / Broken Supercurrent Multiplets:} The $\frac12$-BPS displacement multiplet decomposes into the $\frac14$-BPS displacement multiplet \eqref{4dn2dis} as well as a $\frac14$-BPS broken supercurrent multiplet \eqref{4dn3R}, 
\ie{}
B_1[0]_1^{(0,1)} \to \left(A_2[0]_1^{(0)},\mathbf{1}\right) \oplus \left(B_1[0]_1^{(1)},\mathbf{2}\right) \,.
\fe
These multiplets account for the operators induced by the broken $\psu(2,2|4)/\osp(4^*|4)$ symmetries. 

\item{\it Broken Supercurrent Multiplet:} The broken $\usp(4)_\cR/(\su(2)_\cR \oplus \su(2)_F)$ currents induce marginal scalars $([0]^{(1)}_{1}, \mathbf{2})$ on the defect, while the broken $\frac12$-BPS supercharges induce fermions $([1]^{(0)}_{\frac32},\mathbf{2})$. Altogether, they form an additional broken supercurrent multiplet $(B_1[0]_1^{(1)}, \mathbf{2})$ \eqref{4dn3R}.

\item The defect must preserve $\su(2)_F$ because $\osp(4^*|2)$ does not admit any F-current multiplets.
 
\item Note that because $\su(2)_{\text{rot}}$ transverse rotations are fully contained within $\osp(4^*|2)$, it is not possible for the defect to break transverse rotation symmetry without breaking any further supersymmetry.

\end{itemize}

\subsection{Superconformal Lines in $d=5$}

\subsubsection*{$\frac12$-BPS Lines (maximal)}

The $\frac12$-BPS line preserves a \textit{maximal} 1d subalgebra
\ie{}
D(2,1;2;0) \oplus \su(2)_{\text{right}} \oplus \mf{G}_f \supset \so(2,1)_\text{conf} \oplus \so(4)_\text{rot} \oplus \su(2)_\cR \oplus \mf{G}_f
\fe
of the flavored 5d $\cN=1$ algebra
\ie{}
F(4;2) \supset \so(2,5)_\text{conf} \oplus \su(2)_R \oplus \mf{G}_f \,,
\fe
where the transverse rotation symmetry $\so(4)_\text{rot} \simeq \su(2)_{\text{left}} \oplus \su(2)_{\text{right}}$ sits inside $\so(1,4) \subset \so(2,5)_{\text{conf}}$, and $\su(2)_\cR \simeq \su(2)_R$. Analyzing the symmetries broken by the defect leads to the following required multiplets. Recall that $\su(2)_{\text{left}}$ irreps are labeled by $[j]$ and $\su(2)_{\text{right}}$ irreps by $\mathbf{j'+1}$. Note that everything transforms trivially under $\mf{G}_f$

\begin{itemize}

\item{\it Displacement Multiplet:} The broken supercharges induce fermions $([0]^{(1)}_{\frac 32},\mathbf{2})$ on the defect. Together with the displacement operator $([1]^{(0)}_2,\mathbf{2})$, they sit inside the displacement multiplet $(B_1[0]_{\frac32}^{(1)},\mathbf{2})$:
\ie{}
\xymatrix{
*++[F-,][F**:Displacement]{{\boldsymbol{ B_1}}}} \qquad \xymatrix @C=3pc @R=3pc @r {
*++[F][F**:BrokenSUSY]{[0]_{\frac32}^{(1)},\bf{2}} \ar[r]|--{{~\cQ~}} 
&*++[F][F**:Displacement]{[1]^{(0)}_{2}, \bf{2}}
}
\fe
 
\item The defect must preserve both $\so(4)_{\text{rot}}$ transverse rotations and the 5d flavor symmetry $\mf{G}_f$ because $D(2,1;2;0)$ does not admit any multiplets with marginal top components (i.e. there are no broken R/F-current multiplets nor TR multiplets).\footnote{This addresses a question in \cite{Assel:2012nf} regarding the possible existence of 5d superconformal lines breaking the transverse rotation symmetry and certain bulk flavor symmetry. The answer is negative.}

\end{itemize}
	
\subsection{Examples of Superconformal Line Defects}
\label{sec:examplelines}

In this section, we briefly review realizations of superconformal lines in SCFTs that appear in the literature. They typically arise when the SCFT has a supersymmetric gauge theory description (either conformal or connected by an RG flow). As discussed in the Introduction, the known line defects are then described by either order-type Wilson lines \eqref{orderdef}, disorder-type line defects \eqref{disorderdef}, or a superposition of the two. We emphasize that all known superconformal line defects preserve the transverse rotation symmetry $\mf{so}(d-1)_{\text{rot}}$.

\subsubsection*{Four Dimensions}
Many 4d SCFTs have superconformal gauge theory descriptions which makes it possible to define manifestly superconformal line defects. 

We start with the most studied case of the 4d $\cN=4$ SYM. Here one can define unitary supersymmetric Wilson lines using the gauge fields $A_\mu$ as well as the adjoint scalars $\Phi^i$ with $i=1,2,\dots,6$ in the vector multiplet \cite{Maldacena:1998im,Rey:1998ik,Zarembo:2002an,Drukker:2006ga,Drukker:2007dw,Drukker:2007qr,Dymarsky:2009si,Wang:2020seq}
\ie{}
W_R^{\rm 4d} = \tr_R {\cal P} \exp \left(i \int {\cal L} \: d t \right), \quad {\cal L} = A_\mu \dot{x}^\mu + |\dot{x}| \Theta^i\Phi^i \,,
\label{4dwl}
\fe
along a general curve specified by $x^\mu(\tau)$, where the trace is taken in a particular representation $R$ of the gauge group. Here $\Theta^i$ is a position dependent coupling constrained by  $x^\mu(\tau)$ to preserve supersymmetry. For a superconformal Wilson line, we can take its worldvolume to be a straight line parametrized as $x^\m=(t,0,0,0)$ up to a conformal transformation. A quick calculation using the supersymmetry transformation rules of the SYM fields then shows that to preserve any supersymmetry requires $\Theta^i(\tau)=(0,\dots,0,1)$ up to a constant $\mf{so}(6)_R$ rotation. The resulting superconformal line is in fact half-BPS and preserves the superconformal subalgebra $\osp(4^*|4)$, which contains as bosonic subalgebras the residual $\so(5) \subset \su(4)_R$ R-symmetry that rotates the 5 scalars $\Phi^i$ with $i\neq 6$, the $\so(3)_{\text{rot}}$ transverse rotations, as well as the $\so(2,1)_{\rm conf}$ conformal symmetry along the line.\footnote{To give an example of a non-unitary superconformal line defect, we take the following scalar Wilson line in the 4d $\cN=4$ abelian SYM
\ie
W^{\rm 4d}= \exp \left( \A \int (\Phi_5+i\Phi_6) dt\right)\,,
\label{4dwlnu}
\fe
with a marginal parameter $\A$. This Wilson line preserves a Borel subalgebra of the complexified $\cN=4$ superconformal algebra.
} 
The generalization to superconformal Wilson lines in 4d $\cN=2$ superconformal gauge theories (such as the $\cN=2$ conformal SQCD) is straightforward. In this case, the scalar $\Phi^6$ that couples to the superconformal Wilson line in $\cN=4$ SYM gets replaced by an adjoint scalar $\sigma$ in the $\cN=2$ vector multiplet, and the corresponding Wilson line is again half-BPS and now preserves $\osp(4^*|2)$ superconformal symmetry. 

There are also supersymmetric line defects of the disorder-type in 4d $\cN=2$ and $\cN=4$ superconformal gauge theories given by 't Hooft line operators. They are specified by singular supersymmetric boundary conditions for the transverse field strength $F_{ij}$ and adjoint scalar $\sigma$ ($\Phi^6$ for $\cN=4$ SYM) along the line $x^\m=(t,0,0,0)$ \cite{Kapustin:2005py},
\ie
F_{ij}(t,\vec x)=&{1\over 2} \ep_{ijk} { x^k \over |\vec x|^3} T+{\rm regular}
,\quad 
\sigma(t,\vec x)= -{ 2\over |\vec x|}T+{\rm regular}
\label{4dtl}
\fe
where $T$ is an element in the cocharacter lattice $\Lambda_{\rm cochar}(G)$ of the gauge group $G$. 
These superconformal 't Hooft lines are related by S-duality to the supersymmetric Wilson lines, and more generally to Wilson-'t Hooft lines under the full duality group (e.g. $SL(2,\mZ)$ for the $\cN=4$ SYM with simply-laced gauge groups), which preserve the isomorphic half-BPS superconformal symmetry 
\cite{Gomis:2009ir,Gomis:2009xg,Drukker:2009tz,Alday:2009fs,Drukker:2010jp}. In large $N$ theories with $AdS_5$ supergravity duals, these  superconformal line defects correspond to either probe brane (string) solutions that wrap an $AdS_2\subset AdS_5$ submanifold \cite{Maldacena:1998im,Rey:1998ik,Skenderis:2002vf} or fully back-reacted geometries with an $AdS_2$ factor \cite{DHoker:2007mci,Gutperle:2019dqf} (e.g. the giant Wilson loop  in the symmetric representation of rank $\cO(N)$ in the $\cN=4$ SYM with gauge group $SU(N)$).
 
Furthermore, the 4d $\cN=4$ SYM with gauge group $G$ of ${\rm rank}(G)>1$ contains ${1\over 4}$-BPS superconformal Wilson-'t Hooft lines preserving the $\mf{osp}(4^*|2)\oplus \mf{su}(2)$ subalgebra, although they have not been studied much in the literature.\footnote{We thank Cumrun Vafa and Xi Yin for discussions related to this point.} Compared to the the half-BPS Wilson-'t Hooft lines in $\cN=4$ SYM, which involve assigning nontrivial boundary conditions to a single adjoint scalar field $\Phi_6$ as in \cite{Kapustin:2005py}, the ${1\over 4}$-BPS lines are specified by boundary conditions for two adjoint scalars $(\Phi_{5},\Phi_{6})$ aligned with two independent directions in the Cartan subalgebra of $G$.\footnote{The residual $\mf{so}(4)$ that rotates the other four adjoint scalars $\Phi_{1,2,3,4}$ corresponds to the 1d R-symmetry in $\mf{osp}(4^*|2)\oplus \mf{su}(2)$.}  On the Coulomb branch of the SYM, such ${1\over 4}$-BPS lines correspond to the world-lines of the well-studied ${1\over 4}$-BPS dyons \cite{Bergman:1997yw,Lee:1998nv}. In IIB string theory where the SYM arises from the worldvolume of a stack of D3 branes, these ${1\over 4}$-BPS superconformal lines can be engineered by three-pronged string webs attached to two probe D3 branes and the D3 stack \cite{Bergman:1997yw,Lee:1998nv}, and by a ${1\over 4}$-BPS D1-D5 system ending on the D3 stack along with its generalizations under $SL(2,\mZ)$ transformations.

With only 4d $\cN=1$ supersymmetry, vanishing of the one-loop beta function for the Yang-Mills coupling does not guarantee that it's exactly marginal,  due to the nontrivial renormalization of the K\"ahler potential \cite{Novikov:1983uc,Novikov:1985rd,Shifman:1986zi,Shifman:1991dz,Seiberg:1994bp,ArkaniHamed:1997mj}. Consequently  4d $\cN=1$ superconformal Lagrangians are rare except for those related by marginal deformations (superpotential deformations) to $\cN>1$ SCFTs.\footnote{For example see \cite{Leigh:1995ep,Aharony:2002hx} for $\cN=1$ exactly marginal deformations of the 4d $\cN=4$ SYM.} Nonetheless $\cN=1$ preserving RG flows from asymptotically free gauge theories in 4d are known to yield a huge zoo of 4d $\cN=1$ SCFTs  \cite{Seiberg:1994pq,Intriligator:1995au,Kutasov:1995ss,Intriligator:2003jj}, many of which have large $N$ holographic duals \cite{Klebanov:1998hh,Gubser:1998ia,Benvenuti:2004dy}. One may hope to construct supersymmetric Wilson ('t Hooft) lines as in the 4d $\cN=2$ and $\cN=4$ cases using the gauge theory description (either conformal or asymptotically free), but this is not possible because the 4d $\cN=1$ vector multiplet does not contain extra scalar fields.\footnote{We emphasize that there are light-like supersymmetric line defects that can be defined using the 4d $\cN=1$ gauge theory. For example, the ordinary Wilson line along the null line $x^\m=(t,t,0,0)$ is half-BPS (and preserves all the superconformal charges in the IR) \cite{Ouyang:2015ada}.} This is in agreement with our classification.

The 4d $\cN=3$ SCFTs do not admit conventional Lagrangian descriptions since the $\cN=3$ vector multiplet is automatically $\cN=4$ supersymmetric \cite{Aharony:2015oyb,Garcia-Etxebarria:2015wns,Aharony:2016kai}. Because of the non-Lagrangian nature, to our knowledge, no supersymmetric line defects are known in 4d $\cN=3$ SCFTs (that are not topological).
Some of these $\cN=3$ SCFTs can be realized by gauging a discrete $\mZ_k$ symmetry  with $k=3,4,6$ in the $\cN=4$ SYM, which is a combination of the $SO(6)$ R-symmetry and a certain $SL(2,\mZ)$ duality transformation that is promoted to a symmetry for special values of the marginal coupling, namely $\tau=e^{\pi i/3}$ or $\tau=i$. Since the $\mZ_k$ generator acts nontrivially on the adjoint scalar $\Phi^6$, the superconformal Wilson and 't Hooft lines of the $\cN=4$ SYM do not survive this discrete gauging.\footnote{More explicitly, up to $SO(6)_R$ conjugation, $\mZ_k$ acts on the pair of scalars $(\Phi_5,\Phi_6)$ by a  $2\pi \over k$ rotation \cite{Garcia-Etxebarria:2015wns}.} It would be interesting to come up with examples of superconformal lines in 4d $\cN=3$ SCFTs.

\subsubsection*{Five Dimensions}
The 5d $\cN=1$ supersymmetric gauge theories contain BPS Wilson loops analogous to the ones in  the 4d $\cN=2$ theories  
\cite{Assel:2012nf},
\ie{}
W^{\rm 5d}_R = \tr_R {\cal P} \exp \left(i \int {\cal L} \: dt \right), \quad {\cal L} = A_\mu \dot{x}^\mu + |\dot{x}| \sigma \,.
\label{5dwl}
\fe
along the straight line $x^\m=(t,0,0,0,0)$. Here the scalar $\sigma$ is a part of the 5d $\cN=1$ vector multiplet.

The Yang-Mills coupling is irrelevant in 5d and so such gauge theory descriptions can arise from a certain supersymmetric mass deformation of 5d $\cN=1$ SCFTs \cite{Seiberg:1996bd,Morrison:1996xf,Intriligator:1997pq,Jefferson:2017ahm}.  The corresponding mass parameter is then identified with the dimensionful Yang-Mills coupling ${1\over 
g_{\mathsf YM}^2}$ of the IR gauge theory. 

We expect the supersymmetric Wilson lines \eqref{5dwl} to be described by a superconformal line defect at the UV fixed point. Due to its half-BPS nature, its enhanced superconformal symmetry is fixed to be $D(2,1;2;0)\oplus \su(2)_{\text{right}}$ according to our classification in Table~\ref{table:defectClass}. In particular, the corresponding  superconformal line preserves the full $\su(2)_R$ R-symmetry of the 5d SCFT and $\so(4)_{\text{rot}} \simeq \su(2)_{\text{left}} \oplus \su(2)_{\text{right}}$ transverse rotations. In large $N$ SCFTs with $AdS_6$ supergravity duals, these superconformal line defects can be described by probe brane (string) solutions wrapping an $AdS_2$ submanifold in $AdS_6$ \cite{Assel:2012nf} or fully-back reacted geometries that involve an $AdS_2$ factor \cite{Chen:2020mtv}.

\subsubsection*{Six Dimensions}
In 6d SCFTs, there are no known superconformal line defects in the literature, partly due to the strongly-coupled nature of these theories. This is suggestive from our classification, which says that 6d $(1,0)$ SCFTs (with no enhanced supersymmetry) admit no superconformal lines, whereas 6d $(2,0)$ SCFTs may admit minimal lines that preserve $\mf{osp}(1|2;\mR)$ while breaking the transverse $\so(5)_{\text{rot}}$ rotation symmetry completely. Many 6d $(1,0)$ SCFTs have supersymmetric Lagrangian descriptions involving 6d gauge theories on their tensor branch moduli space (see for example \cite{Seiberg:1996qx,Intriligator:1997kq,Intriligator:1997dh,Heckman:2013pva,DelZotto:2014hpa,Heckman:2015bfa}), and so one can define Wilson lines as in lower dimensions. However, they are clearly non-supersymmetric, essentially because the 6d $(1,0)$ vector multiplet contains no scalars.\footnote{Once again it is possible to define supersymmetric light-like Wilson lines in 6d SCFTs which correspond to superconformal lightlike line  defects at the fixed point. }

\subsubsection*{Three Dimensions}

The story is significantly richer in 3d, where many different types of superconformal line defects abound. A good playground for the study of such lines is conformal Chern-Simons matter (CSM) theory, which comes with various amounts of supersymmetry, and in particular admits superconformal Wilson lines (for a recent review see \cite{Drukker:2019bev} and references therein). These lines all fall into two general classes, referred to as Gaiotto-Yin (GY) Wilson lines \cite{Gaiotto:2007qi}, and Drukker-Trancanelli (DT) Wilson lines \cite{Drukker:2009hy}.\footnote{Sometimes the GY and DT-type Wilson lines are also referred to as bosonic and fermionic Wilson lines respectively, due to the fermionic couplings in the definition of the DT-type Wilson line.}

The GY-type lines can be defined for any $\cN = 2$ CSM theory in terms of the gauge fields $A_\mu$ and the adjoint scalar $\sigma$ in the 3d vector multiplet  
\cite{Gaiotto:2007qi},
\ie{}
W^{\rm 3d}_{\text{GY}} = \tr_R {\cal P} \exp \left( i \oint {\cal L}_{\text{GY}} \: dt \right), \quad {\cal L}_{\text{GY}} = A_\mu \dot{x}^\mu + |{\dot x}|\sigma \,.
\label{3dwlgy}
\fe
Note the similarities between this construction and the $\frac12$-BPS lines found in 4d and 5d gauge theories. When the contour   is taken to be $x^\m=(t,0,0)$, the Wilson line preserves two real supercharges and their superconformal partners, as well as $\so(2)_{\text{rot}}$ transverse rotations and at least a $\uu(1)_R$'s worth of the R-symmetry, which altogether combine into $\su(1,1|1) \oplus \uu(1)_b$ symmetry. The line also preserves the global flavor symmetry, should any be present.

The $\cN=3$ extension of $\cN=2$ CSM theory includes $\frac13$-BPS GY-type lines, where the coupling is modified to include the scalar component of the $\cN=3$ auxiliary adjoint chiral field \cite{Gaiotto:2007qi}. Furthermore, the $\cN=2$ construction naturally generalizes to lines in $\cN>2$ theories. This includes examples such as $\frac16$-BPS lines  in the $\cN = 6$ $U(N_1)_k \times U(N_2)_{-k}$ theory of Aharony-Bergman-Jafferis-Maldacena (ABJM) and Aharony-Bergman-Jafferis (ABJ) \cite{Aharony:2008ug,Aharony:2008gk,Drukker:2008zx, Chen:2008bp,Rey:2008bh}, $\frac13$-BPS lines in $\cN=3$ flavored ABJM theory \cite{Santamaria:2010dm, Chen:2014gta}, and as special cases of more general $\frac14$-BPS lines in $\cN=4$ orbifold ABJM constructions \cite{Ouyang:2015qma}. The GY-type construction can also be realized in $\cN=8$ Bagger-Lambert-Gustavsson (BLG) theory \cite{Bagger:2007vi,Bagger:2007jr, Bagger:2006sk}, where the line in question is obvious in the $SU(2)_k \times SU(2)_{-k}$ CSM reformulation of the theory \cite{VanRaamsdonk:2008ft, Bandres:2008vf}, although such lines have never been explicitly studied in the literature. A notable feature of the GY-construction  is that it never experiences enhanced SUSY in $\cN > 2$ SCFTs and  always preserves the same $\mf{su}(1,1|1)$ superconformal symmetry (excluding additional bosonic symmetries).

We next consider the DT-type lines, which naturally live in quiver CSM theories. The first example of such lines appeared for $U(N_1)_k\times U(N_2)_{-k}$ ABJ(M) theory in the seminal work of \cite{Drukker:2009hy}, where they were written in terms of a $U(N_1|N_2)$ superconnection involving gauge fields $A_\mu$, as well as the bifundamental scalars $C_I$ and bifundamental fermions $\psi_I$ with $\mf{su}(4)_R$ fundamental indices $I=1,2,3,4$, 
\ie
W^{\rm 3d}_{\text{DT}} =& \tr_R {\cal P} \exp \left(i \int {\cal L}_{\rm DT} \: dt\right), \\  {\cal L}_{\rm DT} =& \begin{pmatrix}
	A_\mu \dot{x}^\mu + {2\pi \over k} |\dot{x}|M_J^{\:\: I} C_I \bar{C}^J&&  \sqrt{2\pi \over k}  |\dot{x}| \bar\zeta_I {\psi}^I \\
 \sqrt{2\pi \over k} 	|\dot{x}| \bar\psi_I {\eta}^I && \widehat{A}_\mu \dot{x}^\mu + {2\pi \over k} |\dot{x}| \widehat  M_J^{\:\: I} \bar{C}^J C_I 
\end{pmatrix} \,.
\label{3dwldt}
\fe
The DT construction contains several free parameters $ M_J^{\:\: I}$ and  $\eta_I^\A$ where $\A=\pm$ is a spinor index, which  can be tuned to preserve either ${1\over 6}$ or ${1\over 2}$ of the $\cN=6$ superconformal symmetries \cite{Ouyang:2015iza,Ouyang:2015bmy}. Hence the corresponding DT Wilson lines in the ABJ(M) theories preserve $\su(1,1|1)\oplus \uu(1)_b\oplus \mf{u}(2)_\cR$ and $\su(1,1|3)\oplus \uu(1)_b$ symmetries respectively.\footnote{The DT-type Wilson lines are invariant under transverse $\so(2)_{\text{rot}}$ rotations despite the fact that the line density $\cL_{\rm DT}$ transforms nontrivially under $\so(2)_{\text{rot}}$ due to the fermion couplings. This is because the rotation of the fermions $\psi \to e^{i\A} \psi, \bar \psi \to e^{-i\A} \bar\psi$ can be undone by a $U(1)$ gauge transformation given by $(e^{i \A\over 2},e^{-{i \A\over 2}})\in U(N_1)\times U(N_2)\subset U(N_1|N_2)$.
}
The DT-type construction can be generalized to a block-diagonal 2-node superconnection in more general quiver CSM theories, namely $\frac12$-BPS lines in $\cN=2$ theories  \cite{Ouyang:2015iza}, $\frac12$-BPS and $\frac14$-BPS lines in $\cN=4$ quiver CSM theories \cite{Ouyang:2015qma, Cooke:2015ila}, and $\frac25$-BPS lines in $\cN=5$ \cite{Lee:2010hk}. Note that such a construction appears to be absent for $\frac13$-BPS lines in $\cN=6$ ABJM \cite{Ouyang:2015bmy}. There are also further generalizations of the DT-type line where the connection cannot be put into block diagonal form consisting of 2-node blocks in the quiver, and sometimes even loses its interpretation as a supergroup connection. This more exotic construction exists for $\frac12$-BPS and $\frac14$-BPS lines in $\cN=4$ circular quivers \cite{Mauri:2017whf}, as well as $\frac12$-BPS lines in $\cN=2$ CSM theories \cite{Mauri:2018fsf}.\footnote{See also \cite{Drukker:2020opf} for recent work on constructing and classifying Wilson loops in $\cN\geq 2$ CSM theories.} 

Apart from the manifestly conformal CSM constructions, a large family of 3d SCFTs arise from supersymmetric RG flows from (abelian and non-abelian) Yang-Mills theories coupled to matter, possibly with additional Chern-Simons couplings \cite{Intriligator:1996ex,deBoer:1996mp,deBoer:1997kr,AdeBoer:1997kr,Kapustin:1999ha,Tong:2000ky,Gaiotto:2008ak,Dimofte:2011ju,Benini:2011mf,Intriligator:2013lca}. For example, a class of 3d $\cN=8$ ABJM theories are believed to describe the IR SCFT of 3d $\cN=8$ SYM with various gauge groups \cite{Gang:2011xp}.
These theories are generally harder to study due to the strongly coupled Yang-Mills dynamics in the IR, but a number of important CFT data such as the superconformal index, F-function, as well as conformal and flavor central charges can be extracted using supersymmetric localization (see \cite{Pestun:2016zxk}, and in particular the chapters \cite{Willett:2016adv} and \cite{Pufu:2016zxm} for a comprehensive review). The UV gauge theories admit supersymmetric Wilson lines of the GY-type \eqref{3dwlgy} and it is generally expected that they flow to superconformal line defects in the IR SCFT. This is supported by the results of localization methods that compute the defect free energy $\log \la \cD \ra$ of a circular Wilson loop on $S^3$ \cite{Kapustin:2009kz} and the defect superconformal index on $S^1\times S^2$ \cite{Drukker:2012sr}.

Finally there are also disorder-type line defects in 3d supersymmetric gauge theories, known as vortex lines \cite{Drukker:2008jm,Kapustin:2012iw}. Analogous to the 't Hooft lines in 4d, a vortex line along $x^\m=(t,0,0)$ is specified by a singularity of the gauge field\footnote{One can also define supersymmetric vortex lines for background gauge fields \cite{Kapustin:2012iw,Drukker:2012sr}.} 
\ie
F_{12}(t,x_i)=H \D_\Sigma(x_\perp) +{\rm regular} \,.
\label{3dvortex}
\fe
Here, $H$ is the vorticity --- it labels an element of the maximal torus of the gauge group $G$,
\ie
H\in {\mf{h}(G)}/2\pi \Lambda_{\rm cochar}(G)\,,
\fe
which induces a nontrivial holonomy for fields charged under $G$. Vortex lines can be generalized by allowing for singularities in the matter fields of the theory \cite{Drukker:2008jm,Drukker:2012sr}. For example the 3d $\cN=6$ ABJM theories contain superconformal vortex lines that are either ${1\over 2}$, ${1\over 3}$ or ${1\over 6}$-BPS, depending on which of the bifundamental scalars $C_I$ develop a scale invariant singularity
\ie
C_I (t,x_i)= {B_I\over \sqrt{z}}+{\rm regular} \,,
\fe
where $z=x_1+ix_2$ is a holomorphic transverse coordinate and $B_I\in \mf{h}(G)$ is in the same block diagonal form as $H$ \cite{Drukker:2008jm}. For CS levels $k=1,2$, the bulk SCFT has enhanced 3d $\cN=8$ superconformal symmetry, and as argued in \cite{Drukker:2008jm} these vortex lines become ${1\over 2}$, ${1\over 4}$ and ${1\over 8}$ BPS respectively.\footnote{Certain ${1\over 3}$-BPS vortex line in the 3d $\cN=6$ ABJM theory can also become $1\over 2$-BPS at $k=1,2$ \cite{Drukker:2008jm}.}  
By 3d mirror symmetry \cite{Intriligator:1996ex,deBoer:1996mp,Hanany:1996ie}, the vortex lines and Wilson lines can describe the same superconformal line in the SCFT, which is explored in \cite{Kapustin:2012iw,Drukker:2012sr,Dimofte:2019zzj}.
	
\section{Deformations of Superconformal Lines}
\label{sec:defor}
In this section, we discuss unitary deformations of a given unitary superconformal line defect $\cD$ by a local operator $\cO$ on its worldvolume $\Sigma$ with coupling $g_{\cO}$,
\ie
S_{\rm  DCFT} \to S_{\rm DCFT}+ g_{\cO}\int_{\Sigma} \cO(x_\parallel)\,,
\fe
that preserves the supersymmetry generators $\cQ\in \mf{g}_s$, the Poincar\`e symmetry (translation along the line), but not necessarily the conformal and superconformal symmetries. This requires $\cO$ to be a top component in a unitary 1d superconformal multiplet with respect to $\mf{g}_s$. Depending on whether the operator $\cO$ resides in a short or long multiplet, we refer to the corresponding deformations as F-term and D-term deformations respectively.

These types of deformations fall into one of three categories based on the scaling dimension $\Delta_\cO$ of $\cO$, namely relevant deformations for $0<\Delta_\cO<1$, marginal deformations for $\Delta_\cO=1$ and irrelevant deformations for $\Delta_\cO>1$. The marginal case  further consists of marginally relevant, marginally irrelevant, and exactly marginal operators. The relevant and irrelevant (as well as marginally relevant and marginally irrelevant) deformations keep track of supersymmetric defect RG flows from and into the superconformal defect $\cD$, whereas the exactly marginal deformations generate a conformal manifold of defects $\cM_{\cD}=\{g_\cO\}$ parametrized by the couplings. 

The D-term deformations come from the unique top component in long multiplets of the defect superconformal algebra $\mf{g}_s$. For a defect $\cD$ that preserves $\cN_\cD$ supercharges $\cQ$ (for line defects $\cN_D=\cN_1$ as in Table~\ref{table:defectClass}), $\cO$ is given schematically by $\cQ^{\cN_\cD}$ acting on the superconformal primary in a long multiplet. Consequently a D-term deformation always has $\Delta_\cO > {{\cN_\cD}\over 2}$, which is irrelevant for ${\cN_\cD}>1$.\footnote{Recall that the ${\cN_\cD}=1$ 1d superconformal algebra is   $\mf{osp}(1|2;\mR)$ which admits no short multiplets.}
The F-term deformations are more intricate as they can show up as either manifest or accidental top components in a short multiplet.\footnote{Recall that a short multiplet may contain multiple manifest top components, some of which can appear in the ``middle'' of the multiplet in terms of scaling dimensions \cite{Cordova:2016emh,Cordova:2016xhm}.} Since the superconformal primaries are annihilated when acted upon by certain combinations of the supercharges $\cQ$, these top components can have lower scaling dimensions. 

From the classification of unitary  superconformal multiplets for admissible 1d superconformal algebras of line defects in Table~\ref{table:defectClass}, we observe that, apart from the minimal superconformal lines:
\begin{itemize}
    \item \textit{Relevant F-term} deformations are only possible for line defects in 3d $\cN>1$ SCFTs preserving the defect superconformal algebra $\mf{su}(1,1|1)$. They arise as the unique manifest top component in short multiplets of the type $A_1 \bar{L}[j]_j$ (and its conjugate) for $0<j<{1\over 2}$.
    \item \textit{Marginal F-term} deformations are only possible for line defects in 3d $\cN>1$ SCFTs preserving the following defect superconformal algebras:  $\mf{su}(1,1|1)$, $\uu(1) \rtimes \psu(1,1|2) \rtimes \uu(1)$,  $\mf{su}(1,1|3)$ and  $\mf{su}(1,1|4)$. For the case of $\mf{su}(1,1|1)$,  the marginal operator  comes from the unique top component of the multiplet $A_1 \bar{L}[{1\over 2}]_{1\over 2}$ (and its conjugate). In all other cases, the marginal operators are associated to manifest top components of the multiplet $A_1 \bar{A}_1[0]^{(R)}_{1\over 2}$ that appears as a level-1 $\cQ$-descendant of the superconformal primary.\footnote{We emphasize that these short multiplets which contain marginal operators are never \textit{absolutely protected}, in the sense that they can recombine with other multiplets into a long multiplet at threshold, and can consequently develop anomalous dimensions as we tune $g_\cO$. To answer whether this happens or not in a particular defect SCFT potentially requires dynamical data of the theory, for example the three-point function of the marginal operator $\cO$.}
    \item \textit{Irrelevant F-term} deformations are possible for all superconformal lines defects. 
\end{itemize}
The minimal superconformal lines (in any dimension)    preserve $\mf{g}_s=\mf{osp}(1|2;\mR)$  and do not admit F-term deformations, but they can have D-term deformations that are either relevant, marginal, or irrelevant.

Let us now discuss in more detail the implications of the above statements for superconformal line defects in relation to their explicit realizations, as reviewed in Section~\ref{sec:examplelines}. 

\subsection{Rigid Superconformal Lines in 4d and 5d SCFTs}
Superconformal line defects that preserve $\cN_1>1$ supercharges in $d=4,5$ SCFTs represent \textit{completely stable} fixed points of supersymmetric defect RG flows. They include all superconformal lines  invariant under the transverse rotation symmetry $\mf{so}(d-1)_{\text{rot}}$ in these spacetime dimensions. Indeed, for Wilson lines in 4d $\cN=4$ SYM of the form \eqref{4dwl}, and the $\cN=2$ generalizations, it is straightforward to check that the superconformal half-BPS line admits no relevant or marginal unitary supersymmetric deformations in terms of the elementary fields (similarly for the 't Hooft loops \eqref{4dtl}).\footnote{We emphasize the importance of unitarity again. Without unitarity, there are marginal deformations of the half-BPS Wilson lines (that couple to $A_\m$ and $\Phi_6$ only) by a coupling to the complex scalar $\Phi_4+i \Phi_5$ along the line as in \eqref{4dwlnu} that preserves a fraction of the supersymmetries.} They do however admit non-supersymmetric deformations by the adjoint scalar $\sigma$ in the $\cN=2$ vector multiplet ($\Phi^6$ in the $\cN=4$ case), which is known to be marginal for the half-BPS Wilson line in $\cN=4$ SYM in the free limit (or in the abelian SYM) but irrelevant at finite coupling \cite{Beccaria:2017rbe}.\footnote{This operator tracks the defect RG flow from the non-BPS conformal Wilson line in $\cN=4$ SYM into the BPS Wilson line \cite{Polchinski:2011im}.} There are also marginal (irrespective of bulk gauge coupling) non-supersymmetric deformations of these half-BPS Wilson loops by the superconformal primary scalars in the displacement multiplet ($A_2[0]_1^{(0)}$ for $\cN=2$ \eqref{4dn2dis} and $B_1[0]_1^{(0,1)}$ for $\cN=4$ \eqref{4dn4dis}).

Wilson lines in 5d $\cN=1$ SCFTs are more intricate due to the non-renormalizable nature of the 5d gauge theory description. As reviewed in Section~\ref{sec:examplelines}, the 5d supersymmetric Wilson lines are defined in a similar fashion to their cousins in 4d gauge theories. The belief is that they describe superconformal line defects of the SCFT in the IR gauge theory phase, which has been verified by supersymmetric localization computations in the field theory, as well as comparisons to string and probe brane solutions in $AdS_6$ supergravity via AdS/CFT \cite{Assel:2012nf,Uhlemann:2020bek}. This matching is still mysterious since the localization result utilizes only the classical two-derivative super-Yang-Mills action coupled to matter hyper-multiplets, while in principle one should also include irrelevant interactions, both in the bulk and on the defect, generated along the supersymmetric RG flow from the UV fixed point by the mass deformation.\footnote{Relatedly, it would be interesting to understand the counter-term ambiguities in the localization formulae that compute defect observables similar to the analysis in \cite{Chang:2017cdx}.} Moreover, if the superconformal line defect were to have an exactly marginal parameter $g$, we would need to specify for what value of $g$ the IR localization formulae applies (i.e. which point on the conformal manifold). The fact that 5d superconformal lines do not admit relevant or marginal deformations, as seen in our earlier analysis, means that such complications do not arise on the defect worldvolume. It furthermore explains, to an extent, why the localization formula works for Wilson line defects in 5d SCFTs.

\subsection{Deformable Superconformal Lines in 3d SCFTs}
The situation is drastically different for 3d SCFTs, where superconformal lines generally have marginal parameters. If the number of supersymmetries preserved by the line defect is small (e.g. $\cN_1\leq 2$), relevant deformations are also possible. A large class of 3d SCFTs are described by conformal Chern-Simons matter theories of various amounts of supersymmetry, which admit rich families of superconformal Wilson lines and vortex lines as reviewed in Section~\ref{sec:examplelines}. While the GY-type Wilson lines \eqref{3dwlgy} do not have marginal parameters, the DT-type Wilson lines \eqref{3dwldt} admit supersymmetric marginal deformations by tuning the parameters $M_I{}^J$ and $\eta_I$ in \eqref{3dwldt} simultaneously and preserve $\mf{g}_s=\mf{su}(1,1|1)$ superconformal symmetry in general. Computations of the defect free energy $F_\cD\equiv-\log \la \cD \ra$ via localization on $S^3$ with the defect along a great $S^1$ \cite{Kapustin:2009kz} suggest that such deformations are exactly marginal \cite{Drukker:2009hy,Ouyang:2015bmy,Ouyang:2015iza},  giving rise to a conformal manifold $\cM_{\rm DT}$ of $\mf{su}(1,1|1)$ preserving superconformal lines, which include the GY-type Wilson lines on a sub-locus. 

To be explicit, let us  consider DT-type Wilson lines in the 3d $\cN=6$ $U(N_1)_k\times U(N_2)_{-k}$ ABJ(M) theories as in \eqref{3dwldt}. Here we have a family of ${1\over 6}$-BPS lines (preserving the same $\su(1,1|1)$ subalgebra) parametrized by \cite{Ouyang:2015bmy,Ouyang:2015iza,Drukker:2019bev}
\ie
\cM^{\rm cplx}_{\rm DT}:~~M={\rm diag}(-1,-1 ,1,1)
-2 \A \otimes \B
,\quad \bar\zeta_I=  \A_I (i,-1),\quad  \eta^I=    \B^I (1,-i)\,,
\label{DTcm}
\fe
where $\A_I=(\A_1,\A_2,0,0)$ and $\B^I=(\B^1,\B^2,0,0)$ are complex parameters subjected to the $\mC^*$ (gauge) identification $(\A \sim  \lambda \A, \B \sim  \lambda^{-1} \B)$ which does not change the Wilson line. The parameter space $\{(\A_a,\B^b)\}/\mC^*$ for $a,b=1,2$ is the usual conifold and the conformal manifold $\cM_{\rm DT}$ of unitary ${1\over 6}$-BPS lines is given by a 3-dimensional real section.\footnote{We thank Nadav Drukker for explaining this to us.}
The GY-type Wilson line in the ABJ(M) theory is located at $\A_a=\B^b=0$  on $\cM_{\rm DT}$. Moreover $\cM_{\rm DT}$ contains isolated points with enhanced $\mf{su}(1,1|3)$ superconformal symmetry, for example at $\A_1=\B^1=0$ which describes the half-BPS DT-type Wilson lines \cite{Drukker:2009hy}.
In the  't Hooft large $N$ limit,  the simplest Wilson lines correspond  to boundary-anchored fundamental strings in the IIA holographic dual on $AdS_4\times \mathbb{CP}^3$. The marginal parameters of the Wilson lines can be understood from the $AdS_2$ world-sheet of the strings. In particular, the analysis of \cite{Correa:2019rdk} suggests that the above marginal deformations (along a one dimensional submanifold of $\cM_{\rm DT}$) are described by a family of mixed boundary conditions for the massless fermions on the $AdS_2$.
We emphasize that although the presence of   marginal deformations for the superconformal Wilson lines in ABJM theories are consistent with our classification at a technical level, the conformal manifold $\cM_{\rm DT}$ in \eqref{DTcm} displays the exotic feature that the deformation is induced by a \textit{nonlocal} defect operator of the schematic form\footnote{This first appeared in  Diego Correa's talk ``Wilson lines as 1d defects'' at IGST 2020. The slides are avaliable at \texttt{https://www.ictp-saifr.org/wp-content/uploads/2020/08/CorreaIGST2020.pdf}.}
\ie
\cO(t)=C_2\bar C^2(t)+\psi^1(t)\int_t^\infty dt' \bar \psi_1(t')\,,
\fe
essentially because of the fermionic off-diagonal components of the superconnection in \eqref{3dwldt}. This nonlocal feature is also present in the proposed holographic dual \cite{Correa:2019rdk}.

Let us now discuss deformations of the superconformal vortex lines, focusing on the  ${1\over 2}$-BPS examples in 3d $\cN=6$ $U(N)_k\times U(N)_{-k}$ ABJM theories \cite{Drukker:2008jm}. For even $k$ these vortex lines come with  discrete families labeled by Levi subgroups of $U(N)$,\footnote{A similar family of vortex lines exist for ABJM theories at odd $k$ \cite{Drukker:2008jm}.}
\ie
L=U(N_0)\times U(N_1)\times \dots \times U(N_m)
\fe
with $\sum_{i=0}^m N_i=N$. Within each family, the vortex line defect is defined by singularities in the gauge and scalar fields as
\ie
C_1=&~{1\over \sqrt{z}} {\rm diag} (0_{N_0},\B_1 1_{N_1},\dots,\B_m 1_{N_m})
,\quad 
C_2= C_3=C_4=0\,,
\\
A_z=&~\widehat{A}_z=-{i\over 4kz}  {\rm diag} (0_{N_0},\A_1 1_{N_1},\dots,\A_m 1_{N_m})
,\quad 
A_t=\widehat{A}_t=-2\pi \bar C_1 C^1\,.
\fe
The real parameters $(\A_i,\B_i)$ are marginal and so it would be interesting to investigate whether they are exactly marginal, thus defining a conformal manifold $\cM_{\rm VL}$ of superconformal vortex lines. Similarly, ${1\over 4}$ and ${1\over 6}$-BPS vortex lines in the ABJM theories also contain (more) marginal parameters \cite{Drukker:2008jm}, confirming our general argument from superconformal representation theory.

Line defects preserving $\mf{g}_s=\mf{su}(1,1|1)$ superconformal symmetry in 3d SCFTs may also have relevant supersymmetric deformations according to our analysis. This requires a defect superconformal primary $\cV$ of scaling dimension $\Delta<{1\over 2}$ whose level-one $\cQ$-descendant $\cO$ can trigger a supersymmetric RG flow. Clearly $\cV$ has to be an intrinsic defect operator since its dimension is below the unitarity bound of 3d local operators. For the superconformal Wilson lines in CSM theories, such an operator $\cV$ can potentially arise as (dressed) monopole operators that are stuck to the Wilson line (e.g. a monopole operator in the adjoint representation of the gauge group). It would be interesting to analyze such deformations in more detail.

\subsection{Unavoidable Marginal Deformations}
Interestingly, we also find that certain line defects \textit{must} have marginal deformations. One set of examples are the half-BPS superconformal lines in 3d $\cN=8$ SCFTs. Recall that the $\mf{su}(1,1|4)$ displacement multiplet $A_1\bar A_1 [{1\over 2}]^{(1,0,0)}_{1\over 2}$ \eqref{eq:3dN8halfBPSdisplacementMult} contains two top components, one being the displacement operator for the line at level 3, while the other is a marginal operator $\cO$ at level 1 (which is exactly marginal in this case). Thus all half-BPS lines in 3d $\cN=8$ SCFTs are guaranteed to have a complex one-dimensional conformal manifold. This is akin to the case of 4d $\cN=4$ SCFTs, where the stress-tensor multiplet contains a scalar top component of $\Delta=4$ that generates a conformal manifold of $\cN=4$ SCFTs \cite{Cordova:2016emh,Cordova:2016xhm}.

Another example of unavoidable defect marginal parameters are those which arise from superconformal lines that break global symmetries of the bulk SCFT, or more generally any one-form currents that commute with the worldvolume superconformal symmetry $\mf{g}_s$. The modified Ward identities of these currents give rise to marginal operators that commute with the supercharges in $\mf{g}_s$, as follows from Theorem~\ref{thm2}.

\section{Further Applications and Discussions}
 \label{sec:discuss}

In this work, we have focused on exploring kinematic constraints on line defects in SCFTs coming from the (residual) superconformal symmetry and its unitary representations. We have observed that the introduction of supersymmetry leads to nontrivial constraints that greatly narrow down the list of admissible superconformal symmetries for line defects (see Section~\ref{sec:classify} and Table~\ref{table:defectClass}). Additionally, possible supersymmetric deformations of these superconformal lines are also heavily constrained (see Section~\ref{sec:defor}). Below we discuss some additional applications of our results.

\subsubsection*{Defect Conformal Bootstrap}
While some of the superconformal lines in Table~\ref{table:defectClass} have explicit realizations in either superconformal gauge theories or via RG flows that connect to supersymmetric non-conformal gauge theories, others are still elusive (including the minimal superconformal line in $d=3,4,5,6$).
In either case, it would be interesting to study such line defects following the defect conformal bootstrap program \cite{Liendo:2012hy,Billo:2016cpy,Gadde:2016fbj,Lauria:2018klo,Liendo:2018ukf,Bianchi:2020hsz}, which incorporates dynamical information of the defect SCFT, namely the operator spectrum and OPE coefficients. For example, one can start with the four-point function of the universal displacement multiplet on the defect, or with the two-point function of bulk local operators in the presence of the defect. For both setups, the correlator is a function of a single conformal cross-ratio that obeys nontrivial constraints from superconformal symmetry, unitarity, and crossing. For superconformal lines that have known realizations, it would be interesting to combine the bootstrap with other techniques such as supersymmetric localization \cite{Pestun:2016zxk} and integrability methods \cite{Beisert:2010jr}  to solve the defect correlation functions.\footnote{For recent progress along this line for the half-BPS Wilson lines in the 4d $\cN=4$ SYM, see for example \cite{Giombi:2017cqn,Kim:2017sju,Giombi:2018qox,Liendo:2018ukf,Giombi:2018hsx,Kiryu:2018phb,Giombi:2020amn} following earlier works on Wilson line DCFT in \cite{Drukker:2005af,Drukker:2006xg,Cooke:2017qgm}.} For more exotic cases, such as the putative minimal superconformal lines  in 6d $(2,0)$ SCFTs, the bootstrap program holds the promise of either proving their absence or providing evidence for their existence.

\subsubsection*{Defect $C$-function and $C$-theorem}
The $C$-function ($F$-function) that plays an important role in characterizing the monotonic nature of RG flows between CFTs has a close analog in the presence of defects. A central problem in this regard is to identify the defect $C$-function ($F$-function) in terms of defect observables and prove the corresponding $C$-theorem ($F$-theorem) for defect RG flows. Despite a number of conjectures and partial checks \cite{Yamaguchi:2002pa,Nozaki:2012qd,Estes:2014hka,Gaiotto:2014gha,Kobayashi:2018lil}, apart from a proof for line defects in 2d CFTs (also known as $g$-theorem) \cite{Friedan:2003yc,Casini:2016fgb} and surface defects in general CFTs \cite{Jensen:2015swa}, the defect $C$-theorems remain largely open (see \cite{Kobayashi:2018lil} for a summary). For superconformal line defects, as  explained in Section~\ref{sec:defor}, the only cases that admit relevant supersymmetric deformations are the minimal superconformal lines in $d=3,4,5,6$ or the lines preserving $\mf{su}(1,1|1)$ symmetry in 3d $\cN\geq 2$ SCFTs. It would be interesting to investigate these supersymmetric defect RG flows in pursuit of a proof of the defect $C$-theorem in the supersymmetric setting. 
As conjectured in \cite{Kobayashi:2018lil}, one expects the defect free energy  $F_\cD \equiv  -\log \la \cD \ra$, where $\la \cD \ra$ is the vev of the circular line operator (after Wick rotation to Euclidean signature),\footnote{The expectation value of the (Euclidean) conformal defect on a circle differs from that on a straight line despite the two being related by a special conformal transformation. This is expected to be a consequence of certain conformal anomaly of the line defect  \cite{Drukker:2000rr}, however a complete characterization of this anomaly (e.g. in terms of contact terms for the stress-tensor) is still lacking. See  \cite{Bianchi:2019umv} for recent progress in understanding this anomaly.
} to be a monotonic $C$-function (at least in the weak form). For $\mf{su}(1,1|1)$ preserving superconformal 
lines in 3d $\cN\geq 2$ SCFTs, such defect vevs can be computed with the localization method \cite{Kapustin:2009kz}, which provides an explicit and nontrivial testing ground for possible defect RG flows and the conjectured $C$-theorem.
   
\subsubsection*{Higher-form Symmetries}
One notable feature of defects is that they detect higher $p$-form symmetries $G_{[p]}$ \cite{Gaiotto:2014kfa}. In particular, an SCFT has a faithful one-form symmetry $G_{[1]}$ only if it contains line defects that carry charges under $G_{[1]}$. Since the quantized charges (for compact $G_{[1]}$) do not renormalize under (defect) RG flows, we can focus on the conformal line defects that describe IR fixed points of possibly non-conformal line defects charged under $G_{[1]}$ in the CFT.  Recently, numerous examples of SCFTs with nontrivial discrete one-form symmetries have been identified from  field theoretic arguments and string/M/F-theory constructions \cite{Garcia-Etxebarria:2019cnb,Bergman:2020ifi,Morrison:2020ool,Albertini:2020mdx,Bah:2020uev,Closset:2020scj,DelZotto:2020esg,Bhardwaj:2020phs}. In cases where there exists a Lagrangian description, such as in the 4d $\cN=4$ SYM and 3d $\cN=6$ ABJM theories, the familiar (superconformal) Wilson line operators play the role of nontrivial $G_{[1]}$ charged objects. In other non-Lagrangian cases such as 6d SCFTs it is yet to be settled whether the fixed points harbor faithful one-form symmetries.\footnote{See \cite{Apruzzi:2020zot} for relevant recent discussions based on the 6d $(1,0)$ tensor branch effective theory.} Our classification of superconformal lines in Section~\ref{sec:classify} indicates that the 6d $(1,0)$ SCFTs (with no enhanced supersymmetry) do not admit any superconformal line defects, while 6d $(2,0)$ SCFTs only admit the minimal superconformal line, which completely breaks the transverse $\so(5)$ rotation symmetry. This puts strong constraints on the fate of the conformal phase of a putative supersymmetric defect charged under the 6d one-form symmetry  $G_{[1]}$. If the defect that one starts out with preserves $\cN_1>1$ supercharges, it must become a nontrivial topological line in the IR\footnote{It can happen that the one-form symmetries only act non-trivially in a topological sector of the SCFT. To give an example of such a topological sector, let's consider the $d$-dimensional TQFT with the BF action $S_{\rm BF}={N\over 2\pi}\int B\wedge dA $ where  $A$ and $B$ are $U(1)$ 1-form and $(d-2)$-form gauge connections respectively. This TQFT is also known as the $\mZ_N$ discrete gauge theory, as can be seen by integrating out continuous fluctuations of $A$ and $B$. The basic observables in the BF theory are topological defect operators $W=e^{i\oint A}$ and $U=e^{i\oint B}$ defined by integrals over one and $(d-2)$ dimensional submanifolds of the Euclidean spacetime. The BF theory has $(\mZ_N)_{[d-2]}\times (\mZ_N)_{[1]}$ global symmetry with a mixed 't Hooft anomaly. The generators of these higher-form global symmetries are precisely given by $W$ and $U$ respectively. Moreover $W$ is charged under $(\mZ_N)_{[1]}$ while $U$ is charged under $(\mZ_N)_{[d-2]}$, which gives rise to the mixed 't Hooft anomaly.
 } that preserves the full bulk superconformal symmetry.\footnote{Here we assume that there is no spontaneous supersymmetry breaking. It would be interesting to investigate this possibility further. For example, if the Witten index of the theory with the line $\cD$ inserted along the time direction is non-vanishing, such SUSY breaking is forbidden.} For $\cN_1=1$, there is the additional possibility of a minimal superconformal line, but only if the bulk SCFT has $(2,0)$ supersymmetry and the line necessarily breaks transverse rotation symmetry. This illustrates to some degree the difficulty in realizing one-form symmetries in 6d SCFTs. One either has to forgo supersymmetry and look for non-supersymmetric conformal lines charged under $G_{[1]}$ at the cost of losing analytic control, or to look for exotic minimal superconformal lines that break transverse rotation symmetry.
  Similar RG constraints can be easily deduced for line defects in other non-Lagrangian SCFTs, and it would be interesting to understand the implications for one-form symmetries in such theories.

 \subsubsection*{Defect Integrability}
Integrability has proven to be a powerful tool to solve defect observables in the 4d $\cN=4$ SYM and 3d $\cN=6$ ABJM theories in the 't Hooft large $N$ limits (see \cite{Andrei:2018die,Drukker:2019bev,deLeeuw:2019usb} for relevant reviews and \cite{Jiang:2019xdz,Jiang:2019zig,Komatsu:2020sup,Gombor:2020kgu,Kristjansen:2020mhn,Gombor:2020auk} for  more recent developments). The analysis hinges on the assumption that the corresponding defects are integrable, examples of which include BPS Wilson lines and interfaces. A general criterion for integrable defects in integrable field theories was proposed in \cite{Delius:2001he} (see also \cite{Gombor:2019bun}), which states that the   defect superconformal algebra and the bulk superconformal algebra should form a symmetric pair $(\mf{g}_s,\mf{G}_s)$. In other words, $\mf{g}_s$ is a maximal subalgebra defined by the invariant subalgebra of an involution of $\mf{G}_s$. Since $d=5,6$ large $N$ SCFTs are expected to be non-integrable \cite{Wulff:2019tzh}, we focus on $d=3,4$ SCFTs.\footnote{The argument in \cite{Wulff:2019tzh} also rules out integrability for 4d $\cN=2$ SCFTs of class S type which have M-theory duals   \cite{Gaiotto:2009gz}, as well as certain 3d $\cN=3$ SCFTs with massive IIA duals \cite{Pang:2015vna,DeLuca:2018buk}.} Clearly the symmetric pair condition requires the superconformal defect to be half-BPS. 
Among the admissible superconformal lines, this holds for half-BPS lines in 4d $\cN=2,4$ SCFTs with symmetric pairs $(\mf{g}_s,\mf{G}_s)$:
\ie
(\mf{osp}(4^*|2),\mf{su}(2,2|2))\,,~ (\mf{osp}(4^*|4),\mf{psu}(2,2|4))\,,
\fe
and for  half-BPS lines in 3d $\cN=2,4,6,8$ SCFTs with symmetric pairs $(\mf{g}_s,\mf{G}_s)$:
\ie
&(\mf{su}(1,1|1),\mf{osp}(2|4;\mR))\,,~ (\mf{u}(1) \rtimes\mf{psu}(1,1|2)\rtimes \mf{u}(1),\mf{osp}(4|4;\mR))\,,~
\\
&(\mf{su}(1,1|3),\mf{osp}(6|4;\mR))\,,~(\mf{su}(1,1|4),\mf{osp}(8|4;\mR))\,.
\fe
It would be interesting to investigate and apply integrability methods to these line defects beyond the ones in the 4d $\cN=4$ SYM and 3d $\cN=6$ ABJM theories. 
	
\section*{Acknowledgements}
We thank Simone Giombi,  Ken Intriligator, Daniel Jafferis, Shota Komatsu, Bruno Le Floch, David Tong, Cumrun Vafa, and Xi Yin for useful discussions and correspondences. We are also grateful to Simone Giombi for helpful comments on the draft and to Nadav Drukker for helpful comments on the first version of the manuscript. 
The work of YW is  supported in part by the Center for Mathematical Sciences and Applications and the Center for the Fundamental Laws of Nature at Harvard University.

\appendix

\section{Exceptional Isomorphisms}

\subsection{Lie Algebras}

In this section, we list the exceptional isomorphisms of the low-lying real forms of Lie algebras. We refer the reader to e.g. Appendix A of \cite{DHoker:2008wvd} for a complete list of working definitions. The relevant real forms (with anti-Hermtian generators) can be defined as the matrix subalgebras
\ie{}
\su(p,q) &\equiv \{ A \in \msl(p+q,\bC)~|~ A^\dagger I_{p,q} + I_{p,q} A = 0\} \,, \\
\so(p,q) &\equiv \{ A \in \msl(p+q,\bR) ~|~  A^t I_{p,q} + I_{p,q} A = 0\} \,, \\
\usp(2p,2q) &\equiv \msp(2p+2q,\bC) \cap \su(2p,2q) \,, \\
\so(2m^*) &\equiv \{A \in \so(2m,\bC) ~|~  A^\dagger J_{2m} + J_{2m} A = 0 \} \,, \\ 
\su(2m^*)&\equiv \{A \in \msl(2m,\bC) ~|~  A J_{2m} - J_{2m} A^* = 0 \} \,,\\
\fe
where the matrices $I_{p,q}$ and $J_{2m}$ are given by
\ie{}
I_{p,q} = 
\begin{pmatrix}
-I_p & 0 \\
0 & I_q
\end{pmatrix} \,, \quad 
J_{2m} = 
\begin{pmatrix}
0 & I_m \\
-I_m & 0
\end{pmatrix} \,.
\fe
Here, $I_n$ is the $n \times n$ identity matrix.
It follows that the isomorphisms of compact real forms are given by
\ie{}
\so(2) &\simeq \uu(1) \,, \\
\so(3) &\simeq \su(2) \simeq \usp(2) \,, \\
\so(4) &\simeq \su(2) \oplus \su(2) \,, \\
\so(5) &\simeq \usp(4) \,, \\
\so(6) &\simeq \su(4) \,, \\
\fe
while those of the noncompact real forms are given by
\ie{}
\so(2,1) &\simeq \msl(2,\bR)  \simeq \su(1,1) \simeq \msp(2,\bR) \,, \\
\so(3,1) &\simeq \msl(2,\bC) \,, \\
\so(2,2) &\simeq \so(1,2) \oplus \so(1,2) \,, \\
\so(4^*) &\simeq \msl(2,\bR) \oplus \so(3) \,, \\
\so(4,1) &\simeq \usp(2,2) \,, \\
\so(3,2) &\simeq \msp(4,\bR) \,, \\
\so(5,1) &\simeq \su(4^*) \,, \\
\so(4,2) &\simeq \su(2,2) \,, \\
\so(6^*) &\simeq \su(1,3) \,, \\
\so(3,3) &\simeq \msl(4,\bR) \,, \\
\so(6,2) &\simeq \so(8^*) \,.
\fe
Note that $\so(2^*) \simeq \so(2)$ is compact.

\subsection{Lie Superalgebras}

There are several exceptional isomorphisms between the low-lying real forms of Lie superalgebras. Those relevant to this work are (see \cite{DHoker:2008wvd} for a more complete list):
\ie{}
\osp(2^*|2) &\simeq \su(2|1) \,, \\
\osp(2|2;\bR) &\simeq \su(1,1|1) \,, \\
D(2,1;\lambda;0) &\simeq \osp(4^*|2) \,, \quad \lambda = -2, 1 \,, \\
D(2,1;\lambda;0) &\simeq \osp(4|2;\bR) \,, \quad \lambda = -\tfrac12 \,, \\
D(2,1;\lambda;0) &\simeq D(2,1;-1-\lambda;0)\,, \quad \lambda \in \bR \,.
\fe

\section{Superconformal Algebras and Broken Generators}
\label{sec:1dSCA}

In this section, we establish our conventions for the 1d superconformal algebras under consideration and present explicit formulae that may be useful for future work on the topic. In particular, for each bulk spacetime dimension $d=3,4,5,6$, we provide an explicit realization of the maximal 1d superconformal subalgebras, only listing the most important details of further 1d subalgebras. We also describe the broken generators that enter into the modified Ward identities. In most cases, the details of the bulk superconformal algebra are irrelevant and so we need only list the branching rules that give the decomposition of the bulk R-symmetry currents and supercurrents under the 1d subalgebra (as well as their subalgebras). There is a single exception, namely the $\su(1,1|N)$ subalgebra of $\osp(\cN|4;\bR)$, where the $\uu(1)_{\cR_N}$ charges of the broken bulk currents come from a combination of their $\uu(1)_R \subset \so(\cN)_R$ charge as well as their transformation properties under transverse rotations, and so their overall normalization is not fixed from the naive branching rules alone. For this case, we provide an explicit embedding of the 1d algebra into the 3d superconformal algebra.

In all cases, we work with the Euclidean conformal algebra of the radially quantized theory (after Wick rotation), with generators $\{D, P, K\}$ that satisfy 
\ie{}
[D,P] = P\,, \quad [D,K] = -K\,, \quad [K,P] = 2D \,.
\fe
and have the Hermiticity properties\footnote{In this appendix, to avoid factors of $i$'s in the (anti)commutation relations, we work with a different dilatation generator from the main text, namely $D = -iD_E$, whose eigenvalues are given by the real scaling dimension $\Delta$. }
\ie{}
D^\dagger = D\,, \quad P^\dagger = K\,, \quad K^\dagger = P \,.
\fe

\subsection{$\su(1,1|N)$}
\label{sec:su11NAlgebra}

\subsubsection*{Bulk 3d Algebra}

We begin by laying out the conventions for $\osp(\cN|4;\bR)$, the 3d superconformal symmetry algebra. Its bosonic part is isomorphic to $\so(2,3)_{\text{conf}} \oplus \so(\cN)_R$. The conformal algebra consists of dilatations $D$, translations $P_{\alpha\beta}$, special conformal transformations $K^{\alpha\beta}$, and Lorentz rotations $M_\alpha^{\:\:\:\beta}$, where the generators are written in the $\so(1,2)$ spinorial basis with $\alpha,\beta = \pm$. In this basis, the conformal algebra takes the form\footnote{Here, we use parentheses to denote symmetrization over indices, i.e. $A_{(ab)} = \frac12(A_{ab} + A_{ba})$.}
\ie{}
[M_\alpha^{\:\: \beta}, P_{\gamma \delta}] &= \delta_\gamma^\beta P_{\alpha \delta} + \delta^\beta_\delta P_{\alpha \gamma} - \delta^\beta_\alpha P_{\gamma\delta} \,, \\
[M_\alpha^{\:\: \beta}, K^{\gamma \delta}] &= -\delta_\alpha^\gamma K^{\beta\delta} -\delta_\alpha^\delta K^{\gamma\beta} + \delta_\alpha^\beta K^{\gamma \delta} \,, \\
[M_\alpha^{\:\: \beta}, M_\gamma^{\:\: \delta}] &= -\delta_\alpha^{\delta} M_\gamma^{\:\: \beta} + \delta_\gamma^{\beta} M_\alpha^{\:\: \delta} \,, \quad [D,P_{\alpha\beta}] = P_{\alpha\beta}, \quad [D,K^{\alpha\beta}] = -K^{\alpha \beta}\,, \\
[K^{\alpha\beta}, P_{\alpha\beta}] &= 4\delta_{(\gamma}^{(\alpha} M_{\delta)}^{\:\: \beta)} + 4 \delta_{(\gamma}^{\alpha} \delta_{\delta)}^{\beta} D \,,
\fe
The $\so(\cN)_R$ R-symmetry generators $R_{IJ} = -R_{JI}$ satisfy
\ie{}
[R_{IJ}, R_{KL}] = \delta_{IK} R_{JL} + \delta_{JL} R_{IK} - \delta_{JK} R_{IL} - \delta_{IL} R_{JK} \,.
\fe
The odd generators $Q_{I\alpha}$ and $S^\alpha_I$ obey the anti-commutation relations
\ie{}
\{Q_{I\alpha}, Q_{J\beta} \} &= \delta_{IJ} P_{\alpha \beta} \,, \\
\{S^\alpha_I, S^\beta_J \} &= - \delta_{IJ} K^{\alpha \beta} \,, \\
\{Q_{I\alpha}, S^\beta_J \} &= \delta_{IJ} (M_\alpha^{\:\: \beta} + \delta_\alpha^{\: \beta} D) + \tfrac12 \delta_\alpha^{\:\beta} R_{IJ} \,.
\fe
Finally, the remaining commutation relations between the bosonic and fermionic generators are given by
\ie{}\label{eq:ospEvenOdd}
&[K^{\alpha\beta}, Q_{I\gamma}]= \delta^{\:\alpha}_\gamma S_I^\beta + \delta^{\:\beta}_\gamma S^\alpha_I \,, 
&&[P_{\alpha\beta}, S^\gamma_i]= -\delta^{\:\gamma}_\alpha Q_{I\beta} - \delta^{\:\gamma}_\beta Q_{I\alpha} \,, \\
&[M_\alpha^{\:\:\beta}, Q_{I\gamma}] = \delta_\gamma^{\: \beta} Q_{I\alpha} - \tfrac12 \delta^{\:\beta}_\alpha Q_{I\gamma}\,,
&&[M_\alpha^{\:\:\beta}, S^\gamma_I] = -\delta_\alpha^{\: \gamma} S^\beta_I + \tfrac12 \delta^{\:\beta}_\alpha S^\gamma_I \,, \\
&[D,Q_{I\alpha}] = \tfrac12 Q_{I\alpha} \,, 
&&[D,S^\alpha_I] = -\tfrac12 S^\alpha_I \,, \\
&[R_{IJ},Q_{K\alpha}] = \delta_{IK} Q_{J\alpha} - \delta_{JK} Q_{I\alpha} \,, 
&&[R_{IJ},S^\alpha_{K}] = \delta_{IK} S^\alpha_{J} - \delta_{JK} S^\alpha_{I} \,,
\fe
The appropriate real form is specified by the Hermiticity properties of the generators, namely
\ie{}
D^\dagger = D\,, \quad (P_{\alpha\beta})^\dagger = K^{\alpha\beta}\,, \quad (M_\alpha^{\:\:\beta})^\dagger = M_\beta^{\:\:\alpha}\,, \quad (R_{IJ})^\dagger = R_{JI}\,, \quad (Q_{I\alpha})^\dagger = S^\alpha_I \,.
\fe

\subsubsection*{Maximal 1d Subalgebra}

The 3d superconformal algebra $\osp(\cN|4;\bR)$ admits a maximal 1d subalgebra $\su(1,1|N)\oplus\uu(1)_b$ with $N = \frac12 \cN_1 = \lfloor \frac12 \cN \rfloor$. For $N=2$, $\su(1,1|2)$ contains a nontrivial central ideal and admits an extension by its $\uu(1)_b$ outer-automorphism, and so we instead consider the subalgebra $\uu(1)_{\cR_2} \rtimes \psu(1,1|2) \rtimes \uu(1)_b$, where we have introduced the quotient $\psu(1,1|2) \simeq \su(1,1|2)/\uu(1)_{\cR_2}$. We can choose an embedding of the superalgebras such that the 1d conformal generators are given by
\ie{}
D\,, \quad P = P_{+-}\,, \quad K = K^{+-} \,.
\fe
The generators of the $\su(N)_\cR \subset \so(\cN)_R$ R-symmetry subalgebra satisfy 
\ie{}
[\cR_a^{\:\: b} , \cR_c^{\:\: d}] = \delta^b_c \cR_a^{\:\: d} - \delta^d_a \cR_c^{\:\:b} \,,
\fe
and are related to 3d generators as
\ie{}
\cR_a^{\:\:\ b} \equiv -\frac12 \left( R_{2a,2b} + R_{2a-1,2b-1} + i R_{2a,2b-1} + i R_{2b,2a-1} \right) - \frac{1}{N} \delta^b_a r \,.
\fe
Here, the $\uu(1)_r$ commutant of $\su(N)_\cR \subset \so(\cN)_R$ is given by
\ie{}
r = i \sum_{a=1}^{N} R_{2a-1,2a} \,.
\fe 
It follows that the commutant $\uu(1)_{\cR_N}$ of $\su(1,1) \oplus \su(N)_\cR \subset \su(1,1|N)$ is
\ie{}
\cR_N = 
	\frac{N}{N-2} \left( M_+^{\:\:+} - \frac{r}{N} \right)\,,
\fe
where for $N=2$ we should replace $\cR_N \to \frac{N}{N-2} \cR_N$ above and in all the formulas that follow. The residual supercharges organize into two independent sets, namely $\{\cQ_a, \cS_a\}$ and $\{\bar{\cQ}^a, \bar{\cS}^a\}$, which can be defined in terms of the 3d generators as
\ie{}
&\cQ_a \equiv Q_{+}^{2a-1} + i Q_{+}^{ 2a}\,, \quad &&\bar{\cQ}^a \equiv Q_{-}^{ 2a-1} - i Q_{-}^{ 2a}\,, \\
&\cS_a \equiv S^-_{2a-1} + i S^-_{2a}\,, \quad &&\bar\cS^a \equiv S^+_{2a-1} - i S^+_{2a}\,.
\fe
It is straightforward to show that
\ie{}
&\{{\cQ}_a, \bar{\cQ}^b\} = \delta^{\:b}_a P \,,
&&\{{\cS}_a, \bar{\cS}^b\} = -\delta^{\:b}_a K \,, \\
&\{{\cQ}_a, \bar{\cS}^b \} = \delta^b_a \left(D + \tfrac{N-2}{N} \cR_N \right) - \cR_a^{\:\:\: b}\,,
&&\{{\cS}_a,\bar{\cQ}^b \} = \delta^b_a \left(D - \tfrac{N-2}{N} \cR_N \right) + \cR_a^{\:\:\: b}\,.
\fe
The remaining algebraic relations are\footnote{Because $\cR_2$ is central for $\psu(1,1|2)$, the $\uu(1)_b$ factor acts nontrivially on the supercharges in such a way as to maintain the same complex conjugation properties as for $\su(1,1|N)$.}
\ie{}
&[\cR_a^{\:\: b}, \cQ_c]  = \delta^b_c \cQ_a - \tfrac1N \delta^b_a \cQ_c \,, 
&&[\cR_a^{\:\: b}, \cS_c]  = \delta^b_c \cS_a - \tfrac1N \delta^b_a \cS_c \,, \\
&[\cR_a^{\:\: b}, \bar\cQ^c]  = -\delta^c_a \bar{\cQ}^b + \tfrac1N \delta^b_a \bar\cQ^c \,, 
&&[\cR_a^{\:\: b}, \bar\cS^c]  = -\delta^c_a \bar{\cS}^b + \tfrac1N \delta^b_a \bar\cS^c \,, \\
&[\cR_N, \cQ_a] = \tfrac12 \cQ_a \,,  
&&[\cR_N, \cS_a] = \tfrac12 \cS_a \,, \\
&[\cR_N, \bar\cQ^a] = -\tfrac12 \cQ^a \,, 
&&[\cR_N, \bar\cS^a] = -\tfrac12 \bar\cS^a \,,
\fe
where $\cR_2$ is central (i.e. it commutes with all of the supercharges). The Hermiticity properties follow from 3d as
\ie{}
(\cR_{a}^{\:\:b})^\dagger = \cR_{b}^{\:\:a}\,, \quad \cR_N^\dagger = \cR_N \,, \quad (\cQ_{a})^\dagger = \bar{\cS}^a\,, \quad (\bar{\cQ}^a)^\dagger = \cS_a \,. 
\fe

\subsubsection*{Broken 3d Generators}

The set of broken charges in $\osp(\cN|4;\mR)/\su(1,1|N)$  that induce operators on the defect consists of the transverse translations
\ie{}
&{\mathbf P}_\perp \equiv P_{++} \,, \quad
&&\bar{{\mathbf P}}_\perp \equiv P_{--} \,,
\fe
with $M_+^{\:\:+} = \pm 1$, the broken supercharges  
\ie{}
{\mathbf{Q}}^a \equiv Q^{2a-1}_{+} - i Q_{+}^{2a}\,, \quad \bar{{\mathbf{Q}}}_a \equiv Q^{2a-1}_{-} + i Q_{-}^{2a}
\fe
transforming in the (anti-)fundamental of $\su(N)_\cR$ with $M_+^{\:\:+} = \pm \frac12$ and $r = \mp 1$, respectively, and the broken R-symmetry generators
\ie{}
&{\mathbf R}_{ab} \equiv R_{2a-1,2b-1} - R_{2a,2b} - i (R_{2a-1,2b} +  R_{2a, 2b-1} ) \,, \\
&\bar{{\mathbf R}}^{ab} \equiv R_{2a-1,2b-1} - R_{2a,2b} + i (R_{2a-1,2b} + R_{2a, 2b-1}) \,,
\fe
transforming in conjugate rank-2 antisymmetric representations of $\su(N)_\cR$ with $r = \mp 2$.  It follows that their $\uu(1)_{\cR_N}$ charges are
\ie{}\label{eq:suBrokenEven}
\cR_N[{\mathbf P}_\perp] = \frac{N}{N-2} \,, \quad
\cR_N[{\mathbf{Q}}^a] = \frac{N+2}{2(N-2)} \,, \quad
\cR_N[{\mathbf R}_{ab}] = \frac{2}{N-2} \,,
\fe
where the barred generators have opposite $\cR_N$ charges. For odd $\cN$, there are additional broken R-symmetries 
\ie{}
{\mathbf R}^a \equiv R_{\cN,2a-1} - i R_{\cN,2a} \,, \quad \bar{{\mathbf R}}_a \equiv R_{\cN,2a-1 } + i R_{\cN, 2a-1}
\fe
transforming in the (anti-)fundamental with $r = \mp 1$, as well as additional broken supercharges
\ie{}
{\mathbf Q}\equiv Q_{+\cN} \,, \quad {\bar{\mathbf Q}} \equiv Q_{- \cN} 
\fe
with $M_+^{\:\:+} = \pm \frac12$ and $r=0$. Their $\uu(1)_{\cR_N}$ charges are given by
\ie{}\label{eq:suBrokenOdd}
\cR_N[{\mathbf R}^a] = \frac{1}{N-2}\,, \quad \cR_N[{\mathbf Q}] = \frac{N}{2(N-2)} \,,
\fe
where again the generators with $(+) \leftrightarrow (-)$ have opposite $\uu(1)_{\cR_N}$ charges. For $N=2$, we rescale everything in \eqref{eq:suBrokenEven} and \eqref{eq:suBrokenOdd} by $\frac{N-2}{N}$ such that
\ie{}
&\cR_2[{\mathbf P}_\perp] = 1 \,, \quad
\cR_2[{\mathbf{Q}}^a] =  1\,, \quad
\cR_2[{\mathbf R}_{ab}] = 1 \,, \\
&\qquad\quad\cR_2[{\mathbf R}^a] = \tfrac12 \,, \quad \cR_2[{\mathbf Q}] = \tfrac12 \,.
\fe

\subsubsection*{1d Superconformal Subalgebras}

There is a family of maximal 1d subalgebras $\su(1,1|m)\oplus\su(N-m)_F\oplus\uu(1)_b$ of $\su(1,1|N)$, where the 1d R-symmetry of $\su(1,1|m)$ is given by $\su(m)_{\cR} \oplus \uu(1)_{\cR_m}$. The subalgebra is specified by partitioning the $\su(N)_{\cR}$ fundamental indices into two sets, namely ${\cal S}_{\cR} =  \{1, \ldots, m\}$ and $\cS_F = \{m+1, \ldots, N\}$. The new R-symmetry generators and supercharges follow from restricting the indices of $\cR_a^{\:\: b}$  and the supercharges to lie in $\cS_{\cR}$, while the flavor symmetry generators are given by restricting the indices of $\cR_a^{\:\: b}$ to $\cS_F$, with the provision that we remove the trace from the two sets of bosonic generators. This trace generates the commutant $\uu(1)_r$ of $\su(m)\oplus\su(N-m) \subset \su(N)$ with charge $r_N$. We work in the conventions where the fundamentals decompose as
\ie{}\label{eq:suFund}
&\su(4) \supset \su(3) \oplus \uu(1)_r: &&(1,0,0) \to (1,0)_1 \oplus (0)_{-3} \,, \\
&\su(4) \supset \su(2) \oplus \su(2) \oplus \uu(1)_r: &&(1,0,0) \to ((1),(0))_1 \oplus ((0),(1))_{-1} \,, \\
&\su(3) \supset \su(2) \oplus \uu(1)_r: &&(1,0) \to (1)_1 \oplus (0)_{-2} \,, \\
&\su(2) \supset \uu(1)_r &&(1) \to \pm 1 \,.
\fe
The $\uu(1)$ charges are then related as 
\ie{}\label{eq:suSubCharges}
\cR_m &= \frac{1}{N} \left( (N-2)\cR_N - \frac{N-m}{m}  r_m \right) \,, \quad (m \ge 2)\\
\cR_1 &= - \frac12 \cR_2 \,.
\fe
The R-symmetry generators decompose under $\su(m)_{\cR} \oplus \su(N-m)_F \oplus \uu(1)_r$ as  
\ie{}
&\su(4) \supset \su(3) \oplus \uu(1)_r: &&(1,0,1) \to (1,1)_0 \oplus (0,0)_0 \oplus (1,0)_4 \oplus (0,1)_{-4} \,, \\
&\su(4) \supset \su(2) \oplus \su(2) \oplus \uu(1)_r: &&(1,0,1) \to(2,0)_0 \oplus (0,2)_0 \oplus (0,0)_0 \oplus (1,1)_{\pm 2}\,, \\
&\su(3) \supset \su(2) \oplus \uu(1)_r: &&(1,1) \to (2)_0 \oplus (0)_0 \oplus (1)_{\pm 3} \,, \\
&\su(2) \supset \uu(1)_r &&(2) \to 0 \oplus \pm 2 \,.
\fe
The quantum numbers of the preserved and broken supercharged follow from \eqref{eq:suFund} and \eqref{eq:suSubCharges}, and are listed in Table~\ref{table:su(1,1|n)superCharge}.
\begin{table}[htbp]
\footnotesize
\renewcommand{\arraystretch}{1.5}
\begin{center}
\begin{tabular}{|c|c|c|c|c||c|c|c|||c|c||c|c||c|c|}
\hline
$\cQ^{(N)}_a$ & $R_1$ & $R_2$ & $R_3$ & $\cR_N$ & $r_1$ & $r_2$ & $r_3$ & $\cR_1$ & $\su(1,1|1)$ & $\cR_2$ & $\psu(1,1|2)$ & $\cR_3$ & $\su(1,1|3)$ \\  \hline\hline 
$\cQ^{(2)}_{1}$ & $+1$ & & & $0$ & $+1$ & & & ${\color{red} - \frac12}$ & & & & & \\
$\cQ^{(2)}_{2}$ & $-1$ & & &  $0$ & $-1$ & & & $+\frac12$  & $\cQ^{(1)}$ & &  & &\\ \hline \hline
$\cQ^{(3)}_{1}$ & $+1$ & $0$ & & $+\frac12$ & $+1$ & $+1$ & & ${\color{red} - \frac12}$  & & $0$ & $\cQ^{(2)}_1$ & &\\
$\cQ^{(3)}_{2}$ & $-1$ & $+1$ & &  $+\frac12$ & $+1$ & $+1$ & & ${\color{red} - \frac12}$   & & $0$ & $\cQ^{(2)}_2$ & & \\
$\cQ^{(3)}_{3}$ & $0$ & $-1$ & &  $+\frac12$ & $-2$ & $-2$ & & $+ \frac12$  & $\cQ^{(1)}$ & ${\color{red} + \frac12}$  & & & \\ \hline \hline
$\cQ^{(4)}_{1}$ & $+1$ & $0$ & $0$ & $+\frac12$ & $+1$ & $+1$ & $+1$ & ${\color{red} - \frac12}$ & & $0$ & $\cQ^{(2)}_1$ & $+\frac12$ & $\cQ_1^{(3)}$ \\
$\cQ^{(4)}_{2}$ & $-1$ & $+1$ & $0$ &  $+\frac12$ & $+1$ & $+1$ & $+1$ & ${\color{red} - \frac12}$   & & $0$ & $\cQ^{(2)}_2$ & $+\frac12$ & $\cQ_2^{(3)}$\\
$\cQ^{(4)}_{3}$ & $0$ & $-1$ & $+1$ &  $+\frac12$ & $+1$ & $-1$ & $+1$ & ${\color{red} - \frac12}$  & & ${\color{red} + \frac12}$ & & $+\frac12$ & $\cQ_3^{(3)}$ \\
$\cQ^{(4)}_{4}$ & $0$ & $0$ & $-1$ &  $+\frac12$ & $-3$ & $-1$ & $-3$ & $+\frac12$ & $\cQ^{(1)}$ & ${\color{red} + \frac12}$ & & ${\color{red} + \frac32}$ &   \\ \hline 
\end{tabular}
\caption{Decomposition of the $\su(1,1|N)$ supercharges under $\su(1,1|m)$. Here, the $R_i$ function as weights both for $\su(N)$ as well as for $\su(m) \oplus \su(N - m)$. For the right half, black weights denote preserved supercharges (and are accompanied with the associated generators), while red weights denote broken supercharges.  The same decomposition holds for the charges $\bar{\cQ}^a$ given that we reverse the signs of all the weights.}
\label{table:su(1,1|n)superCharge}
\end{center}
\end{table}

Another family of maximal 1d superconformal subalgebras is given by $\osp(N|2;\bR) \subset \su(1,1|N)$, where the bosonic part $\so(N)_\cR$ is maximal in $\su(N)_{\cR}$. The R-symmetry generators decompose under the branching rules for the adjoint,
\ie{}
(1,0,\dots,0,1) \to (2,0,\dots,0) \oplus (0,1,0,\dots,0)  \,,
\fe
where in particular the $\su(N)/\so(N)$ generators transform in the $(2,0,\dots,0)$, i.e. the rank 2 traceless symmetric representation. The preserved (broken) supercharges transform in the vector representation and are given by real linear combinations of the conjugate supercharges in $\su(1,1|N)$.

\subsection{$\osp(4^*|2N)$}
\label{sec:osp4star2NAlgebra}

The maximal 1d superconformal subalgebra in 4d is given by $\osp(4^*|2N) \subset \su(2,2|\cN)$ where $N = \frac14 \cN_1 = \lfloor \frac12 \cN \rfloor$. Its maximal bosonic subalgbera is isomorphic to $\so(2,1)_{\text{conf}} \oplus \su(2)_{\text{rot}} \oplus \usp(2N)_{\cR}$. The 1d R-symmetry algebra consists of $\su(2)_{\text{rot}}$ transverse rotations, generated by $\cM_\alpha^{\:\:\: \beta}$ with $\alpha,\beta = \pm$, and $\usp(2N)_{\cR} \subset \su(\cN)_R$, generated by $\cR_{ab}$. Altogether the bosonic generators satisfy
\ie{}
&[\cM_\alpha^{\:\:\beta}, \cM_\gamma^{\:\:\delta}] =  \delta^\beta_\gamma \cM_\alpha^{\:\:\delta} - \delta_\alpha^\delta \cM_\gamma^{\:\:\beta} \,,\\
&[\cR_{ab},\cR_{cd}] = \Omega_{ac} \cR_{bd} + \Omega_{bc}\cR_{ad} + \Omega_{ad} \cR_{bc} + \Omega_{bd} \cR_{ac} \,,
\fe
where $\Omega_{ab} = -\Omega_{ba}$ is the usual $\usp(2N)$ symplectic form.\footnote{For $N=1$ we take $\Omega_{ab} = \epsilon_{ab}$ to explicitly establish the isomorphism $\usp(2) \simeq \su(2)$, and for $N=2$ we take the nonzero entries of the symplectic form to be $\Omega_{14} = \Omega_{23} = 1$.} There are $8N$ odd generators, which separate into $4N$ Poincar\`e supercharges $\cQ_{a\alpha}$ and $4N$ superconformal charges $\cS^\alpha_a$. They anti-commute among one another as
\ie{}
\{\cQ_{a\alpha}\,, \cQ_{b\beta}\} &= \Omega_{ab} \epsilon_{\alpha\beta} P \,,\\
\{\cS^\alpha_a\,, \cS^\beta_b\} &= \Omega_{ab} \epsilon^{\alpha\beta} K \,,\\
\{\cQ_{a \alpha}, \cS^\beta_b\} &= \Omega_{ab}(\cM_\alpha^{\:\:\beta} + \delta^\beta_\alpha D) + \delta_\alpha^\beta \cR_{ab} \,,
\fe
and transform under the bosonic symmetries as 
\ie{}
&[D, \cQ_{a\alpha}] = \tfrac12 \cQ_{a\alpha}\,, &&[D, \cS^\alpha_a] = -\tfrac12 S^\alpha_a\,, \\
&[K,  \cQ_{a \alpha}] =  \cS^\alpha_a\,, &&[P, \cS^\alpha_a] = - \cQ_{a\alpha}\,, \\
&[\cM_\alpha^{\:\: \beta}, \cQ_{a\gamma}] = \delta_\gamma^\beta \cQ_{a\alpha} - \tfrac12 \delta^\beta_\alpha \cQ_{a\gamma}\,, &&[\cM_\alpha^{\:\: \beta}, \cS^\gamma_a] = -\delta^\gamma_\alpha \cS^\beta_a + \tfrac12 \delta^\beta_\alpha \cS_a^\gamma \,,  \\
&[\cR_{ab}, \cQ_{c\alpha}] = \Omega_{ac} \cQ_{b\alpha} + \Omega_{bc} \cQ_{a\alpha} \,, &&[\cR_{ab}, \cS^\alpha_c] = \Omega_{ac} \cS^\alpha_b+ \Omega_{bc} \cS^\alpha_a \,.
\fe
Their Hermiticity properties are given by
\ie{}
(\cR_{ab})^\dagger = \cR_{ba} \,, \quad (\cQ_{a\alpha})^\dagger = \Omega_{ab} \cS^\alpha_b \,.
\fe

\subsubsection*{Broken 4d Generators}

The broken $\su(2,2|\cN)/\osp(4^*|2N)$ generators relevant for the modified Ward identities consist of transverse translations, broken supercharges, and broken R-symmetries. Since there are so few cases, we will handle each bulk superalgebra with $\cN = 2,3,4$ individually. 

We first consider $\cN=2$ with bulk symmetry $\su(2,2|2)$. The transverse translations transform in the $\mathbf{3} = (2)$ of $\su(2)_{\text{rot}}$ and are neutral under the residual bulk R-symmetries. The $\su(2)_R \simeq \usp(2)_\cR$ R-symmetry is preserved, while $\uu(1)_R$ is broken. The associated generator is neutral under all of the 1d bosonic symmetries. The broken supercharges transform in the $(1)$ of $\su(2)_{\text{rot}}$, which follows directly from the branching rules of $\su(2)_{\text{rot}} \subset \so(1,3) \simeq \mf{sl}(2,\bC)$, namely
\ie{}
(1,1) \to 2(1) \,.
\fe
Here, the other instance of $(1)$ corresponds to the preserved supercharges. 

Next we consider $\cN=3$, whose bulk symmetry $\su(2,2|3)$ is broken to $\osp(4^*|2)$. The broken generators include the broken $\cN=2$ currents, as well as any $\cN=3$ currents which do not belong to the $\cN=2$ bulk symmetry (either R-symmetry or flavor symmetry). In particular, $\su(3)_R$ is broken to $\su(2)_R$, and so there are additional broken currents whose transformation properties follow from the $\su(3)_R \supset \su(2)_R \oplus \uu(1)_R$ branching rules for the adjoint
\ie{}
(1,1) \to (0)_0 \oplus (1)_{\pm 3} \oplus (2)_0 \,.
\fe 
That is, the broken currents transform as the $(0)$  and the $2(1)$ under $\usp(2)_\cR$. Analogously, the additional broken $\cN=3$ supercharges transform in the $2(1)$ of $\su(2)_{\text{rot}}$ and are neutral under $\usp(2)_\cR$.  

At last we consider $\cN=4$, whose bulk symmetry $\psu(2,2|4)$ is broken to $\osp(4^*|4)$. The transverse translation generators transform identically to the other cases. The R-symmetry $\su(4)_R$ is broken to $\usp(4)_\cR$. Given the $\su(4)_R \supset \usp(4)_\cR$ branching rules for the adjoint
\ie{}
(1,0,1) \to (0,1) \oplus (2,0)
\fe
it follows that the broken R-symmetry generators transform in the $(0,1)$ under $\usp(4)_\cR$. The broken supercharges again transform in the $(1)$ of $\su(2)_{\text{rot}} \subset \so(1,3)$, but now transform under $\usp(4)_\cR$ as the $(1,0)$, which can be easily seen from the decomposition of the $\su(4)_R$ (anti-)fundamental.

\subsubsection*{1d Superconformal Subalgebras}

The $N=1$ superalgebra $\osp(4^*|2)$ has a maximal subalgebra $\su(1,1|1) \oplus \uu(1)_b$. Its R-symmetry subalgebra $\uu(1)_{\cR_1}$, generated by $\cR_1$, is a Cartan element in the diagonal $\su(2)$ subalgebra of $\su(2)_{\text{rot}}\oplus \su(2)_\cR$, where $\su(2)_\cR \simeq \usp(2)_\cR$. Given the respective $\su(2)_{\text{rot}}$ and $\su(2)_{\cR}$ Cartans, $\cM$ and $\cR$, with $\pm 1$ eigenvalues acting on the doublet representation, we have that
\ie{}\label{eq:SU(1,1|1)inOSP(4star|2)}
\cR_1 = \cR - \tfrac12\cM \,.
\fe
The transformation properties of the $\osp(4^*|2)$ supercharges under $\su(1,1|1)$ are given in Table~\ref{table:osp(4star|2)superCharge}.
\begin{table}[htbp]
\renewcommand{\arraystretch}{1.5}
\begin{center}
\begin{tabular}{|c|c|c||c|c|}
\hline
$\cQ_{a\alpha}$ & $\cM$ & $\cR$ & $\cR_1$ & $\su(1,1|1)$\\  \hline\hline 
$\cQ_{1+}$ & $+$ & $+$ &  $+\frac12$ & $\cQ$\\
$\cQ_{2+}$ & $+$ & $-$ & ${\color{red} - \frac32}$ & \\
$\cQ_{1-}$ & $-$ & $+$ & ${\color{red} + \frac32}$ & \\
$\cQ_{2-}$ & $-$ & $-$  & $-\frac12$ & $\bar{\cQ}$ \\ \hline
\end{tabular}
\caption{Decomposition of the $\osp(4^*|2)$ supercharges under $\su(1,1|1)$. In the right half, black weights denote preserved supercharges, while red weights denote broken supercharges.}
\label{table:osp(4star|2)superCharge}
\end{center}
\end{table}

The $N=2$ superalgebra $\osp(4^*|4)$ has a maximal subalgebra $\uu(1)_{\cR_2} \rtimes \psu(1,1|2) \rtimes \uu(1)_b$. Its R-symmetry subalgebra is $\su(2)_{\cR} \oplus \uu(1)_{\cR_2} \oplus \uu(1)_b$. The bosonic algebra $\su(2)_\cR \oplus \uu(1)_r$ is maximal in $\usp(4)_\cR$, which we take to be generated by $\{\cR_\pm, \cR, r\}$,
\ie{}
\cR_+ = \cR_{12} \,, \quad \cR_- = \cR_{34}\,, \quad \cR = -(\cR_{14} + \cR_{23}) \,, \quad r =  \cR_{14} - \cR_{23} \,.
\fe
It follows that the $\uu(1)_{\cR_2}$ R-charge can be written as
\ie{}
\cR_2 = \tfrac12(r + \cM) \,.
\fe
The transformation properties of the $\osp(4^*|4)$ supercharges under $\uu(1)_{\cR_2} \rtimes \psu(1,1|2) \rtimes \uu(1)_b$ are given in Table~\ref{table:osp(4star|4)superCharge}.
\begin{table}[htbp]
\renewcommand{\arraystretch}{1.5}
\begin{center}
\begin{tabular}{|c|c|c||c|c|c|}
\hline
$\cQ$ & $\cM$ & $r$ & $\cR$ & $\cR_2$ & $\psu(1,1|2)$ \\  \hline\hline 
$\cQ_{1+}$ & $+$ & $+$ & ${\color{red}+}$ &  ${\color{red}+1}$ & \\
$\cQ_{2+}$ & $+$  & $-$  & $+$ & $0$ & $\cQ_1$ \\
$\cQ_{3+}$ & $+$ &  $+$ & $-$ & $0$ & $\cQ_2$\\
$\cQ_{4+}$ & $+$  & $-$  & ${\color{red}-}$ &${\color{red}+1}$ & \\ \hline \hline
$\cQ_{1-}$ & $-$ &   $+$ & $+$ & $0$ & $\bar{\cQ}^2$ \\
$\cQ_{2-}$ & $-$ &  $-$ & ${\color{red}+}$ &  ${\color{red}-1}$ &  \\
$\cQ_{3-}$ & $-$ &  $+$ &  $-$ &  $0$ & $\bar{\cQ}^1$\\
$\cQ_{4-}$ & $-$ &  $-$ & ${\color{red}-}$ & ${\color{red}-1}$ & \\ \hline
\end{tabular}
\caption{Decomposition of the $\osp(4^*|4)$ supercharges under $\uu(1)_{\cR_2} \rtimes \psu(1,1|2) \rtimes \uu(1)_b$ (by convention, all charges are integers). In the right half, black weights denote preserved supercharges, while red weights denote broken supercharges.}
\label{table:osp(4star|4)superCharge}
\end{center}
\end{table}

\subsection{$D(2,1;\lambda,0)$}
\label{sec:D21lambdaAlgebra}

The maximal 1d superconformal subalgebra in 5d is given by $D(2,1;\lambda,0)\oplus \su(2)_{\text{right}} \subset F(4;2)$ with $\lambda=2$, but to keep the discussion sufficiently general we allow $\lambda \in \bR$ and drop the $\su(2)_{\text{right}}$ factor. The superalgebra $D(2,1;\lambda,0)$ can be viewed as a deformation of $\osp(4^*|2)$ (as well as $\osp(4|2;\bR)$), where we take the $\su(2)_{\text{left}} \oplus \su(2)_{\cR}$ R-symmetry to be generated by $\cM_\alpha^{\:\: \beta}$ and $\cR_a^{\:\: b}$, respectively, where $\alpha = \pm$ and $a,b=1,2$. The R-symmetry generators satisfy
\ie{}
&[\cM_\alpha^{\:\:\beta}, \cM_\gamma^{\:\:\delta}] = \delta^\beta_\gamma \cM_\alpha^{\:\:\delta} - \delta_\alpha^\delta \cM_\gamma^{\:\:\beta} \,,\\
&[\cR_a^{\:\:b},\cR_c^{\:\: d}] = \delta^b_c \cR_a^{\:\: d} - \delta^d_a \cR_c^{\:\: b} \,.
\fe
There are eight odd generators, which separate into the Poincar\`e supercharges $\cQ_{a\alpha}$ and superconformal charges $\cS^{a\alpha}$. They anti-commute among one another as\footnote{The complex superalgebra $D(2,1;\lambda)$ with $\lambda \in \bC$ is isomorphic to $D(2,1,\lambda')$ for nontrivial $\lambda' \in \{\lambda^{-1}, -(1+\lambda), -\lambda(1+\lambda)^{-1}\}$. Its real form $D(2,1;\lambda;0)$ with $\lambda \in \bR$ is isomorphic to $D(2,1;\lambda';0)$, but only for a single nontrivial choice of $\lambda'$, which depends in particular on which of the three $\msl(2,\bC)$ factors in the bosonic part of $D(2,1;\lambda)$ is selected as the conformal algebra $\msl(2,\bR)$ (the other two then have real forms given by $ \su(2) \oplus \su(2) \simeq \so(4)$). In our choice of conventions, the two algebras are isomorphic for $\lambda' = -(1+\lambda)$ (see e.g. \cite{Ivanov:2002pc}). This is contrast to those of \cite{DHoker:2008wvd}, which specify the real form $D(2,1;c;0)$ to be invariant under $c \to c^{-1}$. Their choice of parameter is related to ours by $c=-\lambda(1+\lambda)^{-1}$, as expected.}
\ie{}\label{eq:D21lambdaRelations}
\{\cQ_{a\alpha}\,, \cQ_{b\beta}\} &=  \epsilon_{ab} \epsilon_{\alpha\beta} P \,,\\
\{\cS^{a\alpha}\,, \cS^{b\beta}\} &=  -\epsilon^{ab} \epsilon^{\alpha\beta}{} K \,,\\
\{\cQ_{a \alpha}, \cS^{b\beta}\} &= \delta_a^b(\lambda \cM_\alpha^{\:\: \beta} + \delta^\beta_\alpha D) - (1+\lambda) \cR_a^{\:\: b} \,,
\fe
and transform under the bosonic symmetries as 
\ie{}
&[D, \cQ_{a\alpha}] = \tfrac12 \cQ_{a\alpha}\,, &&[D, \cS^{a\alpha}] = -\tfrac12 \cS^{a\alpha}\,, \\
&[K,  \cQ_{a \alpha}] =  \cS^{a\alpha}\,, &&[P, \cS^{a\alpha}] = - \cQ_{a\alpha}\,, \\
&[\cM_\alpha^{\:\: \beta}, \cQ_{a\gamma}] = \delta_\gamma^\beta \cQ_{a\alpha} - \tfrac12 \delta^\beta_\alpha \cQ_{a\gamma}\,, &&[\cM_\alpha^{\:\: \beta}, \cS^{a\gamma}] = -\delta^\gamma_\alpha \cS^{a\beta} + \tfrac12 \delta^\beta_\alpha \cS^{a\gamma} \,,  \\
&[\cR_a^{\:\: b}, \cQ_{c\alpha}] = \delta^b_c \cQ_{a\alpha}- \tfrac12 \delta^b_a \cQ_{c\alpha} ,  &&[\cR_a^{\:\: b}, \cS^{c\alpha}] = -\delta^c_a \cS^{b\alpha} + \tfrac12 \delta^b_a \cS^{c \alpha} \,.
\fe
The relevant Hermiticity properties are
\ie{}
(\cR_a^{\:\: b})^\dagger = \cR_b^{\:\:a} \,, \quad (\cQ_{a\alpha})^\dagger =  \cS^{a\alpha}\,.
\fe
A few comments are in order:
\begin{itemize}

    \item The superconformal algebra $D(2,1;\lambda,0)$ is isomorphic to $D(2,1;\lambda',0)$ for $\lambda' = -(1+\lambda)$, where the map interchanges $\su(2)_{\text{left}}$ and $\su(2)_\cR$. The isomorphism reduces to a $\bZ_2$ outer-automorphism of $D(2,1;\lambda;0) \simeq \osp(4|2;\bR)$ for $\lambda = -\frac12$. We may therefore restrict to values of $\lambda \ge -\frac12$ without loss of generality.
    
    \item There are several exceptional isomorphisms for certain values of $\lambda$ that follow from the definition of $D(2,1;\lambda)$ in terms of $\osp(4|2;\bC)$, namely  $D(2,1;0;0) \simeq \psu(1,1|2) \rtimes \su(2)$, where the $\su(2)$ acts as an outer-automorphism, and  $D(2,1;1;0) \simeq \osp(4^*|2)$. This can be verified explicitly from \eqref{eq:D21lambdaRelations}.
    
    \item The unitarity bounds depend \textit{discontinuously} on the value of $\lambda$. This can already been seen as a consequence of unitarity on generic multiplets. Consider some state $\ket{\Delta, j,r}$ in the SCP of a long multiplet of $D(2,1;\lambda;0)$, i.e.
    \ie{}
    \cS^{a\alpha} \ket{\Delta, j,R} = \ket{\Delta, j,R} = 0 \,,
    \fe
    where $\cM_-$ and $\cR_-$ are the standard raising operators of $\su(2)_{\text{left}}$ and $\su(2)_{\cR}$, respectively. Here, 
    $\Delta > 0$ labels the state's eigenvalue under $D$ and $j,R   \in \bZ_{\ge 0}$ label its eigenvalues under the $\su(2)_{\text{left}}$ and $\su(2)_{\cR}$ Cartans, respectively.  Unitarity requires that the inner products be positive-definite, i.e. 
    \ie{}\label{eq:innerProd}
    \left\langle\Delta,j',R'\left|\left\{\cQ_{a \alpha}, {\cQ^\dagger}^{b\beta}\right\}\right|\Delta,j,R\right\rangle > 0 \,,
    \fe 
    where the anti-commutator is given by
    \ie{}
    \left\{\cQ_{a \alpha}, {\cQ^\dagger}^{b\beta}\right\} &= \delta_a^b \delta^\beta_\alpha D + \lambda \delta_a^b \cM_\alpha^{\:\:\beta} - (1+\lambda) \delta^\beta_\alpha \cR_a^{\:\: b} \,.
    \fe
    Next, we use the the trick of \cite{Minwalla:1997ka} to relate the the eigenvalues of the $\su(2)_{\rm left}$ generator  acting on $|\Delta,j,R\ra$  to its quadratic Casimir invariants,
     \ie{}\label{eq:quadCasimir}
    \frac12 \left[C_2(J) - C_2(j) - C_2(1) \right] \,,
    \fe 
    where $C_2(j) = \frac12 j(j+2)$ and $(J)$ labels an irrep in the tensor product $(j) \otimes (1)$, with an analogous result for the $\su(2)_R$ generators. Consequently \eqref{eq:innerProd} implies a lower bound on $\Delta$ by the $\su(2)_{\rm rot}$ and $\su(2)_R$ quantum numbers,
    \ie{}
    \Delta 
    &> \max\bigg[-\tfrac12 \lambda j, \: \lambda(1+ \tfrac12 j)\bigg]+  \max\bigg[ \tfrac12 (1+\lambda)R, \: -(1+\lambda)(1+ \tfrac12 R)\bigg] \,.
    \fe
    The unitarity bounds which result clearly depend on the sign of $\lambda \in [-\frac12, \infty)$, with
    \ie{}\label{eq:D21lambdaClasses}
    \Delta > \begin{cases}
    \frac{\lambda}{2} j + \frac{1+\lambda}{2} R \,, & -\frac12 \leq  \lambda < 0 \\
    \frac{1+\lambda}{2} R \,, & \lambda = 0 \\
        \lambda (1+\tfrac12 j) + \frac{1+\lambda}{2} R \,, & \lambda > 0 \\
    \end{cases}
    \fe
These three cases incorporate the well-known examples of $\osp(4|2;\bR)$ ($\lambda =-\frac12$), $\psu(1,1|2) \rtimes \mf{su}(2)$ ($\lambda = 0$), and $\osp(4^*|2)$ ($\lambda =1$).
In particular, $D(2,1;2;0)$ falls into the third class, and so has a multiplet structure most similar to $\osp(4^*|2)$.\footnote{The representations of the (global part of) the large 2d $\cN=4$ superconformal algebra, as considered in \cite{deBoer:1999gea,Gukov:2004ym}, naturally belong to the first class.}
\end{itemize}

\subsubsection*{Broken 5d Generators}

The broken $F(4;2)/D(2,1;2;0)$ generators relevant for the modified Ward identities consist of transverse translations and broken supercharges (the full 5d R-symmetry is preserved). The transverse translations transform in the $\mathbf{4} = (1,1)$ of $\su(2)_{\text{left}} \oplus \su(2)_{\text{right}}$ and are neutral under $\su(2)_\cR$, as expected. The broken supercharges transform as the $(1)$ under $\su(2)_\cR$ and as the $(0,1)$ under $\su(2)_{\text{left}} \oplus \su(2)_{\text{right}}$, which follows from the branching rules of $\su(2)_{\text{left}} \oplus \su(2)_{\text{right}} \subset \so(1,4)$, namely
\ie{}
(0,1) \to (1,0) \oplus (0,1) \,.
\fe
Here, the preserved supercharges transform as the $(1,0)$ factor in the RHS above.

\subsubsection*{1d Superconformal Subalgebras}

The superalgebra $D(2,1;\lambda;0)$ has a maximal 1d superconformal subalgebra $\su(1,1|1) \oplus \uu(1)_b$. Its R-symmetry subalgebra $\uu(1)_{\cR_1}$, generated by $\cR_1$, is a Cartan element in the diagonal subalgebra of $\su(2)_{\text{left}} \oplus \su(2)_\cR$. Given the respective $\su(2)_{\text{left}}$ and $\su(2)_{\cR}$ Cartans $\cM$ and $\cR$, with $\pm 1$ eigenvalues acting on the doublet representation, we have that
\ie{}\label{eq:SU(1,1|1)inD(2,1;lambda;0)}
\cR_1 =  \tfrac{1+\lambda}{2} \cR - \tfrac{\lambda}{2} \cM \,.
\fe
The transformation properties of the $D(2,1;\lambda;0)$ supercharges under $\uu(1)_{\cR_2} \rtimes \psu(1,1|2) \rtimes \uu(1)_b$ are given in Table~\ref{table:D(2,1;lambda;0)superCharge}.
\begin{table}[htbp]
\renewcommand{\arraystretch}{1.5}
\begin{center}
\begin{tabular}{|c|c|c||c|c|}
\hline
$\cQ_{a\alpha}$ & $\cM$ & $\cR$ & $\cR_1$ & $\su(1,1|1)$\\  \hline\hline 
$\cQ_{1+}$ & $+$ & $+$ &  $+\frac12$ & $\cQ$ \\
$\cQ_{2+}$ & $+$ & $-$ & ${\color{red}-(\lambda+\frac12)}$ & \\
$\cQ_{1-}$ & $-$ & $+$ & ${\color{red}\lambda+\frac12}$ & \\
$\cQ_{2-}$ & $-$ & $-$  & $-\frac12$ & $\bar{\cQ}$ \\ \hline
\end{tabular}
\caption{Decomposition of the $D(2,1;\lambda;0)$ supercharges under $\uu(1)_{\cR_2} \rtimes \psu(1,1|2) \rtimes \uu(1)_b$. In the right half, black weights denote preserved supercharges, while red weights denote broken supercharges.}
\label{table:D(2,1;lambda;0)superCharge}
\end{center}
\end{table}

\section{Forbidden Superconformal Lines}
\label{sec:forbiddenLines}

In this section, we discuss 1d superconformal subalgebras that cannot be realized as the symmetry of a unitary superconformal line defect. They appear in general as further subalgebras $\mf{g}'$ of the 1d maximal subalgebras in \eqref{eq:1dSCAs}. As we explain below, apart from the $\mf{osp}(1|2;\mR)$ symmetry of minimal superconformal lines, only the superconformal algebras $\mf{g}_s$ in \eqref{eq:1dSCAs} are admissible symmetries of superconformal lines in each spacetime dimensions.\footnote{In a rough sense, the superconformal lines prefer to be either maximal or minimal in terms of the 1d superconformal symmetries they preserve.}

Recall that due to the modified Ward identities in the bulk, each broken current induces certain distinguished local operators on the defect. In particular, a broken R-symmetry current $J^\mu_R$ leads to a marginal scalar CP $\mathsf{J}_R$ which transforms under the preserved defect subalgebra according to the associated branching rules. We are interested in the case where this R-symmetry current is preserved by a defect $\cD$ described by some maximal 1d subalgebra $\mf{g} \subset \mf{G}$ (with $\mf{g}=\mf{g}_s\oplus \mf{g}_b$ in \eqref{eq:1dSCAs}), but is broken by a putative defect $\cD'$ described by a subalgebra $\mf{g}' \subset \mf{g}$. By Theorem~\ref{thm4} of Section~\ref{sec:struc}, $\mathsf{J}_R$ must reside within a broken R-current multiplet with respect to $\mf{g}'$, either as a top component or one level below a top component. For the 1d superconformal subalgebras considered below, we find that their unitary multiplet structure, as described in Section~\ref{sec:ureps}, cannot accommodate such operators. It follows that no such defect $\cD'$ preserving $\mf{g}'$ (but not $\mf{g}$) can exist. By extension, this also excludes  subalgebras of SCFTs with larger amounts of supersymmetry, i.e. $\mf{g}' \subset \mf{g} \subset \mf{G} \subset \mf{G}'$, where  $\mf{G}'$ can be associated with a higher spacetime dimension than $\mf{G}$. In what follows, we provide explicit arguments for the independent cases which are colored in blue in \eqref{forbidcases}  (and rely on the previous statement for the rest). In summary, the \textit{forbidden line defect algebras} are: 
\ie\label{forbidcases}
&d=3: &&\begin{cases} {\color{blue}\osp(3|2;\bR)\subset \su(1,1|3) \subset \osp(6|4;\bR)} \,, \\
{\color{blue}\osp(4|2;\bR)  \subset \su(1,1|4) \subset \osp(8|4;\bR) }\,, \\
\osp(3|2;\bR) \subset \su(1,1|4)\subset \osp(8|4;\bR) \,, 
\end{cases} \\
&d=4: &&\begin{cases}
{\color{blue}\su(1,1|1) \subset \osp(4^*|2) \subset \su(2,2|2)} \,, \\
\su(1,1|1) \subset \osp(4^*|2) \subset \su(2,2|3) \,, \\
{\color{blue}\uu(1)\rtimes \psu(1,1|2) \rtimes \uu(1) \subset \osp(4^*|4) \subset \psu(2,2|4)} \,, \\
\su(1,1|1) \subset \osp(4^*|4) \subset \psu(2,2|4) \,,
\end{cases} \\
&d=5: &&\:\:\:\:\:{\color{blue}\su(1,1|1) \subset D(2,1;2;0) \subset F(4;2) }\,, \\
&d=6: &&\begin{cases}
{\color{blue}\osp(4^*|2) \subset \osp(8^*|4)} \,, \\
\su(1,1|1) \subset \osp(4^*|2) \subset \osp(8^*|4) \,,
\end{cases}
\fe
where we have suppressed commutants for clarity.\footnote{We emphasize that
while superconformal lines preserving these superconformal symmetries $\mf{g}'$ are forbidden as standalone defects, they can and in many cases do exist as superconformal lines inside a higher dimensional superconformal defect in the SCFT. One example is given by the supersymmetric flavor Wilson-'t Hooft lines on a half-BPS codimension-2 superconformal defect of the 6d $(2,0)$ SCFT \cite{Tachikawa:2011dz}, corresponding to the sequence of subalgebras $\osp(4^*|2) \subset \su(2,2|2)\subset \osp(8^*|4)$. 
}

\subsection{$d=3$, $\cN=6$}

Recall that $\su(1,1|3) \oplus \uu(1)_b$ is maximal in the 3d $\cN=6$ superconformal algebra $\osp(6|4;\bR)$, with bosonic subalgebra
\ie{}
\so(2,1)_\text{conf} \oplus \su(3)_{\cR} \oplus \uu(1)_{\cR_3} \oplus \uu(1)_b \,.
\fe
We now consider the 1d superconformal subalgebras of $\su(1,1|3) \oplus \uu(1)_b$.

\subsubsection*{\underline{$\osp(3|2;\bR)$}}

There is a maximal $\frac14$-BPS subalgebra $\osp(3|2;\bR) \oplus \uu(1)_b$, where its R-symmetry $\so(3)$ is maximal inside $\su(3)_{\cR}$. The $\su(3)_{\cR}$ currents decompose under $\so(3) \simeq \su(2)$ as
\ie{}
(1,1) \to (2) \oplus (4)\,,
\fe
where $\mathbf{3} = (2)$ is the rank-2 antisymmetric representation (adjoint) of $\so(3)$, and $\mathbf{5} = (4)$ is the rank-2 symmetric traceless representation (spin 2). The broken R-symmetry currents transform in the $(4)$ and therefore induce marginal scalars in the $[-]_1^{(4)}$ on the defect. Given that $\osp(3|2;\bR)$ does not have any unitary multiplets with such states, either as a top component or one level below a top component, line defects with $\osp(3|2;\bR)$ symmetry in 3d $\cN=6$ SCFTs are forbidden.

\subsection{$d=3$, $\cN=8$}

Recall that $\su(1,1|4) \oplus \uu(1)_b$ is maximal in the 3d $\cN=8$ superconformal algebra $\osp(8|4;\bR)$, with bosonic subalgebra
\ie{}
\so(2,1)_\text{conf} \oplus \su(4)_{\cR} \oplus \uu(1)_{\cR_4} \oplus \uu(1)_b \,.
\fe
We now consider the 1d superconformal subalgebras of $\su(1,1|4) \oplus \uu(1)_b$.

\subsubsection*{\underline{$\osp(4|2;\bR)$}}

There is a maximal $\frac14$-BPS subalgebra $\osp(4|2;\bR) \oplus \uu(1)_b$, where its R-symmetry $\so(4)$ is maximal inside $\su(4)_{\cR}$. The $\su(4)_{\cR}$ currents decompose under $\so(4) \simeq \su(2) \oplus \su(2)$ as
\ie{}\label{eq:su4toso4}
(1,0,1) \to (2,0) \oplus (0,2) \oplus (2,2)\,,
\fe
where $\mathbf{6} = (2,0)\oplus(0,2)$ is the rank-2 antisymmetric tensor (adjoint) of $\so(4)$, and $\mathbf{9} = (2,2)$ is the rank-2 traceless symmetric tensor (spin 2). The broken R-symmetry currents transform in the $(2,2)$ and therefore induce marginal scalars in the $[-]_1^{(2,2)}$ on the defect.  Given that $\osp(4|2;\bR)$ does not admit any unitary multiplets with such states, either as a top component or one level below a top component, line defects with $\osp(4|2;\bR)$ symmetry in 3d $\cN=8$ SCFTs are forbidden.

\subsection{$d=4$, $\cN=2$}

Recall that $\osp(4^*|2)$ is maximal in the 4d $\cN=2$ superconformal algebra $\su(2,2|2)$, with bosonic subalgebra
\ie{}
\so(2,1)_\text{conf} \oplus \su(2)_{\text{rot}} \oplus \su(2)_{\cR} \,.
\fe
We now consider the 1d superconformal subalgebras of $\osp(4^*|2)$.

\subsubsection*{\underline{$\su(1,1|1)$}}

There is a maximal $\frac14$-BPS subalgebra $\su(1,1|1) \oplus \uu(1)_b$ whose R-symmetry $\uu(1)_{\cR_1}$ sits inside $\text{diag}[\su(2)_{\text{rot}}, \su(2)_\cR]$. The precise relation between the $\uu(1)_{\cR_1}$ charge $\cR_1$ and the respective $\su(2)_{\text{rot}} \oplus \su(2)_\cR$ Cartans, which we take to be $\cM$ and $\cR$ (with $\pm1$ eigenvalues acting on the doublet representation), is given by
\ie{}
\cR_1 = \cR - \tfrac12 \cM \,.
\fe
From this, it follows that the broken $\su(2)_\cR$ R-symmetry currents (with $\cR = \pm 2$) induce marginal scalars in the $[\pm 2]_1$ on the defect. Given that $\su(1,1|1)$ does not admit any unitary multiplets containing such states, as all states obey $\Delta \ge |\cR_1|$, line defects with $\su(1,1|1)$ symmetry in 4d $\cN=2$ SCFTs are forbidden.

\subsection{$d=4$, $\cN=4$}

Recall that $\osp(4^*|4)$ is maximal in the 4d $\cN=4$ superconformal algebra $\psu(2,2|4)$, with bosonic subalgebra
\ie{}
\so(2,1)_\text{conf} \oplus \su(2)_{\text{rot}} \oplus \usp(4)_{\cR} \,.
\fe
We now consider the 1d superconformal subalgebras of $\osp(4^*|4)$.

\subsubsection*{\underline{$\uu(1) \rtimes \psu(1,1|2) \rtimes \uu(1)$}}

There is a maximal $\frac14$-BPS subalgebra $\uu(1)_{\cR_2} \rtimes \psu(1,1|2) \rtimes \uu(1)_b$ whose R-symmetry $\su(2)_{\cR} \oplus \uu(1)_{\cR_2} \oplus \uu(1)_b$ sits inside $\su(2)_{\text{rot}} \oplus \usp(4)_{\cR}$. The bosonic algebras are related as follows. The algebra $\usp(4)_{\cR}$ has a maximal subalgebra $\su(2)_{\cR} \oplus \uu(1)_r$. The $\uu(1)_{\cR_2}$ charge is taken to be
\ie{}
\cR_2 = \tfrac12(r + \cM) \,,
\fe
where $r \in \bZ$ is the $\uu(1)_r$ charge and $\cM$ is the $\su(2)_{\text{rot}}$ Cartan as in previous cases. The $\usp(4)_R$ currents decompose under $\su(2)_{\cR} \oplus \uu(1)_r$ as
\ie{}
(2,0) \to (0)_0 \oplus (2)_0 \oplus (2)_{\pm 2} \,.
\fe
The broken $\usp(4)_{\cR}$ R-symmetry currents transform in the $(2)_{\pm 2}$ and thus induce marginal scalars in the $[\pm 1]^{(2)}_1$ on the defect. Given that $\uu(1)_{\cR_2} \rtimes \psu(1,1|2) \rtimes \uu(1)_b$ does not admit any unitary multiplets containing such states, either as a top component or one level below a top component, line defects with $\uu(1)_{\cR_2} \rtimes \psu(1,1|2) \rtimes \uu(1)_b$ symmetry in 4d $\cN=4$ SCFTs are forbidden.
 
\subsection{$d=5$, $\cN=1$}

Recall that $D(2,1;2;0)\oplus \su(2)_\text{right}$ is maximal in the 5d $\cN=1$ superconformal algebra $F(4;2)$, with bosonic subalgebra
\ie{}
\so(2,1)_\text{conf} \oplus \so(4)_{\text{rot}} \oplus \su(2)_{\cR} \,,
\fe
where $\so(4)_{\text{rot}} \simeq \su(2)_{\text{left}} \oplus \su(2)_{\text{right}}$. We now consider the 1d superconformal subalgebras of $D(2,1;2;0)\oplus \su(2)_\text{right}$.

\subsubsection*{\underline{$\su(1,1|1)$}}

There is a maximal $\frac14$-BPS subalgebra $\su(1,1|1)\oplus \uu(1)_b \oplus \su(2)_\text{right}$ with R-symmetry $\uu(1)_{\cR_1}$, where $\uu(1)_{\cR_1}$ sits inside $\text{diag}[\su(2)_{\text{left}}, \su(2)_{\cR}]$, and is taken to be
\ie{}
\cR_1 = \tfrac32 \cR - \cM \,,
\fe
where $\cR$ and $\cM$ are the Cartans of $\su(2)_{\cR}$ and $\su(2)_{\text{left}}$ respectively as before. From this, it follows that the broken $\su(2)_\cR$ R-symmetry currents (with $\cR = \pm 2$) induce marginal scalars in the $[\pm 3]_1$ on the defect. Given that $\su(1,1|1)$ does not admit any unitary multiplets containing such states, as all states obey $\Delta \ge |\cR_1|$, line defects with $\su(1,1|1)$ symmetry in 5d $\cN=1$ SCFTs are forbidden.

\subsection{$d=6$, $\cN=(2,0)$}

\subsubsection*{\underline{$\osp(4^*|2)$}}

The 6d $\cN=(2,0)$ superconformal algebra $\osp(8^*|4)$ does not possess any maximal subalgebras describing the symmetries of superconformal lines, let alone any odd-dimensional superconformal defects. In this case, the best we can do is to consider a maximal 1d superconformal subalgebra which is maximal in some subalgebra of $\osp(8^*|4)$. There are two possibilities, namely $\osp(4^*|2)\oplus \uu(1)_b$ in $\su(2,2|2)\oplus\uu(1)_b$ and the diagonal subalgebra of $\osp(4^*|2) \oplus \osp(4^*|2)$. Both are $\frac14$-BPS, while the former is a maximal 1d subalgebra. Its bosonic part consists of
\ie{}
\so(2,1)_\text{conf} \oplus \su(2)_{\text{rot}} \oplus \su(2)_{\cR} \oplus \uu(1)_b \,.
\fe
Let us briefly describe the relations among the various bosonic algebras. We have that $\su(2)_{\cR} \oplus \uu(1)_r$ is maximal in the 6d R-symmetry $\usp(4)_R$, where $\uu(1)_b$ is a combination of $\uu(1)_r$ and the \textit{four-dimensional} transverse rotation symmetry of $\bR^{1,3} \subset \bR^{1,5}$. This is in contrast to $\su(2)_{\text{rot}} \simeq \so(3)$, which is the one-dimensional transverse rotation symmetry of $\bR^{1} \subset \bR^{1,3}$. In any case, the $\usp(4)_R$ R-symmetry currents decompose under $\su(2)_{\cR} \oplus \uu(1)_r$ as
\ie{}
(2,0) \to (0)_0 \oplus (2)_0 \oplus (2)_{\pm 2} \,.
\fe
Part of the broken $\usp(4)_R$ currents, as given by $(2)_{\pm 2}$, induce marginal scalars $[0]_1^{(2)}\oplus [0]_1^{(2)}$ on the defect. Given that $\osp(4^*|2)$ does not admit any unitary multiplets containing such states, either as a top component or one level below a top component, line defects with $\osp(4^*|2)$ symmetry in 6d $\cN=(2,0)$ SCFTs are forbidden.

\bibliographystyle{JHEP}
\bibliography{defREF}

\end{document}